\documentclass[fleqn,usenatbib]{mnras}

\usepackage{newtxtext,newtxmath}
\usepackage{caption, subcaption}
\usepackage[T1]{fontenc}
\usepackage{ae,aecompl}

\usepackage{graphicx}	% Including figure files
\usepackage{amsmath}	% Advanced maths commands
\usepackage[normalem]{ulem}
\usepackage{lastpage}
\usepackage{soul}
\usepackage{lastpage}

\newcommand{\Lya}[0]{Ly$\alpha$ }

\newcommand{\sm}[1]{_{\text{#1}}}
\newcommand{\Ghomog}[0]{{\bf{G}}$\sm{homog}$}
\newcommand{\Grecov}[0]{{\bf{G}}$\sm{patchy}$}

    \newcommand{\krange}{$k\leq$0.03 km$^{-1}$ s}  
        \newcommand{\kms}{km$^{-1}$ s}  
 \newcommand{\ditto}{{\tt "}}
 \newcommand{\VpecDx}{\partial v\sm{pec}/\partial v\sm{H}} %% technically this is a partial derivative
 
 \newcommand{\HI}{\hbox{H$\,\rm \scriptstyle I\ $}}
  \newcommand{\HII}{\hbox{H$\,\rm \scriptstyle II\ $}}
 \newcommand{\HeII}{\hbox{He$\,\rm \scriptstyle II\ $}}
 \newcommand{\HeI}{\hbox{He$\,\rm \scriptstyle I\ $}}
 
\title[]{The effect of inhomogeneous reionisation on the Lyman-$\alpha$ forest power spectrum at redshift ${\boldmath z>4}$: implications for thermal parameter recovery}

\author[M. Molaro et al.] {Margherita Molaro$^{1}$\thanks{E-mail:
    margherita.molaro@nottingham.ac.uk}, Vid Ir\v{s}i\v{c}$^{2}$, James S. Bolton$^{1}$, Laura C. Keating$^{3}$, Ewald Puchwein$^{3}$, \newauthor Prakash Gaikwad$^{2}$, Martin G. Haehnelt$^{2}$, Girish Kulkarni$^{4}$ \& Matteo Viel$^{5,6,7,8}$
 \\$^1$School of Physics and Astronomy, University of Nottingham, University Park, Nottingham, NG7 2RD, UK
 \\$^{2}$Kavli Institute for Cosmology and Institute of Astronomy, Madingley Road, Cambridge, CB3 0HA, UK
 \\$^{3}$Leibniz-Institut f\"ur Astrophysik Potsdam, An der Sternwarte 16, 14482 Potsdam, Germany\\$^{4}$Tata Institute of Fundamental Research, Homi Bhabha Road, Mumbai 400005, India  \\$^{5}$SISSA - International School for Advanced Studies, Via Bonomea 265, I-34136 Trieste, Italy
  \\$^{6}$IFPU, Institute for Fundamental Physics of the Universe, Via Beirut 2, I-34151 Trieste, Italy
  \\$^{7}$INAF - Osservatorio Astronomico di Trieste, Via G.B. Tiepolo 11, I-34131 Trieste, Italy 
  \\$^{8}$INFN - National Institute for Nuclear Physics, Via Valerio 2, I-34127 Trieste, Italy}

\pubyear{2021}

\begin{document}
\label{firstpage}
\pagerange{\pageref{firstpage}--\pageref{lastpage}}
\maketitle

\begin{abstract} 
We use the Sherwood-Relics suite of hybrid hydrodynamical and radiative transfer simulations to model the effect of inhomogeneous reionisation on the 1D power spectrum of the \Lya forest transmitted flux at redshifts $4.2\leq z \leq 5$.  Relative to models that assume a homogeneous UV background, reionisation suppresses the power spectrum at small scales, $k \sim 0.1$ \kms{}, by $\sim 10$ per cent because of spatial variations in the thermal broadening kernel and the divergent peculiar velocity field associated with over-pressurised intergalactic gas. On larger scales, $k<0.03\rm\,km^{-1}\,s$,  the power spectrum is instead enhanced by $10$--$50$ per cent by large scale spatial variations in the neutral hydrogen fraction. The effect of inhomogeneous reionisation must therefore be accounted for in analyses of forthcoming high precision measurements.  We provide a correction for the \Lya forest power spectrum at $4.1\leq z \leq 5.4$ that can be easily applied within other parameter inference frameworks using similar reionisation models.  We perform a Bayesian analysis of mock data to assess the extent of systematic biases that may arise in measurements of the intergalactic medium if ignoring this correction. At the scales probed by current high resolution \Lya forest data at $z>4$, $0.006 \rm \,km^{-1}\,s\leq k \leq 0.2 \rm\, km^{-1}\,s$, we find inhomogeneous reionisation does not introduce any significant bias in thermal parameter recovery for the current measurement uncertainties of $\sim 10$ per cent.  However, for $5$ per cent uncertainties, $\sim 1\sigma$ shifts between the estimated and true parameters occur.   

\end{abstract}


%%%%%%%%%%%%%%%%%%%%%%%%%%%%%%%%%%%%%%%%%%%%%%%%%%%%%%%%%%%%%%%%%%%%%	
%%%%%%%%%%%%%%%%%%%%%%%%%% SECTION 1 %%%%%%%%%%%%%%%%%%%%%%%%%%%%%%%%
%%%%%%%%%%%%%%%%%%%%%%%%%%%%%%%%%%%%%%%%%%%%%%%%%%%%%%%%%%%%%%%%%%%%%

\section{Introduction}

The study of \Lya absorption features in the spectra of bright, high-redshift quasars offers a valuable insight into the physical properties of the intergalactic medium (IGM) in the early Universe \citep{Fan2006,Mortlock2011,Becker2015,Eilers2017,Bosman2018,Yang2020}. Cumulatively referred to as the ``\Lya forest'', these absorption features provide constraints on the ionisation and thermal state of the IGM during and immediately after the final stages of reionisation at redshifts $z\simeq 5-7$ \citep{Onorbe2017b,Kulkarni2019,Walther2019,Gaikwad2020,Keating2020,Nasir2020,Qin2021}.  Additionally, because the \Lya forest closely tracks the dark matter down to scales of $\sim 100\rm\,ckpc$, it is also sensitive to the suppression of the matter power spectrum on small scales \citep[see e.g.][]{Seljak2006,Viel2008,Boyarsky2009}. The power spectrum of the \Lya forest transmitted flux thus provides one of the tightest lower limits on the mass of a putative warm dark matter thermal relic \citep{Viel2013,Irsic2017,Garzilli2019,Palanque2020,Rogers2021}. 

Existing constraints on reionisation and warm dark matter from the \Lya forest typically rely on high quality quasar spectra combined with numerical models for the distribution of matter in the IGM \citep[see e.g.][]{McQuinn2016}.  On the observational side, the number of known quasars at $z>5$ has significantly increased in recent years \citep[e.g][]{Banados2016,Matsuoka2018,Reed2019,Wang2019}, and the latest measurements of the \Lya forest power spectrum exhibit improved precision and extend toward smaller scales \citep{Irsic2017_XQ100,Chabanier2019,Boera2019,Bayu2021,Naim2021}.  The computational demands on state-of-the-art simulations of the high redshift \Lya forest are, however, still formidable. The simulations must capture the patchy thermal and ionisation state of the IGM following reionisation, have a large dynamic range that resolves gas at the Jeans scale \citep{Theuns2000,BoltonBecker2009} while simultaneously sampling a volume large enough to follow the percolation of ionised bubbles \citep{Iliev2014}, and they must furthermore span a large and uncertain parameter space.   

One of the most expensive physical processes to implement is the prescription for the radiative transfer of UV photons during inhomogeneous reionisation \citep{Gnedin2000,Razoumov2002,Ciardi2003,Mellema2006,Finlator2018,Molaro2019}. A common approximation used in hydrodynamical simulations of the \Lya forest that side-steps this requirement are pre-computed, spatially homogeneous ultraviolet background (UVB) models \citep{Haardt2012,Puchwein2019,Khaire2019,FaucherGiguere2020}. While these ``UVB synthesis'' models provide an excellent description of the IGM ionisation state following reionisation, when the mean free path for Lyman continuum photons is $\gtrsim 100\rm\,cMpc$ \citep[see e.g.][]{Lukic2015,Bolton2017,Rossi2020, Villasenor2021}, they neglect the large-scale fluctuations in the ionisation and thermal state of the IGM that exist immediately following the completion of reionisation \citep{Becker2015,DAloisio2015,Davies2016,Chardin2017,Kulkarni2019}.   It has been suggested that ignoring these fluctuations could weaken existing \Lya forest constraints on the free streaming length of dark matter, leaving the door firmly open for alternatives to cold dark matter \citep[e.g.][]{Hui2017}.   

Further progress in this area therefore necessitates the development of efficient and accurate numerical schemes that capture the effect of inhomogeneous hydrogen reionisation on the \Lya forest at $z>4$. Ideally, any such scheme should also be straightforward to incorporate into existing \Lya forest power spectrum parameter estimation frameworks.  While significant progress has been made on this problem using hydrodynamically decoupled radiative transfer simulations \citep[e.g.][]{Cen2009,Keating2018,Daloisio2019} or semi-numerical reionisation models \citep[e.g.][]{Lidz2014,Montero2019}, these neglect the dynamical effect of patchy heating on the small scale structure of the \Lya forest (i.e. the scales at wavenumbers $k\sim 0.1\,\rm km^{-1}\,s$ that are most sensitive to the IGM thermal state and the coldness of dark matter at $z>4$).  Recent efforts toward addressing this deficiency have been presented by \citet{Onorbe2019} and \citet{Wu2019}, using independent approaches.   \citet{Onorbe2019} used a hybrid method that couples Eulerian hydrodynamical simulations performed with Nyx \citep{Almgren2013} with a semi-numerical reionisation model where energy is injected into the IGM by hand.     This hybrid approach has the advantage of speed and efficiency, but at the cost of using an approximate treatment for the photo-heating of the IGM.  By contrast, \citet{Wu2019} used AREPO-RT \citep{Kannan2019} to perform multi-frequency radiation hydrodynamical (RHD) simulations.  The RHD simulations self-consistently model photo-heated gas temperatures, but at the expense of increased computational cost. 

In this work, we complement these studies by adopting a third approach that is intermediate between semi-numerical models and full RHD simulations.  The simulations we use here are part of the Sherwood-Relics project (Puchwein et al. in prep),  a large scale set  of IGM simulations that directly builds upon our earlier Sherwood simulation project \citep{Bolton2017}.   In Sherwood-Relics, we model the effect of inhomogeneous reionisation on the high redshift \Lya forest using a hybrid approach that combines radiative transfer calculations performed using ATON \citep{Aubert2008, Aubert2010} with P-GADGET-3 cosmological hydrodynamical simulations \citep{Springel2005}.  By using a two step approach, where ionisation maps produced by empirically calibrated radiative transfer (RT) calculations are applied on-the-fly to the hydrodynamical simulations (see Section \ref{section::simulation} for details), we are able to capture the patchy ionisation and thermal state of the IGM \emph{and} self consistently model the hydrodynamical response of gas following inhomogeneous heating \citep[see also][for other recent applications of this approach]{Gaikwad2020,Soltinsky2021}.  We then use the Sherwood-Relics simulations to construct and test a generalised ``patchy reionisation'' correction to the \Lya forest power spectrum at $z>4$ predicted by (homogeneous UVB) hydrodynamical simulations.  A key advantage of this approach is that the correction can be straightforwardly applied to existing grids of hydrodynamical simulations used in cosmological parameter inference frameworks \citep[e.g.][]{Boera2019,Walther2019,Bird2019,Rossi2020}, thus avoiding the need to perform large numbers of additional RHD simulations.

This paper is structured as follows.  In Section~\ref{section::simulation} we describe the Sherwood-Relics simulations and introduce the reionisation models used in this work.  We examine the effect of inhomogeneous reionisation on the \Lya forest power spectrum at $4.2\leq z \leq 5$ in our fiducial reionisation model in Section~\ref{section:changes_Lya_forest}, and discuss the physical origin of the generic features we observe: an enhancement of power on large scales, $k<0.03\,\rm km^{-1}\,s$, and $\sim 10$ per cent suppression of power on small scales, $k>0.1\,\rm km^{-1}\,s$ \citep[cf.][]{Onorbe2019,Wu2019}.    In Section \ref{section::comparison_536067} we then expand our analysis to consider different reionisation histories and construct a generalised inhomogeneous reionisation correction to the \Lya forest power spectrum that can be applied to ``traditional'' hydrodynamical simulations of the IGM (for readers wishing to skip the details, this correction is implemented using Eq.~(\ref{eqn:best_fit}) and Table~\ref{table:tabulated_pk}, and a simple python script to compute this correction is available at \url{https://github.com/marghemolaro/RT_1dps_correction.git}). In Section \ref{section:MCMC} we then assess the importance of any biases that may be introduced to measurements of the IGM thermal state from the \Lya forest power spectrum by applying this correction within our existing Monte Carlo Markov Chain (MCMC) analysis framework \citep{Viel2013,Irsic2017}.   Finally, we conclude in Section \ref{section::conclusions}.  Supplementary information is provided in Appendix~\ref{app:example_spec}. 

\section[Simulations of the Lyman-alpha forest]{Simulations of the Lyman-$\alpha$ forest} \label{section::simulation}
\subsection{Hydrodynamical simulations of the IGM during reionisation} 

\begin{table*}
\centering
\caption{List of simulations used in this work. From left to right, the columns list the simulation name, the box size in $h^{-1}\rm\,cMpc$, the number of particles, the dark matter and gas particle mass in $h^{-1}\,M_{\odot}$, the redshift of reionisation (defined as the redshift when the volume averaged ionised fraction $1 - x\sm{HI} \leq 10^{-3}$), the Thomson scattering optical depth $\tau_{\rm e}$, the gas temperature at the mean density, $T_{0}$, the cumulative energy input per proton mass at the mean density, $u_{0}$, for $4.6\leq z \leq 13$ \citep[cf.][]{Boera2019}, and the method used for modelling the photo-ionisation of the IGM by UV photons.  The upper section of the table lists the models in the first set of simulations -- all performed using (variations of) the \citet{Puchwein2019} UV background synthesis model -- that we use for our MCMC analysis (see text for details). The lower section of the table lists our second set of simulations, which includes hybrid radiative transfer simulations and the corresponding ``paired'' homogeneous models that are matched to the thermal and reionisation histories in the hybrid-RT runs.  We do not quote $T_{0}$ or $u_{0}$ for the hybrid-RT models, as this quantity will vary spatially and depend on when any given gas element is reionised.  However, note that (by design) the \emph{average} thermal history will be very similar to the paired homogeneous simulations.}
\label{table:summary_sims}
\begin{tabular}{l|c|c|c|c|c|c|c|c|c}
\hline
Name     & $L_{\rm box}$ & $N_{\rm part}$ & $M_{\rm dm}$ & $M_{\rm gas}$  & $z\sm{R}$ &  $\tau_{\rm e}$ & $T_{0}(z=4.6)$ & $u_{0}(z=4.6)$ &  UVB model\\
 & $[h^{-1}\rm\,cMpc]$ & & $[h^{-1}M_{\odot}]$ & $[h^{-1}M_{\odot}]$ &  & & [$\rm K$] &  $[\rm eV\,m_{\rm p}^{-1}]$ &  \\
\hline 
20-1024      & 20.0 & $2\times 1024^{3}$ & $5.37\times10^{5}$ & $9.97\times10^{4}$  & 6.00 & 0.062 & 10066 & 7.7 & P19 \\
20-1024-zr54 & \ditto & \ditto &\ditto & \ditto &  5.37 & 0.055 & 10069 & 6.6 & Rescaled P19 \\
20-1024-zr67 & \ditto & \ditto &\ditto & \ditto &  6.70 & 0.071 & 10050 & 9.6 & \ditto\\
20-1024-zr74 & \ditto & \ditto &\ditto & \ditto &  7.40 & 0.079 & 10003 & 11.4 & \ditto\\
20-1024-cold & \ditto & \ditto &\ditto & \ditto &  5.98 & 0.062 & 6598 & 4.3 & \ditto\\
20-1024-zr54-cold & \ditto & \ditto &\ditto & \ditto &  5.35 & 0.055 & 6409 & 3.6 & \ditto\\
20-1024-zr67-cold & \ditto & \ditto & \ditto & \ditto &  6.69 & 0.070 & 6803 & 5.4 & \ditto\\
20-1024-zr74-cold & \ditto& \ditto & \ditto & \ditto &  7.39 & 0.079 & 6806 & 6.4 & \ditto\\
20-1024-hot & \ditto & \ditto &\ditto & \ditto &   6.01 & 0.063 & 13957 &14.4 & \ditto\\
20-1024-zr54-hot & \ditto &\ditto&  \ditto & \ditto & 5.38 & 0.055 & 13451 &12.5 & \ditto\\
20-1024-zr67-hot & \ditto &\ditto & \ditto & \ditto &  6.71 & 0.071 & 14369 & 17.8 & \ditto\\
20-1024-zr74-hot & \ditto&\ditto & \ditto & \ditto &  7.41 & 0.080 & 14624 & 21.1 & \ditto\\
40-2048  & 40.0 & $2\times 2048^{3}$ & \ditto & \ditto &  6.00 & 0.062 & 10063 & 7.7 &  P19 \\
\hline
RT-late   & 40.0 & $2\times 2048^{3}$ & $5.37\times10^{5}$ & $9.97\times10^{4}$ &  5.30 & 0.056 &  -- & --&  Hybrid-RT \\
RT-mid    & \ditto & \ditto &\ditto & \ditto &  5.99 & 0.057 & -- & -- & \ditto \\
RT-early  & \ditto & \ditto &\ditto & \ditto &  6.64 & 0.064 & --& -- & \ditto \\
Homog-late   & \ditto & \ditto & \ditto & \ditto & 5.30 & 0.054 & 10545 & 5.4 & Paired homogeneous\\
Homog-mid    & \ditto & \ditto &\ditto & \ditto &  5.99 & 0.055 & 10431 & 5.5 &  \ditto \\
Homog-early  & \ditto & \ditto &\ditto & \ditto & 6.64 & 0.062 & 10245 & 6.2 &  \ditto \\
\hline
\end{tabular}
\end{table*}

   \begin{figure*}
    \centering
    \includegraphics[trim=105 60 105 40,width=5.5cm]{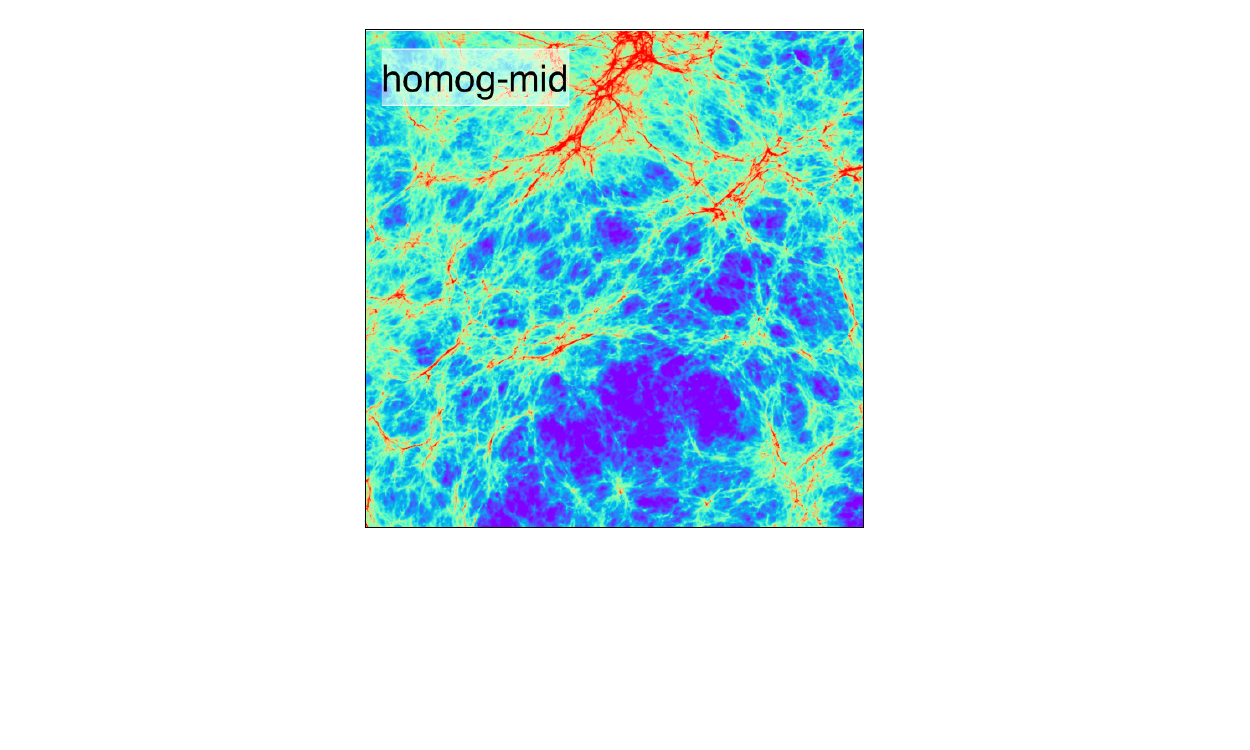}
    \includegraphics[trim=75 5 75 60, width=5.5cm]{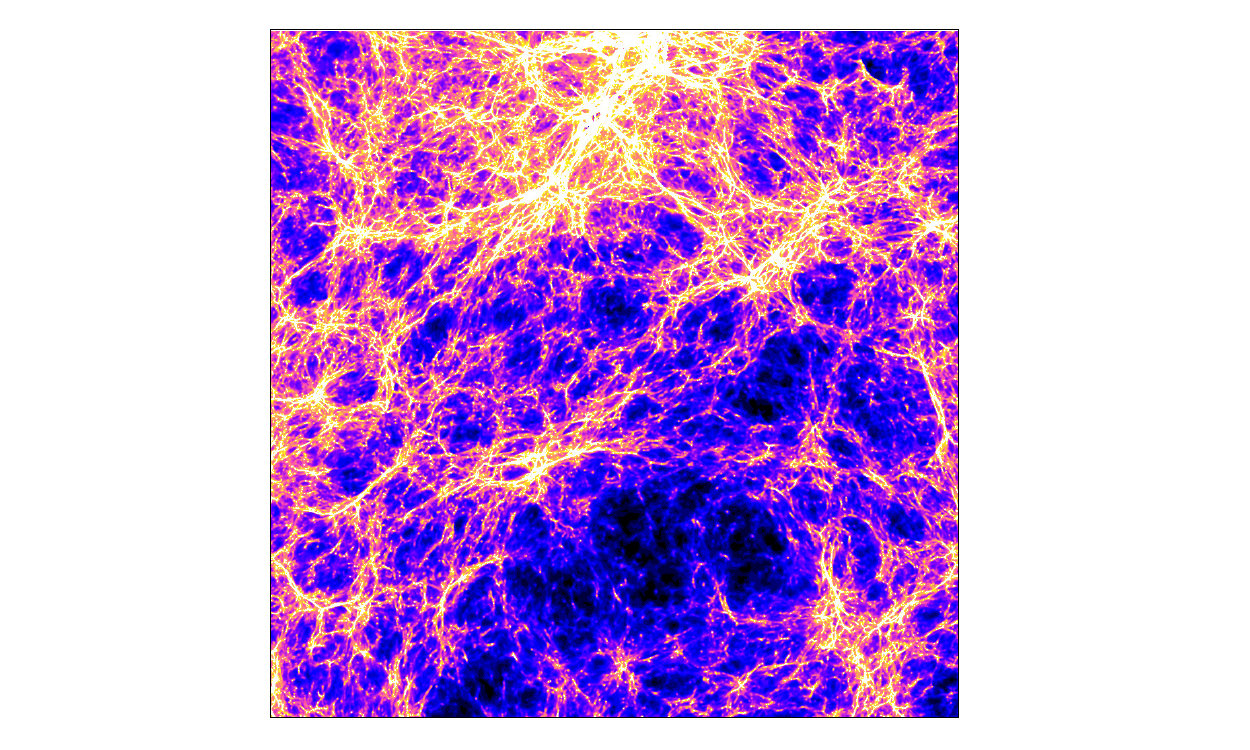}
    \includegraphics[trim=75 5 75 100,width=5.5cm]{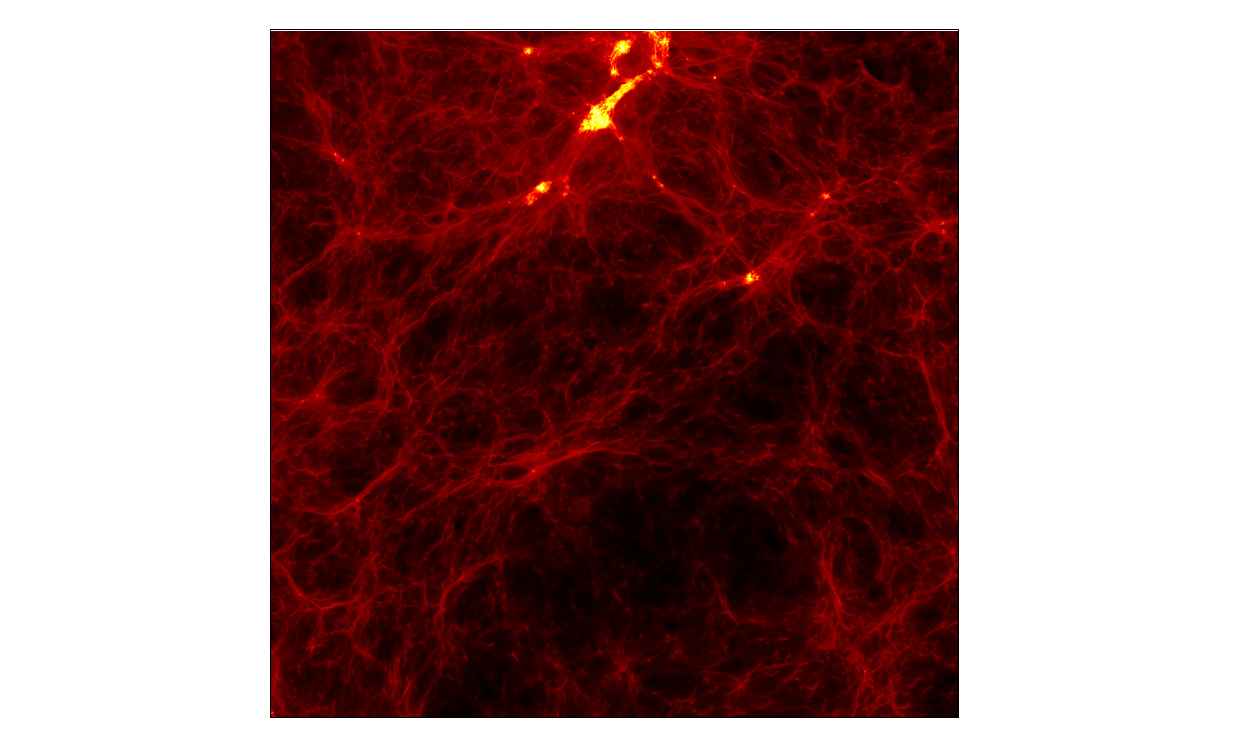}\\
    \includegraphics[trim=105 20 105 0,width=5.5cm]{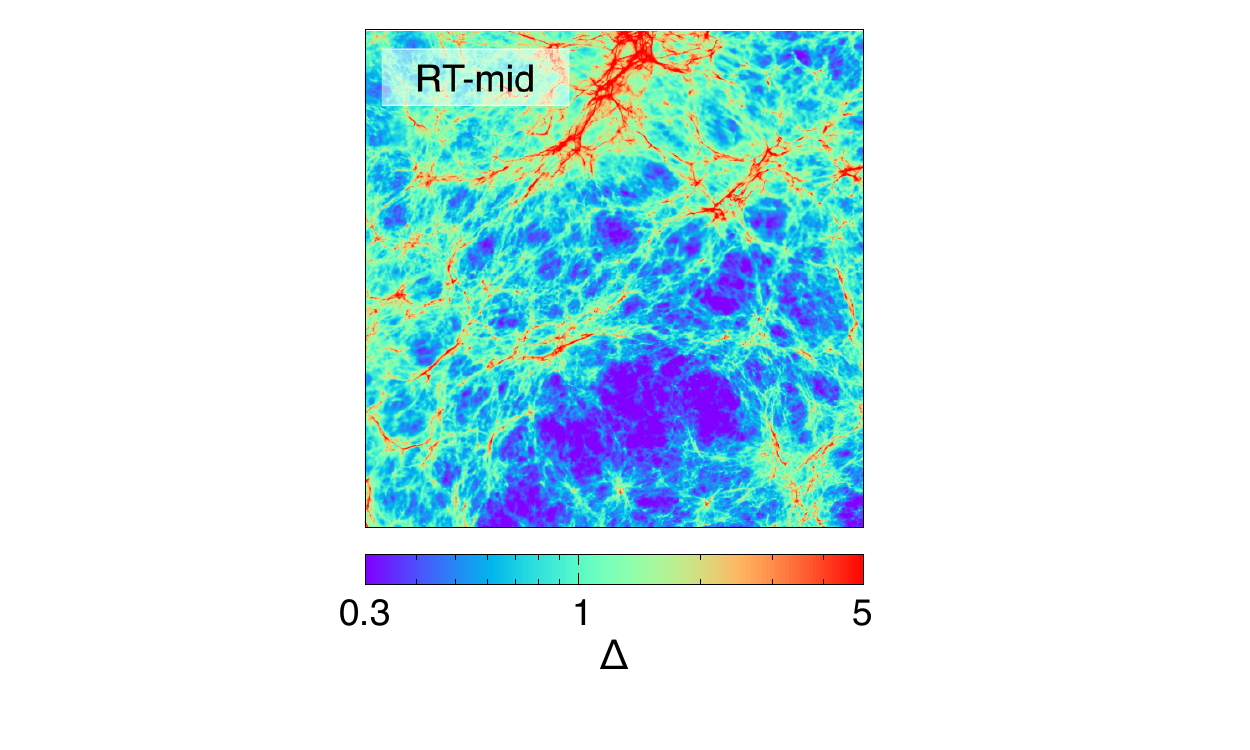}
    \includegraphics[trim=105 20 105 40,width=5.5cm]{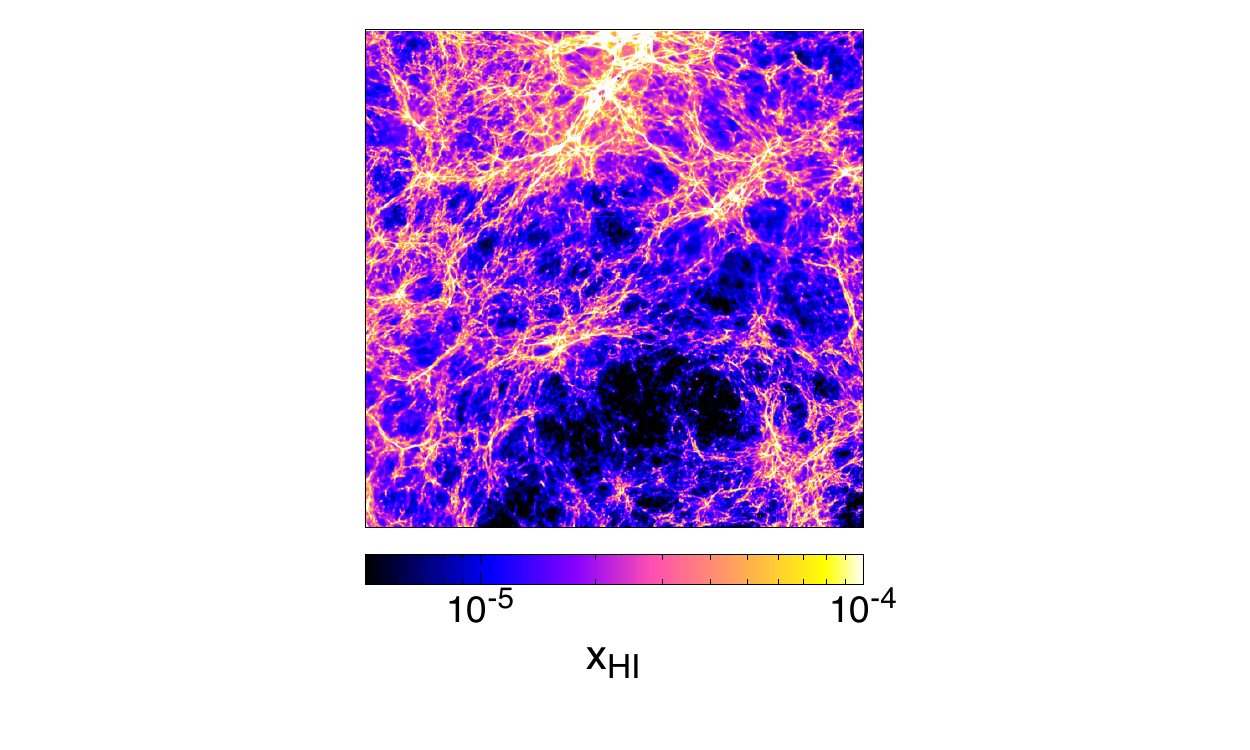}
    \includegraphics[trim=105 20 105 0,width=5.5cm]{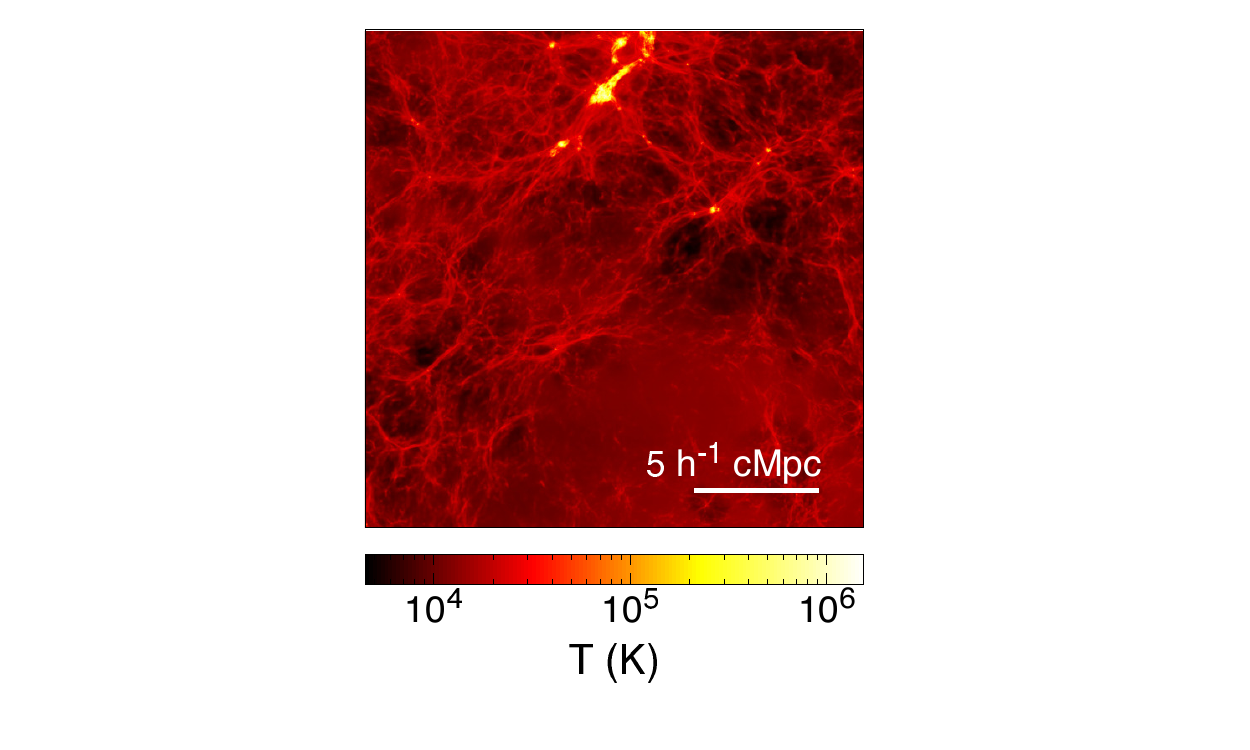}\\    
    \vspace{-0.2cm}
    \caption{The gas density $\Delta = \rho/\langle \rho \rangle$ \textit{(left column)}, neutral hydrogen fraction $x\sm{HI}$ \textit{(middle column)}, and gas temperature $T$ \textit{(right column)} in a $20h^{-2}\rm\,cMpc^{2}$ region of the Homog-mid \textit{(top row)} and RT-mid \textit{(bottom row)} simulation at $z=5.4$.  The slices have been projected over a slab of thickness $2h^{-1}\rm\,cMpc$.  Differences arising from patchy reionisation are particularly visible in the large void in the lower right corner of the slice, where the later reionisation and heating of the low density gas leads to a higher temperatures and lower neutral fractions in the hybrid-RT model.}
    \label{figure:projected_plane}
\end{figure*}

  \begin{figure*}
   \centering
      \includegraphics[width=7cm,trim = 50 0 40 0]{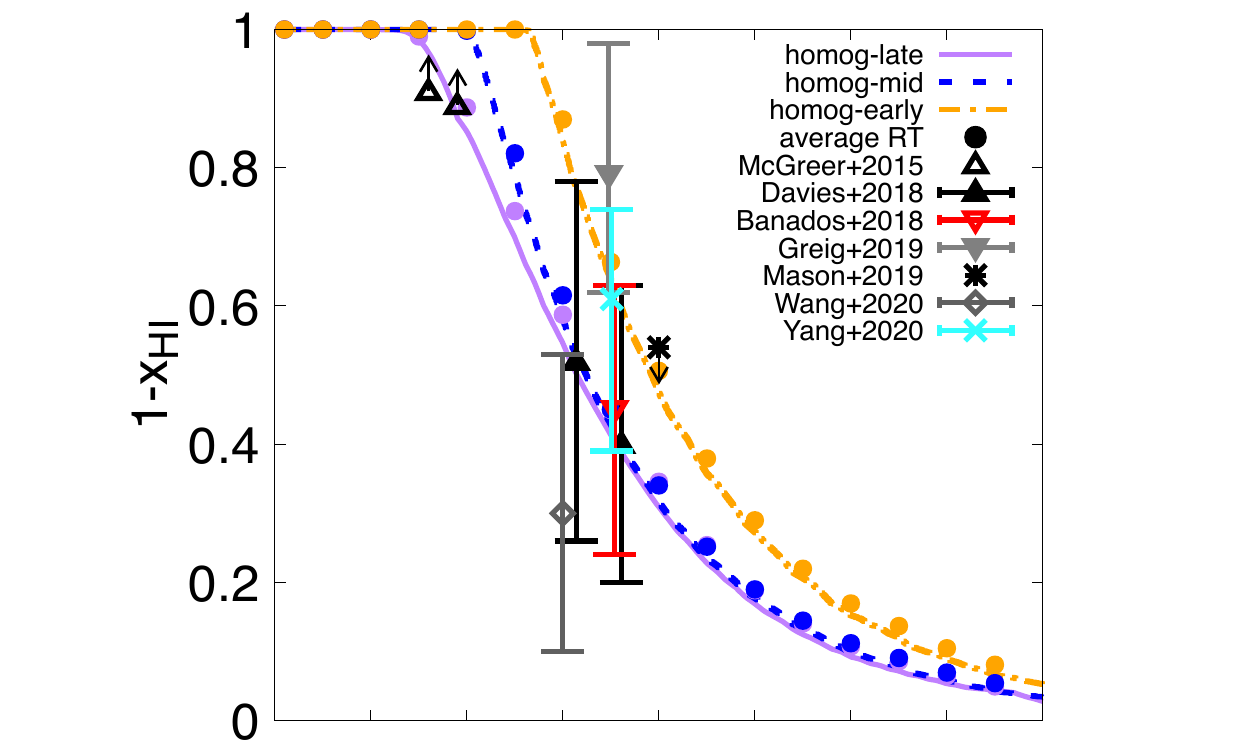} 
      \includegraphics[width=7cm,trim = 50 0 40 0]{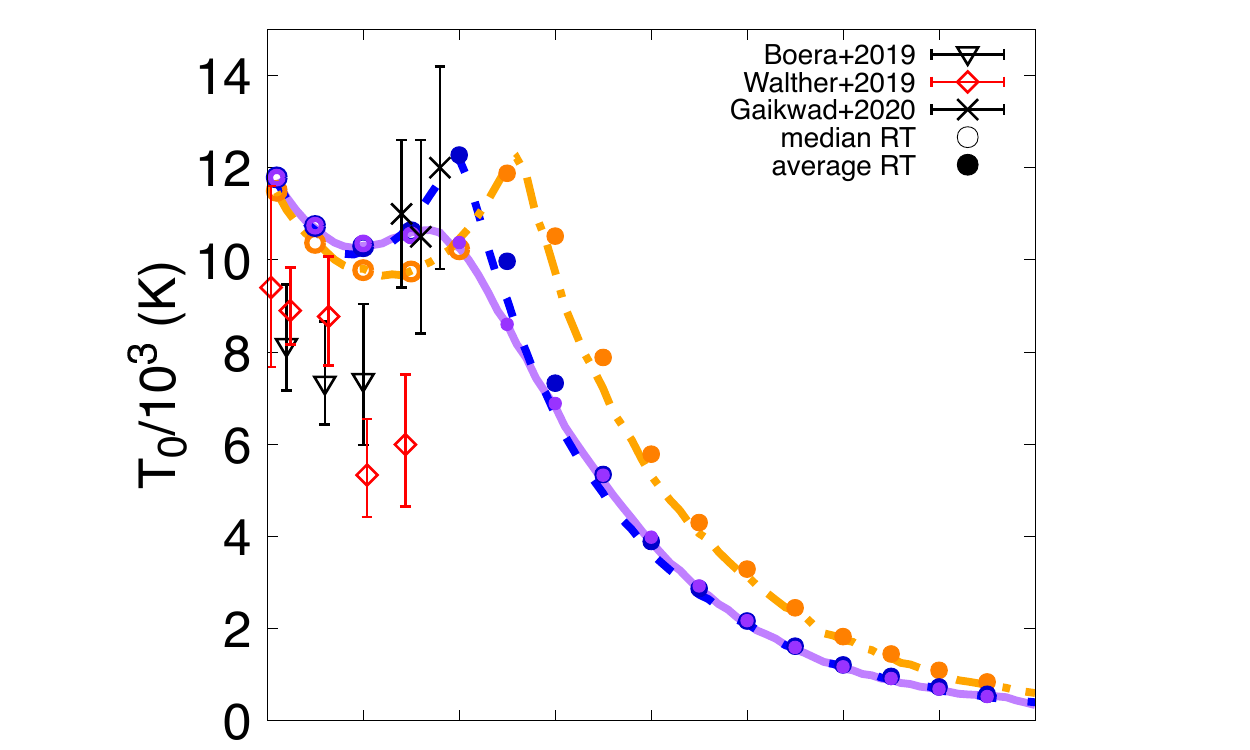}\\
      \includegraphics[width=7.5cm,trim = 45 0 50 0]{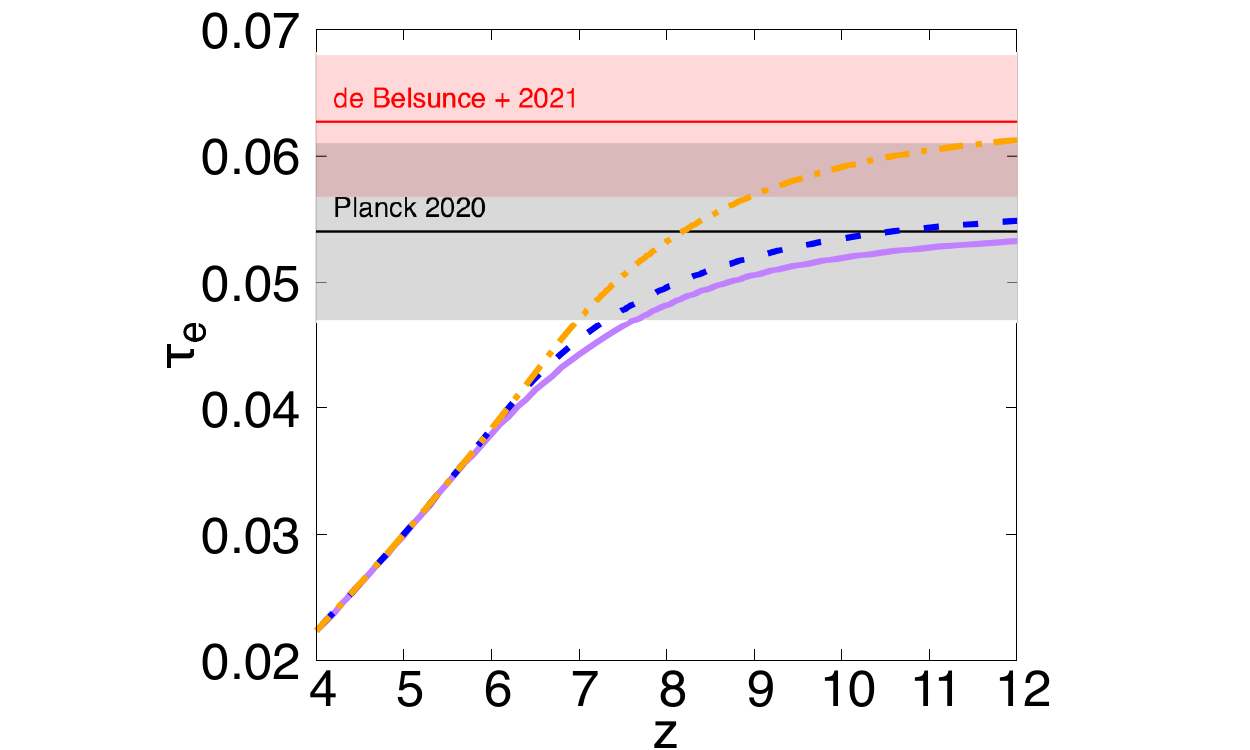} 
       \includegraphics[width=7.5cm,trim = 60 0 40 0]{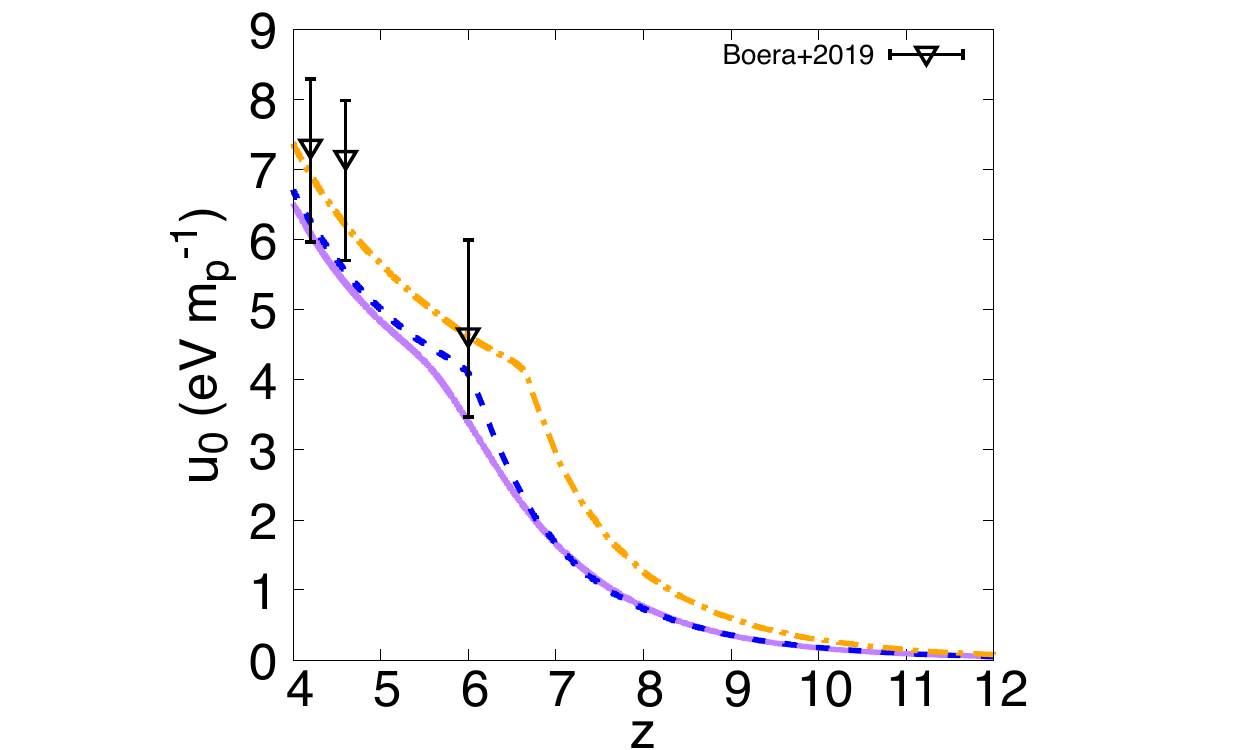} 
       \vspace{-0.1cm}
      \caption{Summary of the hybrid-RT/paired homogeneous reionisation models used in this work.  The curves in each panel display the redshift evolution in the homogeneous simulations for three different reionisation histories ending at $z\sm{R}=5.30$ (late, solid violet curve), $z\sm{R}=5.99$ (mid, dashed blue curve), and $z\sm{R}=6.64$ (early, dot-dashed yellow curve). \textit{Top left}: the volume averaged ionised fraction evolution, $1-x\sm{HI}$, shown against recent observational measurements from \citet{McGreer2015}, \citet{Banados2018}, \citet{Davies2018b}, \citet{Greig2019}, \citet{Mason2019}, \citet{Yang2020} and \citet{Wang2020}. The higher redshift data points from \citet{Greig2019} and \citet{Davies2018b} were shifted in redshift by -0.02 and +0.02 respectively for clarity. The filled circles show the average ionised fraction in the corresponding RT simulations -- note these match by design. \textit{Top right}: the temperature at mean density, $T\sm{0}$, shown against recent measurements from \citet{Boera2019},  \citet{Walther2019} (shifted by +0.04 in redshift for clarity), and \citet{Gaikwad2020}.  The filled (open) circles show the average (median) temperature from the corresponding RT simulations. \textit{Bottom left}:  the Thomson optical depth, $\tau\sm{e}$, shown against the \citet{Planck2020} and the \citet{deBelsunce2021} constraints. \textit{Bottom right}: the cumulative energy per proton mass at the mean density, $u\sm{0}$, shown against the observational estimates from \citet{Boera2019}.}
      \label{figure:global_props}
   \end{figure*}

 We use simulations drawn from the Sherwood-Relics project (see \citet{Gaikwad2020}, \citet{Soltinsky2021} and Puchwein et al. in prep.).  These are a series of high-resolution cosmological hydrodynamical simulations that use a customised version of \textsc{P-GADGET-3} (see \citet{Springel2005} for the original \textsc{GADGET-2} reference).    We use cosmological boxes of size $20h^{-1}$cMpc or $40h^{-1}$cMpc with $2\times1024^3$ or $2\times2048^3$ dark matter and gas particles. The box size and mass resolution have been chosen to adequately resolve the small scale structure that contributes to the power spectrum of the \Lya forest transmitted flux at $z>4$, while still retaining a relatively large cosmological volume \citep{BoltonBecker2009,Lukic2015,Bolton2017}.  Note, however, that coherent ionised or neutral structures on scales larger than the box size will not be present in our simulations \citep{Iliev2014,Kaur2020}.  In all models we use a simple, computationally efficient scheme whereby a gas particle is converted into a collisionless star particle if it reaches an overdensity $\Delta=1+\delta>10^{3}$ and temperature $T<10^5 \rm\, K$ \citep{Viel2004}.    We assume a flat $\Lambda$CDM cosmology with $\Omega_{\Lambda}=0.692$, $\Omega_{\rm m}=0.308$, $\Omega_{\rm b}=0.0482$, $\sigma_8=0.829$, $n_{\rm s} =0.961$, $h=0.678$ \citep{planck2014}, and a primordial helium fraction by mass of $Y_{\rm p}=0.24$ \citep{Hsyu2020}. The initial conditions for the simulations are identical to those used for our earlier Sherwood simulation project \citep{Bolton2017}. 
 
We consider two distinct sets of simulations in this work: (i) a large grid of ``traditional'' hydrodynamical simulations of the IGM that use the UV background synthesis model from \citet{Puchwein2019}, and (ii) a smaller set of simulations, half of which follow the reionisation of the IGM using a hybrid radiative transfer approach and include the hydrodynamical response of the gas to inhomogeneous heating, and half of which use the same spatially uniform UV background approach adopted in set (i), but where this has now been adjusted to match the average ionisation and thermal histories of the hybrid-RT simulations. The models are summarised in Table~\ref{table:summary_sims}, and the details of each set are discussed below.

The first set of simulations are constructed using modifications to the spatially uniform UV background synthesis model introduced by \citet{Puchwein2019}. These simulations will be used to construct the grid of models we use for our Markov-Chain Monte-Carlo (MCMC) analysis in Section~\ref{section:MCMC}, and are similar to models we have used in our earlier work on the \Lya forest power spectrum at high redshift \citep{Viel2013,Nasir2016,Irsic2017}. The main improvements in this study are the larger dynamic range of the simulations, the use of a non-equilibrium thermo-chemistry solver \citep{Puchwein2015} and an improved treatment of the IGM opacity that consistently captures the transition from a neutral to ionised IGM \citep[see also][]{Onorbe2017a}.  In addition to running a model with the fiducial \citet{Puchwein2019} UV background, we also vary the photo-heating rates to achieve models with different gas temperatures and/or end redshifts for reionisation, following the approach described in \citet{Becker2011}. 

In the second set of simulations, half of the models include a hybrid-RT approach, in which we follow the radiative transfer of monochromatic UV photons with the moment-based, M1-closure radiative transfer code ATON \citep{Aubert2008}.  We construct the models as follows (see also Puchwein et al. in prep): ATON is initially run on a base \textsc{P-GADGET-3} simulation, using snapshots that are spaced at time intervals of $t=40\rm\,Myr$.  This produces three dimensional maps of spatially varying \HI photo-ionisation rates, $\Gamma_{\rm HI}$, as a function of redshift.  The luminosity of \HI photo-ionising sources are assumed to be proportional to the total halo mass, with a halo mass threshold of $M\sm{h} > 10^9 h^{-1} M_{\odot}$ and a mean energy for the ionising photons of $18.6\rm\, eV$. Note, however, the ATON simulations do not make a prediction for the \emph{absolute} ionising luminosity associated with each halo.  Instead, we take advantage of the efficiency of ATON to calibrate the ionising emissivity, thus producing specific realisations of the reionisation history. In this work, we construct three different reionisation models that correspond to an ``early'' ($z_{\rm R}\sim 5.3$), ``mid'' ($z_{\rm R}\sim 6.0$) and ``late'' ($z_{\rm R}\sim 6.6$) end redshift for reionisation, $z_{\rm R}$, defined as the redshift when the volume averaged \HI fraction first falls below $10^{-3}$.  Adjusting the emissivity in this way is equivalent to treating the (uncertain) escape fraction from the ionising sources as a free parameter. The resulting \HI photo-ionisation rate maps are then used as input in a re-run of the base \textsc{P-GADGET-3} model, but now incorporating the \emph{hydrodynamical} response of the gas to the inhomogeneous ionisation and heating.  The \HI photo-heating rate is obtained by multiplying through the photo-ionisation maps by the excess energy per \HI photo-ionisation, $18.6\rm\, eV$--$13.6\rm\, eV=5.0\rm\, eV$, yielding IGM gas temperatures consistent with recent observational constraints from \citet{Gaikwad2020}. The \HeI photo-ionisation rate is set equal to the \HI photo-ionisation rate, and the \HeI photo-heating rate is $1.3$ times that of the \HI photo-heating rate, matching the ratios from \citet{Puchwein2019}. Finally, the \HeII photo-ionisation and photo-heating rates are assumed to be spatially homogeneous\footnote{Note that if \HeII reionisation begins at $z>4$ \citep{Bolton2012,Makan2021}, our results may still somewhat underestimate the effect of large scale temperature fluctuations on the high redshift \Lya forest by missing the effect of inhomogeneous \HeII photo-heating \citep[see e.g][]{MeiksinTittley2012,Greig2015b}.} and are adopted from \citet{Puchwein2019}.  The advantage of this multi-step approach is that it consistently models the small-scale structure of the diffuse IGM within different, constrained reionisation histories.  We can then directly contrast the results to simulations with spatially uniform ionisation and heating by a homogeneous UV background \citep[see also][for a related approach]{Onorbe2019}.  

For the other half of the models in our second set of simulations, we therefore construct paired simulations that make use of the same homogeneous UVB approach as our grid of ``traditional'' models, but where this has now been adjusted to reproduce the average reionisation histories in the hybrid-RT models.\footnote{This approach is slightly different to that used by \cite{Onorbe2019} and \citet{Wu2019}, who both instead compared the results of their inhomogeneous reionisation models to hydrodynamical simulations using ``flash'' (i.e. very rapid) reionisation histories. The flash models were constructed to have the same mid-point of reionisation as the inhomogeneous reionisation simulations, rather than matching to the average ionisation and thermal history as we do here.}  This allows us to perform a direct comparison of the hybrid-RT simulations to models with a spatially uniform ionising background.  We ensure that -- for each reionisation history considered -- the evolution of the average \HI fraction and the gas temperature at mean density are consistent across the paired simulations. We achieve this by tuning the spatially uniform UVB model in the paired homogeneous simulations to match the volume averaged \HI fraction, the volume averaged IGM temperature (at mean density) during reionisation, and the median IGM temperature (at mean density) after reionisation.  The median temperature has the advantage that it is less affected by the shock heating of a small fraction of the gas to high temperatures following reionisation.  Further details can also be found in Puchwein et al. (in prep.) This allows us to isolate the effect of the ``patchiness'' of reionisation on the \Lya forest power spectrum from differences that otherwise arise from changes in the spatially averaged ionisation and thermal history.  

An illustration of the gas density, neutral hydrogen fraction and temperature predicted by the hybrid-RT and paired homogeneous simulations (in this case for the RT-mid and Homog-mid models at $z=5.4$) is displayed in Fig.~\ref{figure:projected_plane}. Note in particular the higher temperatures and lower \HI fractions in the hybrid-RT model within the prominent void in the lower half of each panel.   This arises because the void has been reionised recently and is therefore hotter compared to the paired homogeneous model, as there is less time for subsequent cooling \citep[see e.g.][]{Trac2008,Cen2009,Furlanetto2009,Raskutti2012,Lidz2014,DAloisio2015,Keating2018,Davies2019}.  Further discussion of the resulting differences in the $T-{\Delta}$ relation between the two simulations can be found in Puchwein et al. (in prep.) and \citet{Gaikwad2020}.

\subsection{Reionisation histories used in the hybrid-RT simulations}
\label{section:eor_histo}

In Fig.~\ref{figure:global_props} we show the redshift evolution of the volume averaged ionised hydrogen fraction, $1-x\sm{HI}$, the temperature at mean density, $T_0$, the Thomson scattering optical depth, $\tau\sm{e}$, and the cumulative energy per proton mass deposited at the mean density, $u\sm{0}$, for each for the three reionisation histories we use for the hybrid-RT and paired homogeneous simulations.  The results from the homogeneous models are displayed by the curves, while the corresponding $1-x\sm{HI}$ and $T_{0}$ values in the hybrid-RT runs are shown at redshift intervals of $\Delta z=0.5$ by the open and filled circles.    By design, these match very closely.  Note also that in the case of the gas temperature, we show the median $T_{0}$ instead of the mean following reionisation, as the median is less sensitive to high temperature, shock heated gas.  

The reionisation histories are deliberately chosen to span a range of $z_{\rm R}$, rather than calibrated in detail to reproduce observational data.   For example, the large fluctuations in the \Lya forest opacity observed at $z=5.5$ \citep{Becker2015} will only be captured by the late reionisation model \citep[cf][]{Kulkarni2019,Nasir2020,Qin2021,Bosman2021, Choudhury2021}.  Nevertheless, there is good agreement between our simulations and existing constraints on the IGM ionisation history.  In the upper left panel we compare to a selection of measurements from \citet{McGreer2015}, based on dark gaps in the \Lya and Ly$\beta$ forests, from \citet{Banados2018,Davies2018b,Greig2019,Yang2020,Wang2020}, based on \Lya damping wings in high redshift quasars, and from \citet{Mason2019}, based on the visibility of \Lya emitting galaxies.  All three reionisation histories are furthermore consistent with the \citet{Planck2020} Thomson scattering optical depth, displayed in the lower left panel, although only the Homog-early model is within $1\sigma$ of the slightly higher $\tau\sm{e}$ inferred recently by \citet{deBelsunce2021}.  

In the right hand panels the simulations are compared to the gas temperature and cumulative energy per proton mass at the mean density measured by \citet{Boera2019} and \citet{Walther2019} from the \Lya forest power spectrum, and by \citet{Gaikwad2020} using \Lya transmission spike widths.  There is again reasonable agreement between our simulations and the data, although note there is $\sim 2$--$2.5\sigma$ difference between the \citet{Boera2019} and \citet{Walther2019} $T_{0}$ measurement from the \Lya forest power spectrum and the gas temperature evolution in the simulations.  This may indicate the rise in the IGM temperature at $z<5$ due to \HeII photo-heating may occur too early in the simulations \citep[but see][]{Makan2021}.  Alternatively, there may also be systematic differences between the measurements due to a degeneracy between $T_0$ and the pressure smoothing scale, where models with increased smoothing can lead to a systematic decrease in the inferred $T_{0}$ from the power spectrum (cf. the 3$\sigma$ discrepancy between the \citet{Walther2019} and \citet{Gaikwad2020b} measurements at $z=5.5$ -- see section 5.4.3 in \citet{Gaikwad2020b} for further discussion of this point).  We will also return to this point later in Section~\ref{section:MCMC}.

\subsection[Simulating the Lyman-alpha forest power spectrum]{Simulating the Lyman-$\alpha$ forest power spectrum}
\label{section:sim_power_spec}

We obtain the power spectrum of the \Lya forest transmitted flux using mock absorption spectra extracted from the simulations \citep[e.g.][]{Theuns1998}.  We extract 5000 lines of sight, each with 2048 pixels, drawn parallel to the cosmological box boundaries. The line of sight positions are the same for all the simulations, and furthermore all the models use initial conditions generated with the same random seed.  This allows a direct, pixel-by-pixel comparison of the \Lya transmission across different models. As we are primarily interested in comparing different simulations, we do not add noise and instrumental broadening effects to the spectra in this analysis.  

Once we have obtained the \Lya optical depth, $\tau_{\alpha}$, in each pixel, we re-scale the transmitted flux $F = e^{-\tau_{\rm Ly\alpha}}$ in each pixel to match the observed redshift evolution of the \Lya forest optical depth, $\tau\sm{eff} = - \text{ln}\langle F \rangle$, where $\langle F \rangle$ is the mean observed transmission.  Uncertainties in the IGM temperature and background photo-ionisation rate mean a rescaling is commonly used to match the simulation to the data as closely as possible to the observed $\tau_{\rm eff}$ \citep{Bolton2005,Lukic2015}.  It furthermore conveniently allows us to vary the effective optical depth in our MCMC analysis without requiring additional simulations.  Note, however, this scaling is only a good approximation following reionisation, as it implicitly assumes the gas in the low density IGM is in photo-ionisation equilibrium, such that $\tau_{\alpha} \propto x_{\rm HI} \propto \Gamma_{\rm HI}^{-1}$.  It is important to emphasise that, for this reason, we do not apply this optical depth rescaling to our hybrid-RT simulation outputs prior to the redshift at which reionisation ends, $z_{\rm R}$, in the simulation volumes.  The redshift evolution for $\tau\sm{eff}$ we adopt is:
\begin{equation}
%\[
\tau\sm{eff} =
\begin{cases}
\label{eqn:tau_trams}
   -0.132 + 0.751[(1 + z)/4.5]^{2.90},& \text{if } 2.2 \leq z < 4.4\\
   1.142[(1 + z)/5.4]^{4.91}. & \text{if } 4.4 \leq z \leq 5.5
\end{cases}
%\]
\end{equation}
This is taken from \cite{Viel2013} and \cite{Becker2013} for the upper and lower redshift ranges, respectively.

Once this rescaling has been performed, we calculate the power spectrum of the transmitted flux, $P(k)$, using the estimator $\delta\sm{F}=F/\langle F \rangle-1$. Since in this work we are primarily interested in analysing differences arising between the hybrid-RT and paired homogeneous simulations, we will focus mainly on the ratio, $R(k,z)$, of the power spectra, that is:
\begin{equation}
\label{eqn:defR}
    R(k,z) =  \frac{P\sm{RT}(k,z)}{P\sm{homog}(k,z)},
\end{equation}
\noindent
where $P_{\rm RT}(k,z)$ and $P_{\rm homog}(k,z)$ are the power spectrum from the hybrid-RT and paired homogeneous simulations, respectively.  We will concentrate in particular on this ratio in three redshift bins, $z=$ 4.2, 4.6, 5.0, which correspond to those observed by \citet{Boera2019}.  We do not consider a redshift bin at $z=5.4$ within our framework, despite the availability of observational constraints here \citep{Viel2013}.  This is because -- as already discussed above -- our assumption of ionisation equilibrium when rescaling $\tau_{\rm eff}$ will break down for late reionisation models, invalidating our approach. 

%%%%%%%%%%%%%%%%%%%%%%%%%%%%%%%%%%%%%%%%%%%%%%%%%%%%%%%%%%%%%%%%%%%%%
%%%%%%%%%%%%%%%%%%%%%%%%%% SECTION 3 %%%%%%%%%%%%%%%%%%%%%%%%%%%%%%%%
%%%%%%%%%%%%%%%%%%%%%%%%%%%%%%%%%%%%%%%%%%%%%%%%%%%%%%%%%%%%%%%%%%%%%

\section[The effect of inhomogeneous reionisation on the Lyman-alpha forest power spectrum]{The effect of inhomogeneous reionisation on the \Lya forest power spectrum}
\label{section:changes_Lya_forest}

\begin{figure*}

  \centering
\includegraphics[height=6cm,trim = 80 0 70 0]{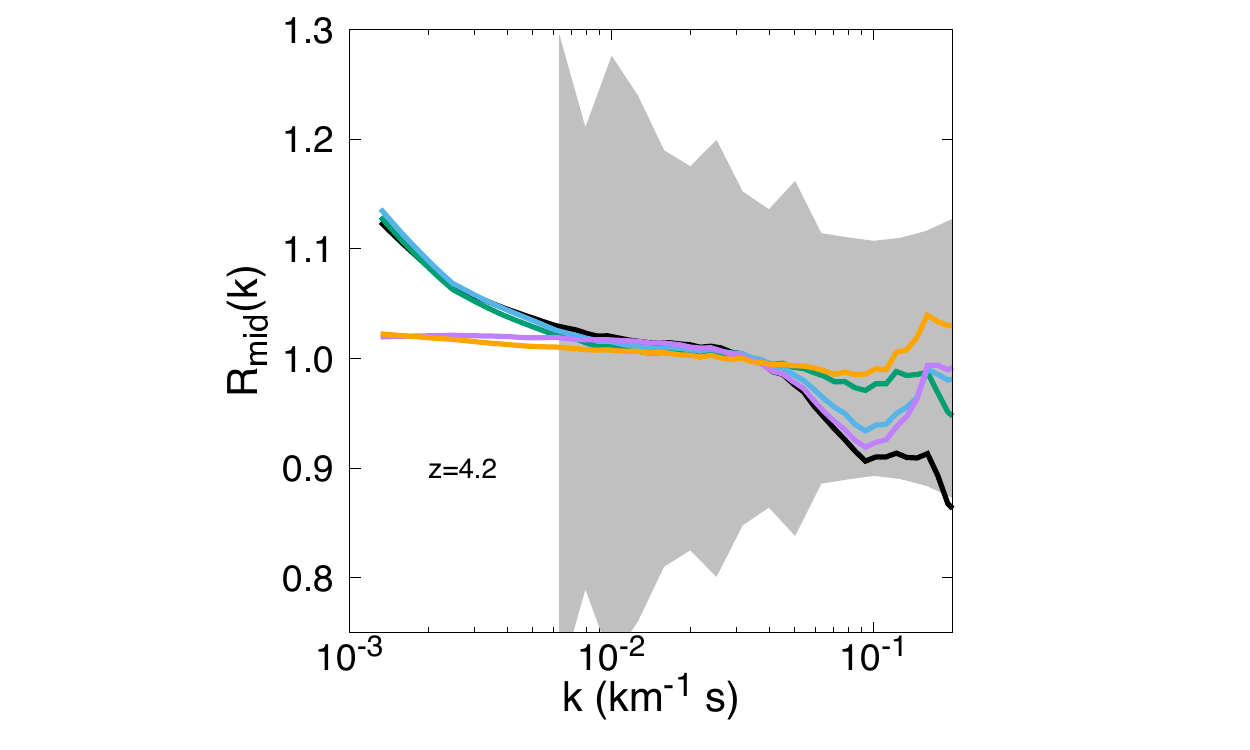}
\includegraphics[height=6cm,trim = 70 0 70 0]{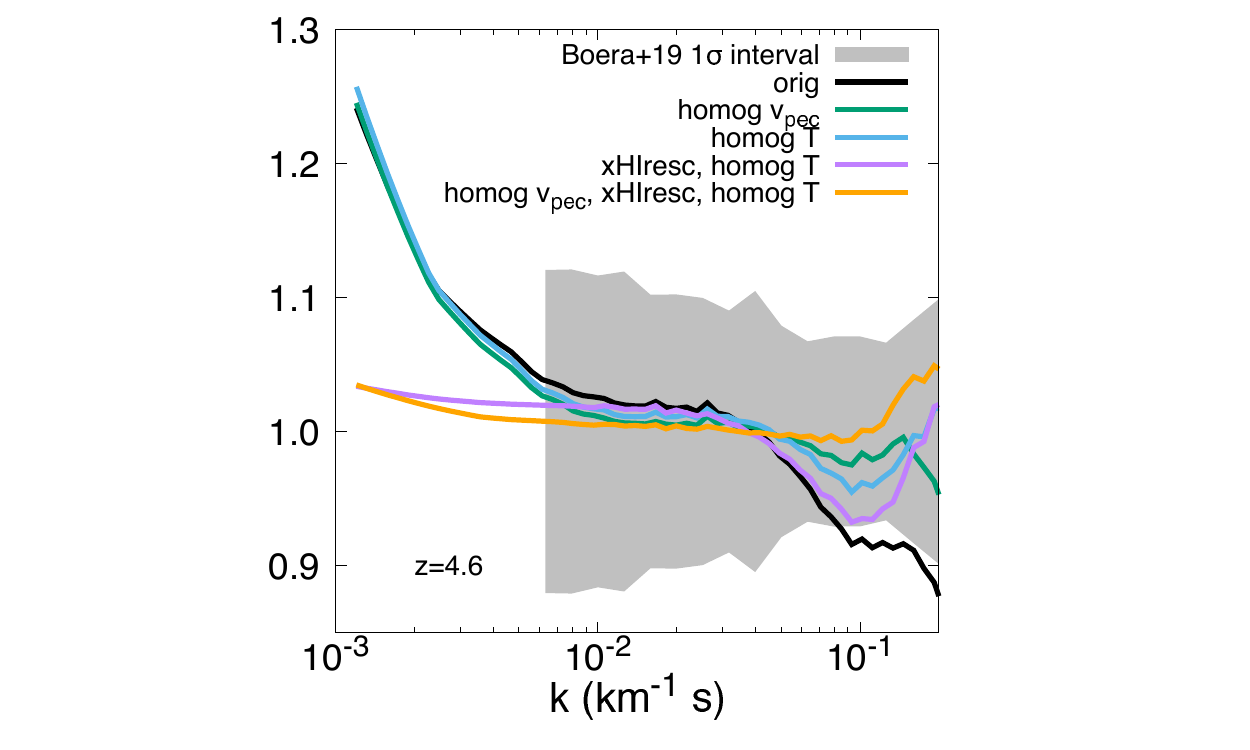}
\includegraphics[height=6cm,trim = 80 0 100 0]{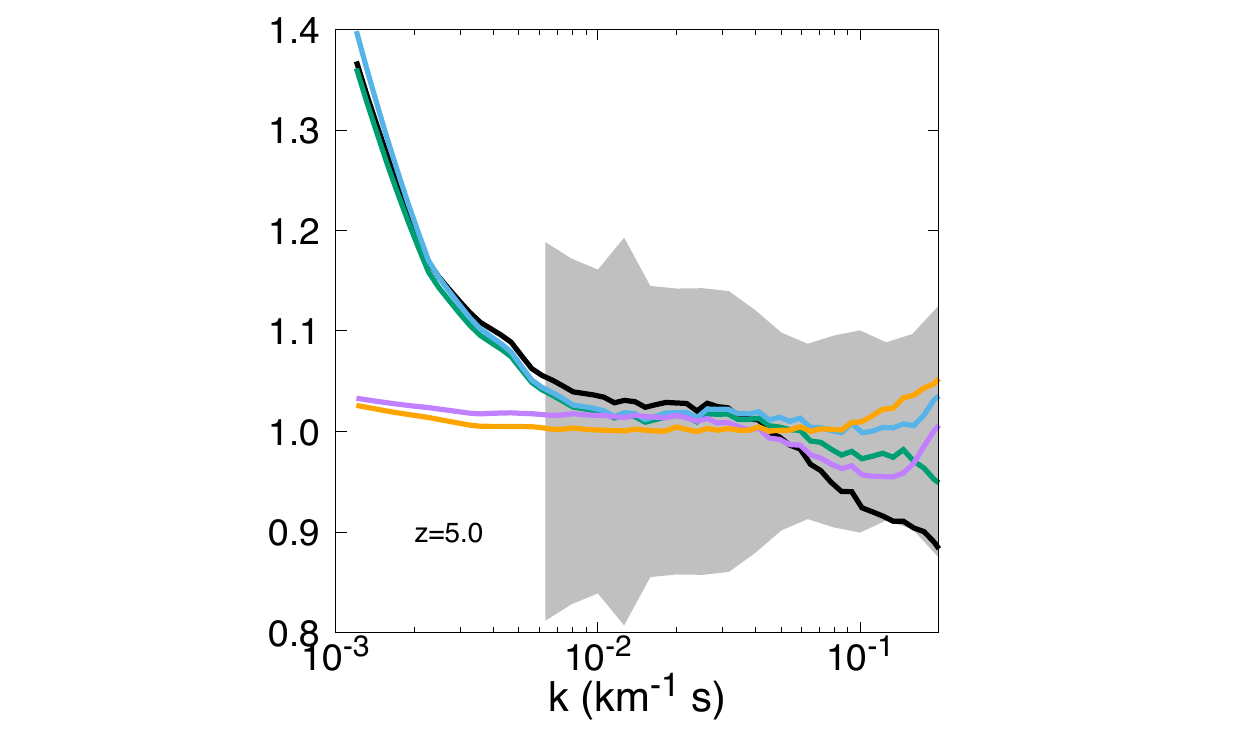}
\vspace{-0.3cm}
\caption{The ratio of \Lya forest transmitted flux power spectrum obtained from the RT-mid and Homog-mid simulations in three redshift bins (from left to right, $z=4.2,4.6,5.0$).   At each redshift, all mock spectra have been rescaled to the same effective optical depth, given by Eq.~(\ref{eqn:tau_trams}).  The ratio taken directly from the simulations is shown by the black curves, while the other curves are obtained by progressively replacing quantities in the RT-mid simulation with those from Homog-mid, and then recomputing the simulated \Lya forest spectra (see text for details).   These are the peculiar velocity field (``homog v$\sm{vpec}$"), the gas temperature (``homog T''), and ionised hydrogen fraction (``xHIresc", obtained by rescaling the ionisation fraction to account for the temperature dependence of the recombination coefficient). The shaded grey region shows the 1$\sigma$ uncertainties from the power spectrum measurements presented by \citet{Boera2019} using high resolution Keck/HIRES and VLT/UVES data, centred around $R(k)=1$.  Note the different scale on the vertical axis of each panel.}
\label{figure:Ratio_Lyalpha_PS_wboot}
\end{figure*}

\begin{figure}
\centering
     \includegraphics[trim =80 0 80 0,height=6.5cm]{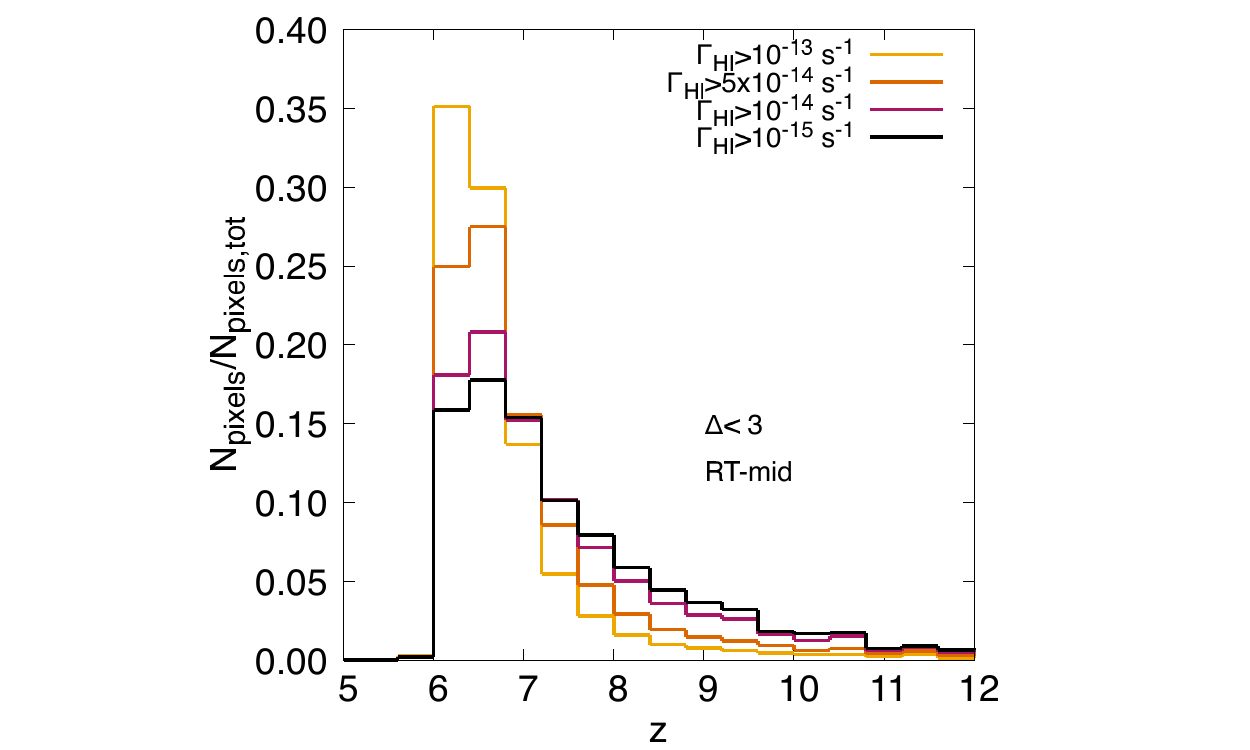} 
     \vspace{-0.2cm}
    \caption{The distribution of redshifts at which pixels with gas densities $\Delta<3$ along the lines of sight drawn from the RT-mid simulation first pass a fixed threshold in the hydrogen photo-ionisation rate, $\Gamma_{\rm HI}$.  These are $\Gamma_{\rm HI}>10^{-15}\rm\,s^{-1}$ (black), $\Gamma_{\rm HI}>10^{-14}\rm\,s^{-1}$ (red), $\Gamma_{\rm HI}>5\times 10^{-14}\rm\,s^{-1}$ (orange) and $\Gamma_{\rm HI}>10^{-13}\rm\,s^{-1}$ (yellow).  The distributions are normalised to the total number of pixels, $N\sm{pixels,tot}$, in the simulated lines of sight.  Note that in the paired homogeneous simulation, Homog-mid, by construction all pixels exceed a given $\Gamma_{\rm HI}$ threshold at the same redshift of $z=7.03$.}
    \label{figure:Px_illumi_zdistr}
\end{figure}

\begin{figure*}
    \includegraphics[trim=85 0 120 0,height=6.5cm]{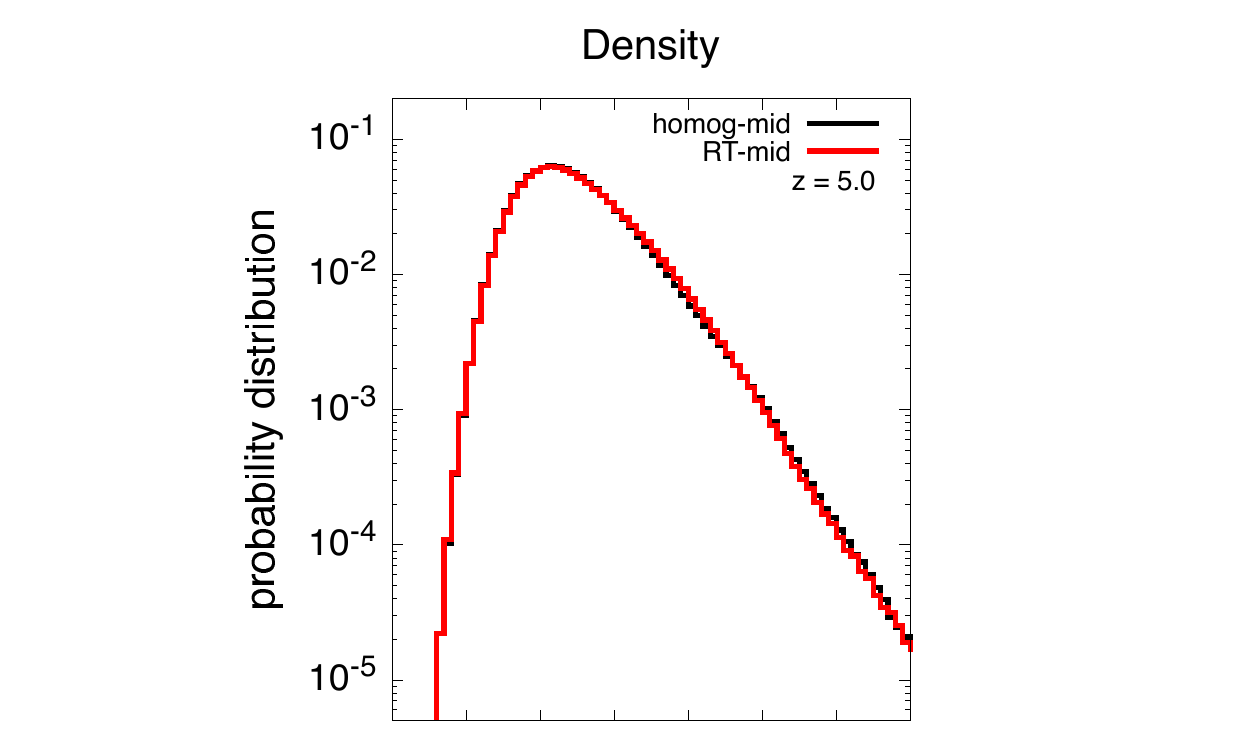}
    \includegraphics[trim=65 0 130 0,height=6.5cm]{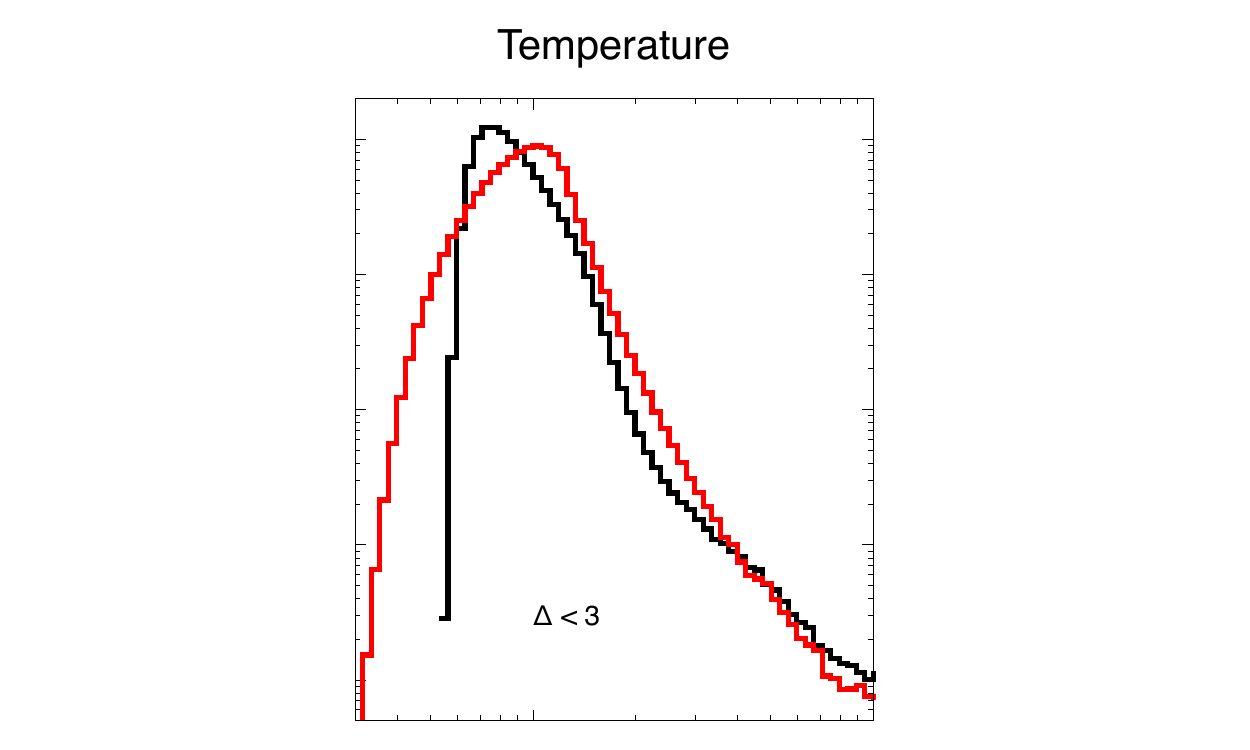}
    \includegraphics[trim=70 0 70 0,height=6.5cm]{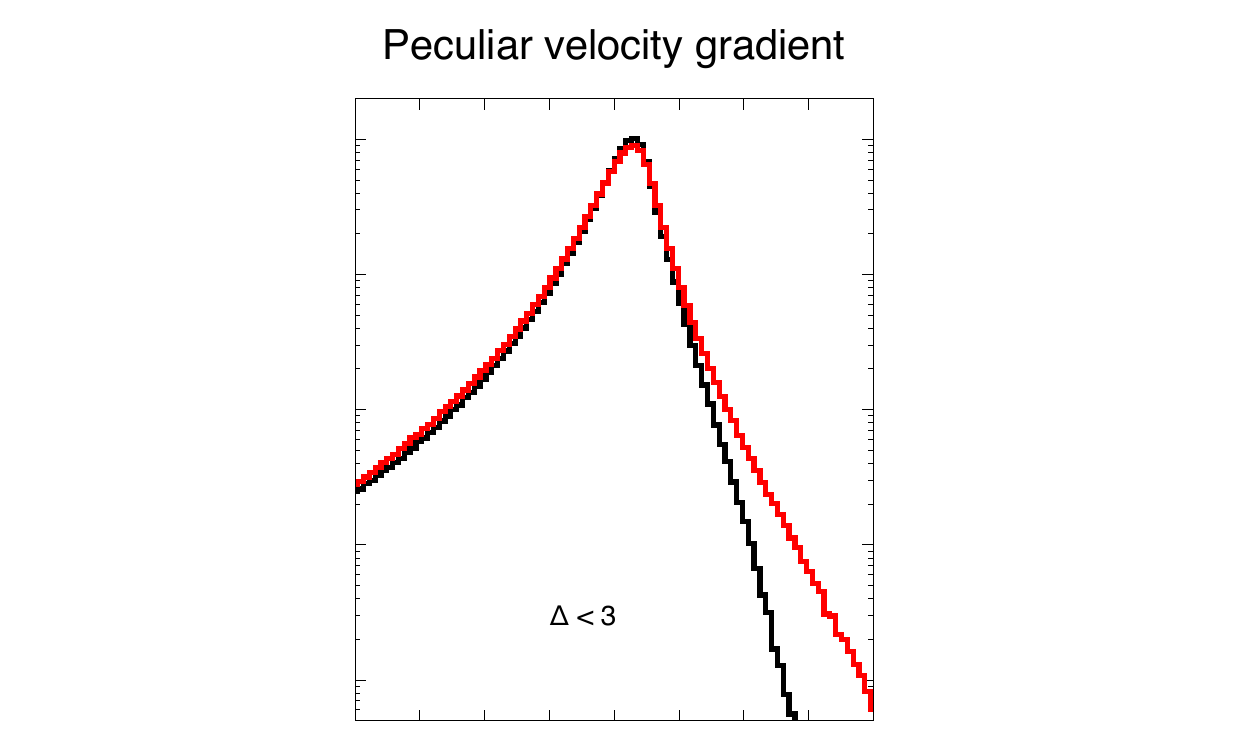}\\
    \includegraphics[trim=80 10 100 0,height=6.5cm]{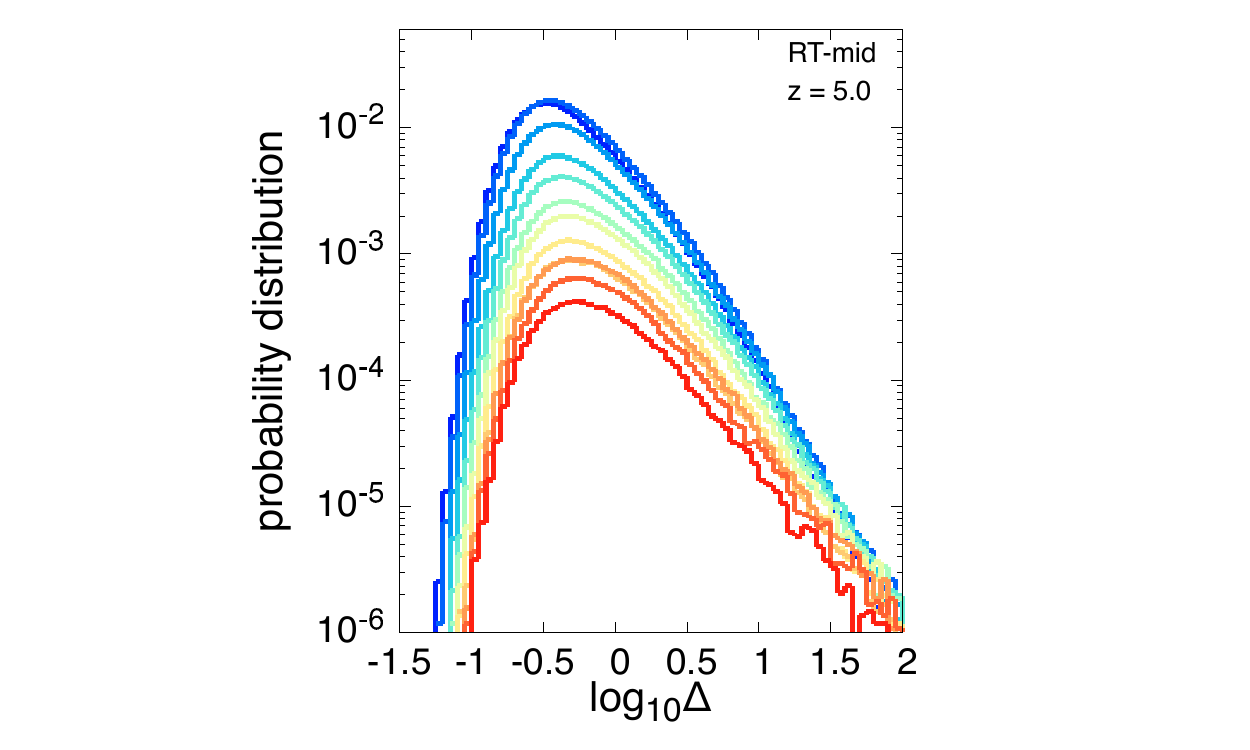}
    \includegraphics[trim=90 0 90 0,height=6.5cm]{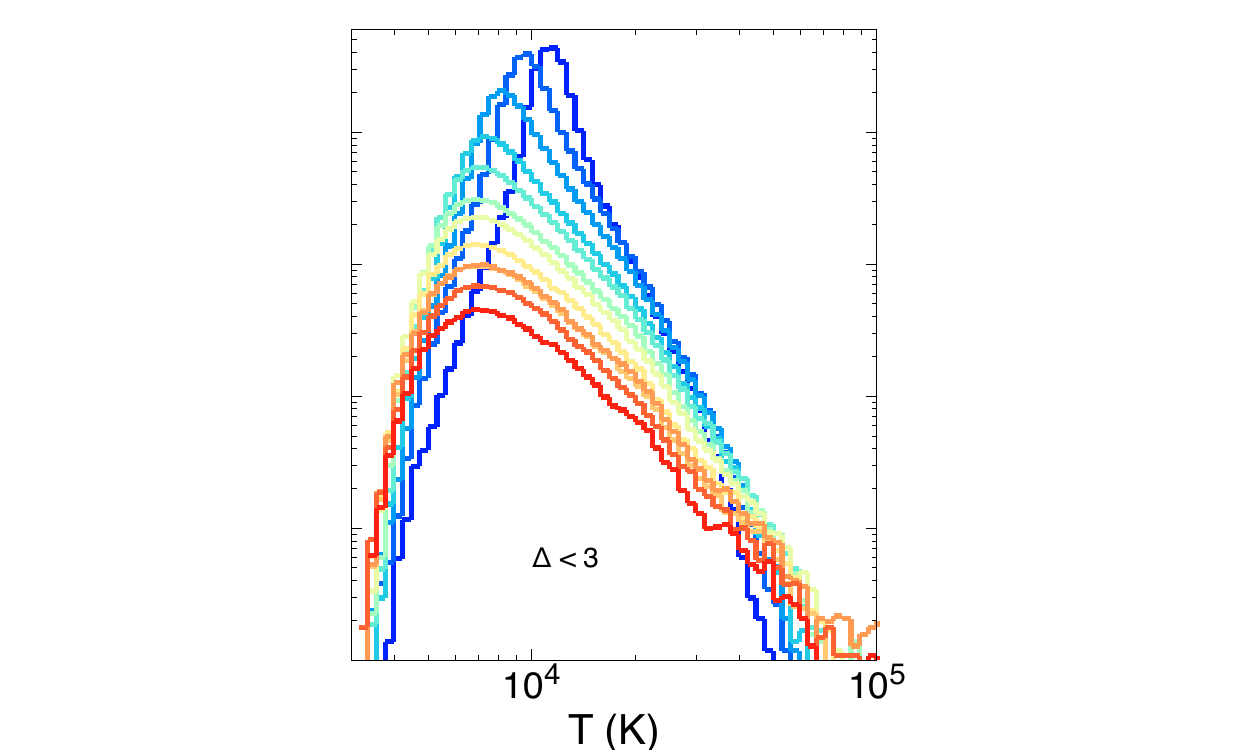}
    \includegraphics[trim=90 10 90 0,height=6.5cm]{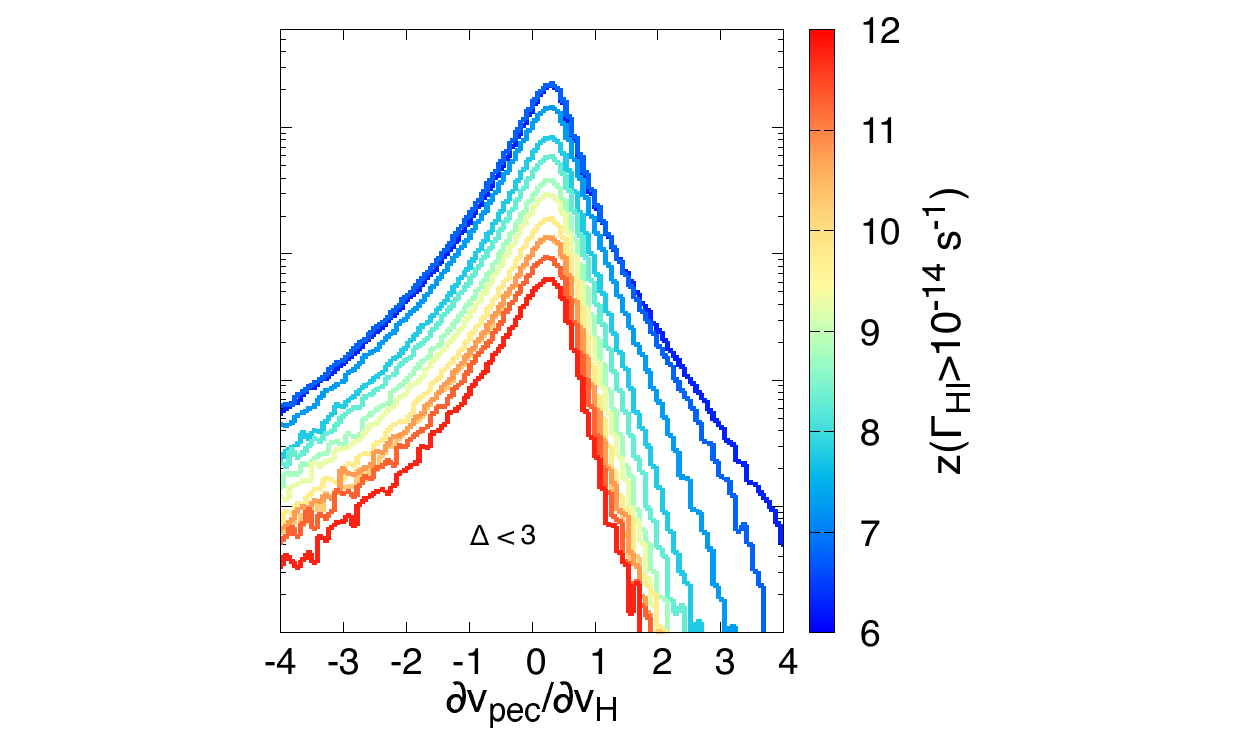}\\ 
    \caption{\textit{Top row}: Comparison between the gas density \textit{(left column)}, temperature \textit{(middle column)}, and peculiar velocity gradient $\VpecDx$ \textit{(right column)} PDFs at redshift $z=5$ in the Homog-mid (black curves) and RT-mid (red curves) simulations. The temperature and $\VpecDx$ distributions are for gas with $\Delta<3$, to exclude the higher density gas that is not visible in transmission at high redshift. \textit{Bottom row}: PDFs from the RT-mid simulation at redshift $z=5$, but now shown only for pixels in which the photoionisation rate, $\Gamma_{\rm HI}$, first exceeds $10^{-14} \text{s}^{-1}$ at a selection of different redshifts. The colour axis shows the central value of the redshift bins we use here, which have width $\Delta z=0.5$.  Note that the normalisation of the PDF in each bin reflects the fraction of the volume that exceeds the $\Gamma_{\rm HI}$ threshold at that redshift (cf. Fig. \ref{figure:Px_illumi_zdistr}), highlighting how the different redshift bins contribute to the shape of the overall distribution shown in the top row.}
    \label{figure:delta_T_dvpec_zr60_distr}
\end{figure*}

We now turn to examine the effect of inhomogeneous reionisation on the shape of the \Lya forest transmitted flux power spectrum.   We first consider the intermediate reionisation history with an end redshift for reionisation at $z_{\rm R}=6.0$.  We refer to the hybrid-RT and paired homogeneous simulations for this model as ``RT-mid'' and ``Homog-mid'', respectively (see Table \ref{table:summary_sims}). We will examine the effect of changing the end redshift of reionisation on the power spectrum ratio, $R(k,z)$, in Section~\ref{section::comparison_536067}.  

The black curves in Fig.~\ref{figure:Ratio_Lyalpha_PS_wboot} show the ratio $R(k,z)$ of the power spectrum compared to the 1$\sigma$ uncertainty in the \citet{Boera2019} power spectrum measurements in three redshift bins ($z=4.2,4.6,5.0$).  Note the \citet{Boera2019} data cover $-2.2 \leq \log(k/\rm km^{-1}\,s)\leq -0.7$, whereas in the simulations we also consider larger scales up to $\log(k/\rm km^{-1}\,s)=-2.9$.  At small $k$ values, the uncertainties in the power spectrum measurements are typically dominated by cosmic variance and the low number of large-scale modes. At high $k$ values, on the other hand, the uncertainties are mainly due to noise, metals and instrumental broadening corrections.
In all three redshift bins the hybrid-RT prescription boosts the power spectrum at large scales (small $k$ values) by $\sim 10-40$ per cent and suppresses it at small scales (large $k$ values) by 10-15 per cent.   The difference between the hybrid-RT and paired homogeneous simulations is comparable to or smaller than the 1 $\sigma$ uncertainties from the \citet{Boera2019} power spectrum measurements at scales $\log(k/\rm km^{-1}\,s)>-2.2$.  This already suggests that analyses of current \Lya forest power spectrum measurements from high resolution ($R\sim 40,\,000$) data at $z\leq 5$ \citep[e.g.][]{Irsic2017fuzzy,Garzilli2021,Rogers2021} should not be strongly biased if ignoring the effect of inhomogeneous reionisation \citep[cf.][]{Hui2017}.  This will change, however, for more precise measurements and/or data that extend to larger scales.  We explore this further in Section~\ref{section:MCMC}.

An explanation for the change in the shape of the power spectrum in the hybrid-RT model may be obtained by successively isolating the physical quantities that influence $R(k,z)$ (i.e. the \HI fraction, gas temperature, gas density, and peculiar velocity).  The other curves in Fig. \ref{figure:Ratio_Lyalpha_PS_wboot} display the results of this process, where we substitute quantities in the hybrid-RT simulation with those from the paired homogeneous run, and then re-extract the \Lya forest spectra we use to calculate the power spectrum.  In particular, in the hybrid-RT model we will replace the peculiar velocity (labelled ``homog $v\sm{pec}$'' in Fig.~\ref{figure:Ratio_Lyalpha_PS_wboot}) and the gas temperature (``homog T'') with the values from the paired homogeneous run. Additionally, after the gas temperature in the hybrid-RT model has been substituted with the one from the homogeneous simulation, we also rescale the neutral hydrogen fraction, $x\sm{HI}$ (``xHIresc''), to account for differences in the temperature dependent recombination rate ($\alpha_{\rm HII}(T) \propto T^{-0.7}$, e.g. \cite{McQuinn2016}).  For further assistance with visualising the changes this process causes in the mock \Lya forest spectra, in the supplementary Appendix~\ref{app:example_spec} we show the transmitted flux along a single line of sight for each case considered in Fig.~\ref{figure:Ratio_Lyalpha_PS_wboot}.

First, the orange curve in Fig. \ref{figure:Ratio_Lyalpha_PS_wboot} shows the case where all quantities ($v\sm{pec},T$, and $x\sm{HI}$) in the hybrid-RT model have been substituted with those from the homogeneous simulation, in the manner described above.  The only remaining difference between the models is the underlying gas distribution, where we find the ratio, $R(k)$, varies by $<5$ per cent at all scales.   Pressure (or Jeans) smoothing of the gas distribution - which, as discussed in section \ref{section::simulation}, is implemented self-consistently in our hybrid-RT approach - will counter gravitational collapse and help push the baryons outward from peaks in the dark matter density \citep[e.g.][]{GnedinHui1998,Kulkarni2015,Rorai2017}. This means that if the IGM has been reionised and heated more recently in the hybrid-RT model, gas will have had less time to dynamically respond to the change in pressure \citep{Daloisio2019}, which will increase power on small scales. Indeed, we find a small excess ($<5$ per cent) in the power spectrum ratio shown by the orange curve at $k>0.1$ km$^{-1}$ s.  This implies that while changes in the pressure smoothing due to differences in the \emph{average} thermal history of the IGM will alter the shape of the power spectrum following reionisation \citep[e.g.][]{Nasir2016,Onorbe2017b,Wu2019}, variations around this average due to the ``patchiness'' of pressure smoothing will have a much smaller effect.

The purple curve in Fig. \ref{figure:Ratio_Lyalpha_PS_wboot} also replaces $T$ and $x\sm{HI}$ in the hybrid-RT model as previously described, however it now relies on the original hybrid-RT $v\sm{pec}$. Comparison to the orange curve in Fig. \ref{figure:Ratio_Lyalpha_PS_wboot} therefore demonstrates the effect that differences between the peculiar velocities in the hybrid-RT and homogeneous simulations have on the power spectrum.  This suggests that the suppression of power by 5-10 per cent on scales of $k\simeq 0.1$km$^{-1}$ partly arises from changes in the IGM peculiar velocity field.  This can also be seen by comparing the black and green curves, where in the latter case only $v_{\rm pec}$ has been replaced in the hybrid-RT model using the homogeneous simulation values. As we demonstrate shortly, this suppression of power occurs because of the divergent peculiar velocity field associated with gas that has been recently heated, resulting in \Lya transmission that is more smoothed out in velocity space \citep[see also e.g.][]{Gaikwad2020}. 

Thermal broadening will also lead to a suppression of the power spectrum at small scales by smoothing the \Lya absorption in velocity space \citep{Zaldarriaga2001,Nasir2016,Walther2019}.  Indeed, this is what is shown by the blue curve in Fig. \ref{figure:Ratio_Lyalpha_PS_wboot}, where the gas temperature in the hybrid-RT model has been replaced with the values from the homogeneous simulation.  The power at small scales is now increased relative to the black curve by up to 10 per cent at the smallest scales.  The higher gas temperatures in the recently reionised regions in the hybrid-RT model therefore act (in combination with the peculiar velocity field) to suppress the power spectrum at small scales.  

We may gain further insight into this behaviour in Fig.~\ref{figure:Ratio_Lyalpha_PS_wboot} by examining differences in the IGM properties in the hybrid-RT and paired homogeneous simulations.  First, in Fig. \ref{figure:Px_illumi_zdistr} we show the fraction of pixels with gas density $\Delta < 3$  in the RT-mid simulation (i.e. the gas typically responsible for transmission in the \Lya forest at high redshift, see fig. 2 in \citet{Nasir2016}) that have crossed different photo-ionisation rate thresholds (in this case, $\Gamma_{\rm HI} > 10^{-15}\rm\,s^{-1},10^{-14}\,s^{-1},5 \times 10^{-14}\,s^{-1}$, and $10^{-13}$ s$^{-1}$) by a given redshift.  Whereas this occurs by construction at a single redshift in the Homog-mid simulation (corresponding to $z=7.03$ for $\Gamma_{\rm HI}>10^{-14}\rm\,s^{-1}$),  inhomogeneous reionisation means this occurs at different redshifts in the RT-mid model. Note also that while the distributions for $\Gamma_{\rm HI}>10^{-14}\rm\,s^{-1}$ and $\Gamma_{\rm HI}>10^{-15}\rm\,s^{-1}$ are similar in Fig.~\ref{figure:Ratio_Lyalpha_PS_wboot}, for the larger thresholds the distribution becomes more strongly peaked at $z\sim 6.2$, reflecting the fact that the amplitude of the ionising background increases toward lower redshift.  

Taking $\Gamma_{\rm HI}>10^{-14}\rm\,s^{-1}$ as a proxy for when a pixel is first ionised, in Fig. \ref{figure:delta_T_dvpec_zr60_distr} we show the probability distribution function (PDF) for the gas density, temperature, and the derivative of the line of sight peculiar velocity  $\VpecDx$ (where $v\sm{H}$ is the Hubble velocity) at redshift $z=5$ in the RT-mid and Homog-mid simulations.   For the latter quantity, positive (negative) values correspond to a divergent (convergent) peculiar velocity field.  We show the PDFs for all pixels in the mock spectra (upper panels) and, in the case of RT-mid only, broken down by the redshift at which the pixels first cross the $\Gamma_{\rm HI}>10^{-14}$s$^{-1}$ threshold (lower panels).  The gas density PDF for all pixels appears similar in both models, consistent with the rather small differences we observe for $R(k)$ when comparing the black and orange curves in Fig.~\ref{figure:Ratio_Lyalpha_PS_wboot}.   However, the temperature and peculiar velocity gradient PDFs show larger differences.  In particular there is a broader distribution of gas temperatures in the RT-hybrid model, with a peak in the distribution that is $\sim 4000\rm\,K$ hotter compared to the homogeneous simulation.  The fraction of pixels with $\VpecDx>1$ is also larger in the hybrid-RT run, due to newly expanding gas that has recently been reionised and heated.

The RT-mid PDF for all pixels (red curve in the upper panels) is an average of the distributions shown in the lower panels, weighted by the relative amplitude of each redshift bin in Fig. \ref{figure:Px_illumi_zdistr}.  First, note that for gas that is reionised earlier, the gas density PDF in the lower left panel of  Fig. \ref{figure:delta_T_dvpec_zr60_distr} shifts toward larger values of $\Delta$, with a high density tail that becomes shallower.   This is consistent with the highest density regions in the simulation reionising first.  Second, we observe for the $T$ and $\VpecDx$ PDFs that higher gas temperatures and divergent peculiar velocities arise from gas that has been recently ionised.  As discussed for Fig.~\ref{figure:Ratio_Lyalpha_PS_wboot}, we argue here that the suppression of power in the RT-mid simulation on small scales, $k> 0.05\rm\,km^{-1}\,s^{-1} $, relative to the Homog-mid model is primarily due to these differences acting together in redshift space, rather than a physical smoothing of the gas density. 

Finally, we note the results displayed in Fig.~\ref{figure:Ratio_Lyalpha_PS_wboot} have also been discussed by a number of other studies using different numerical methods \citep{Cen2009,Keating2018,Daloisio2019,Onorbe2019,Wu2019,Montero2020}.  All these studies find that large-scale fluctuations in the post-reionisation IGM temperature will increase the power spectrum on large scales; our results are in good agreement with this canonical expectation.  Of particular note, however, is the recent study by \citet{Wu2019}.  These authors used fully coupled radiation-hydrodynamical simulations to model the effect of reionisation on the power spectrum at $z\sim 5$, finding that power was enhanced up to $\sim 30$--$40$ per cent at $k\sim 0.002 \rm\,km^{-1}\,s^{-1}$ and suppressed by up to $\sim 10$ per cent on scales $k>0.1\rm\,km^{-1}\,s$ (see their fig. 5).  Our results appear to be remarkably consistent, suggesting that the hybrid approach we adopt is well suited for efficiently modelling the power spectrum.

%%%%%%%%%%%%%%%%%%%%%%%%%%%%%%%%%%%%%%%%%%%%%%%%%%%%%%%%%%%%%%%%%%%%%	
%%%%%%%%%%%%%%%%%%%%%%%%%% SECTION 4 %%%%%%%%%%%%%%%%%%%%%%%%%%%%%%%%
%%%%%%%%%%%%%%%%%%%%%%%%%%%%%%%%%%%%%%%%%%%%%%%%%%%%%%%%%%%%%%%%%%%%%

\section[A correction for the effect of large-scale temperature fluctuations on the Lyman-alpha forest power spectrum]{A correction for the effect of large-scale temperature fluctuations on the Ly$\alpha$ forest power spectrum}
\label{section::comparison_536067}

\begin{figure*}
    \centering
    \includegraphics[height=6cm,trim=80 0 70 0]{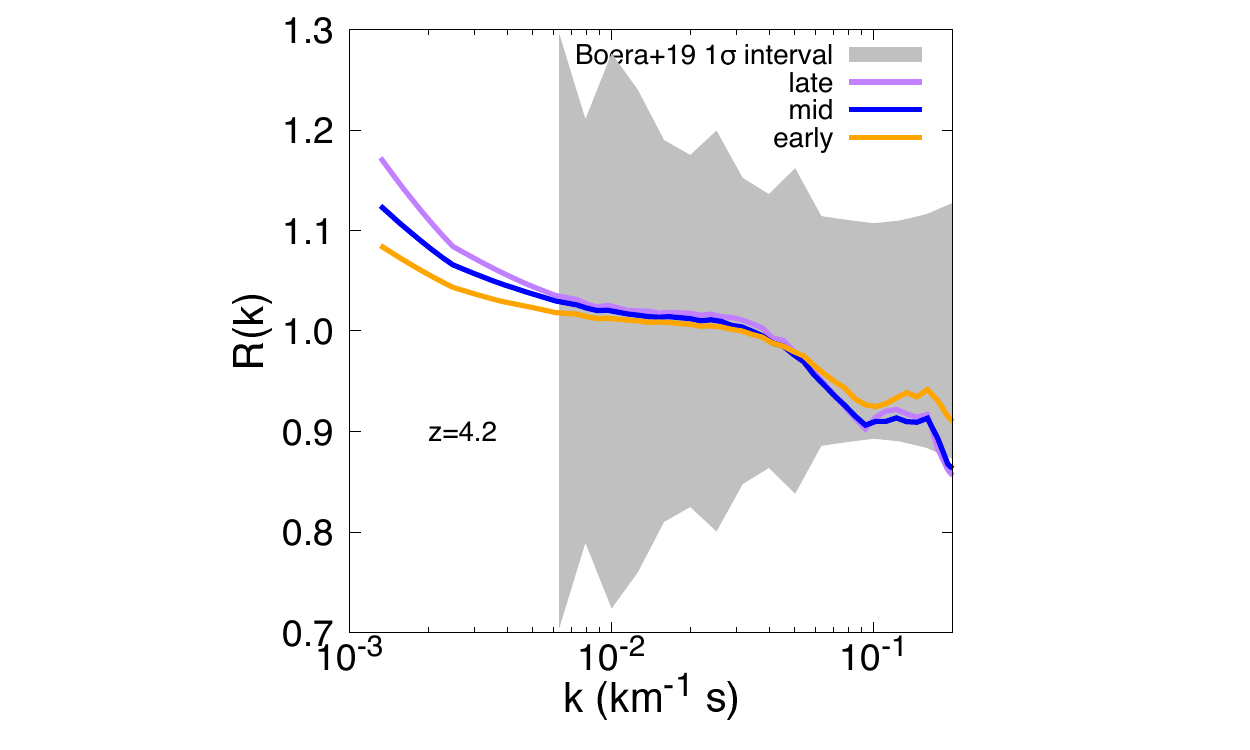}
    \includegraphics[height=6cm,trim=70 0 70 0]{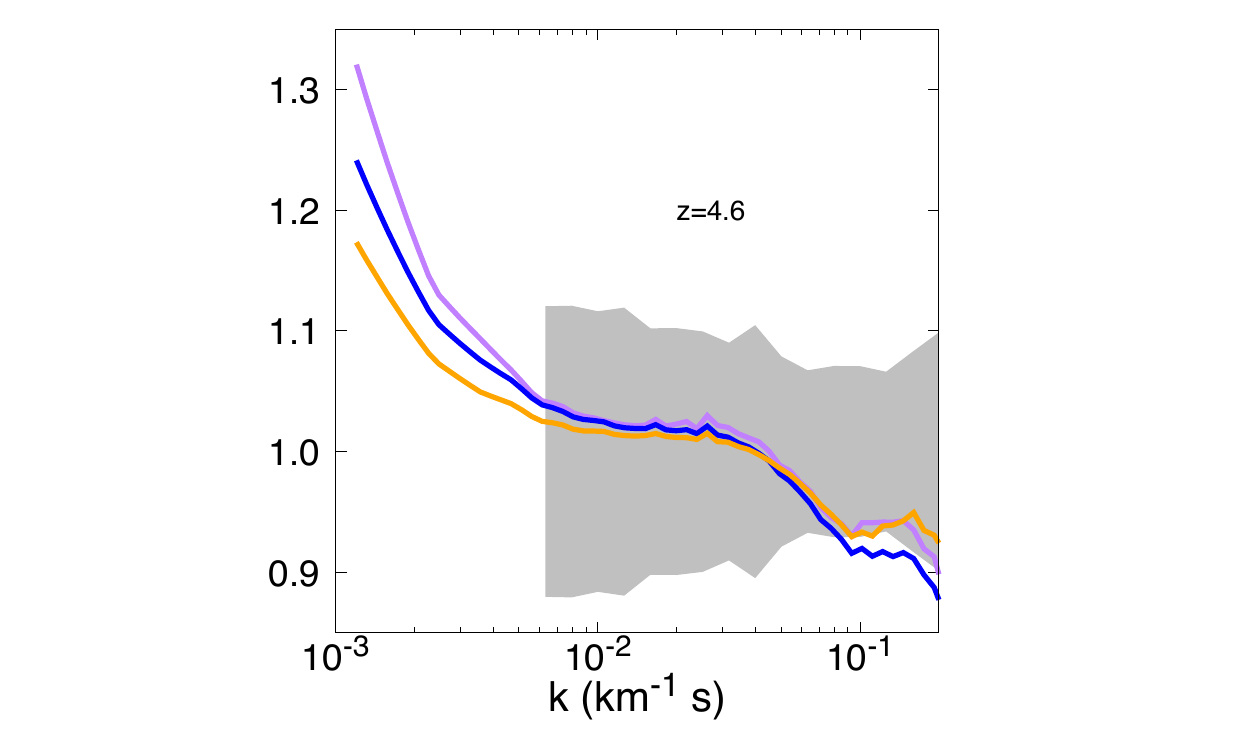}
    \includegraphics[height=6cm,trim=80 0 100 0]{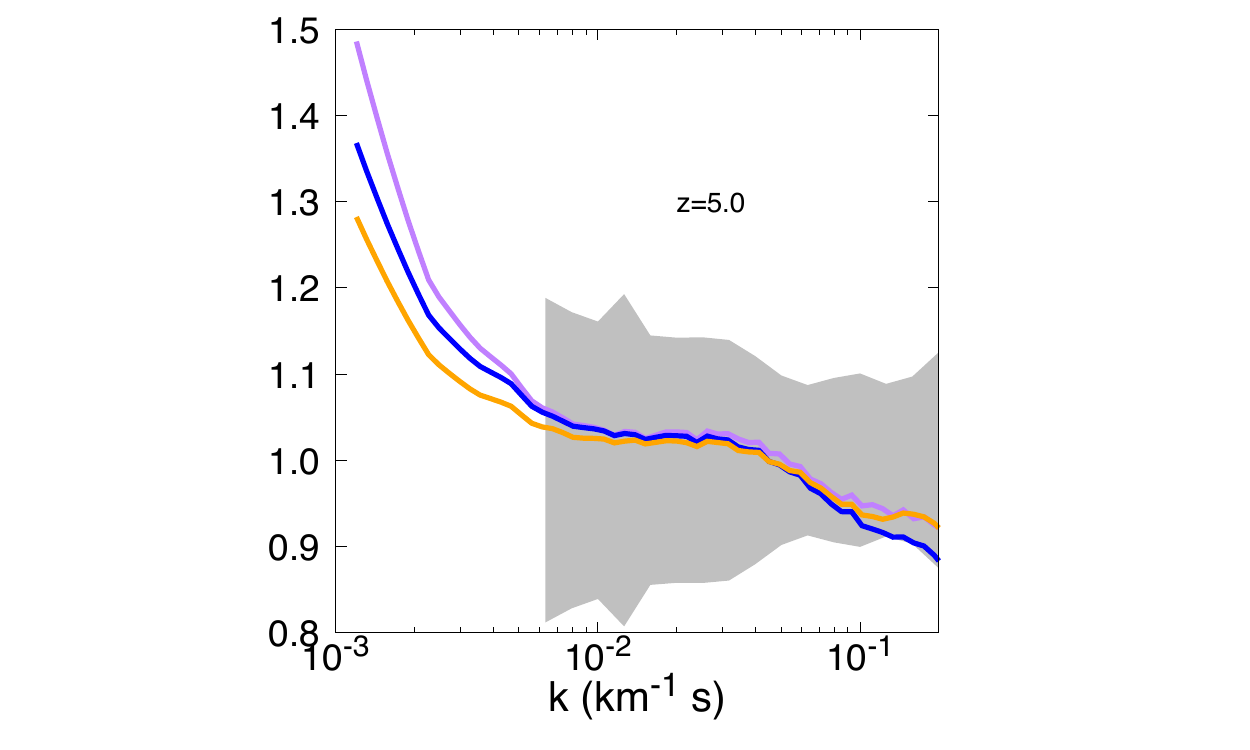}
    \vspace{-0.35cm}
    \caption{The ratio of the \Lya forest power spectrum in the hybrid-RT and paired homogeneous simulations, $R(k)$, at redshifts $z=4.2,4.6$ and $5.0$, for the three reionisation histories shown in Fig.~\ref{figure:global_props}. At each redshift, all mock spectra have been rescaled to the same effective optical depth, given by Eq.~(\ref{eqn:tau_trams}). The shaded grey region shows the 1 $\sigma$ uncertainties on the power spectrum measurements presented by \citet{Boera2019}.  Note the different scale on the vertical axis of each panel.}
    \label{figure:ratio_Lya_zr53_zr60_zr67}
\end{figure*}

\begin{figure*}
    \centering
    \includegraphics[height=6cm,trim = 80 0 70 0]{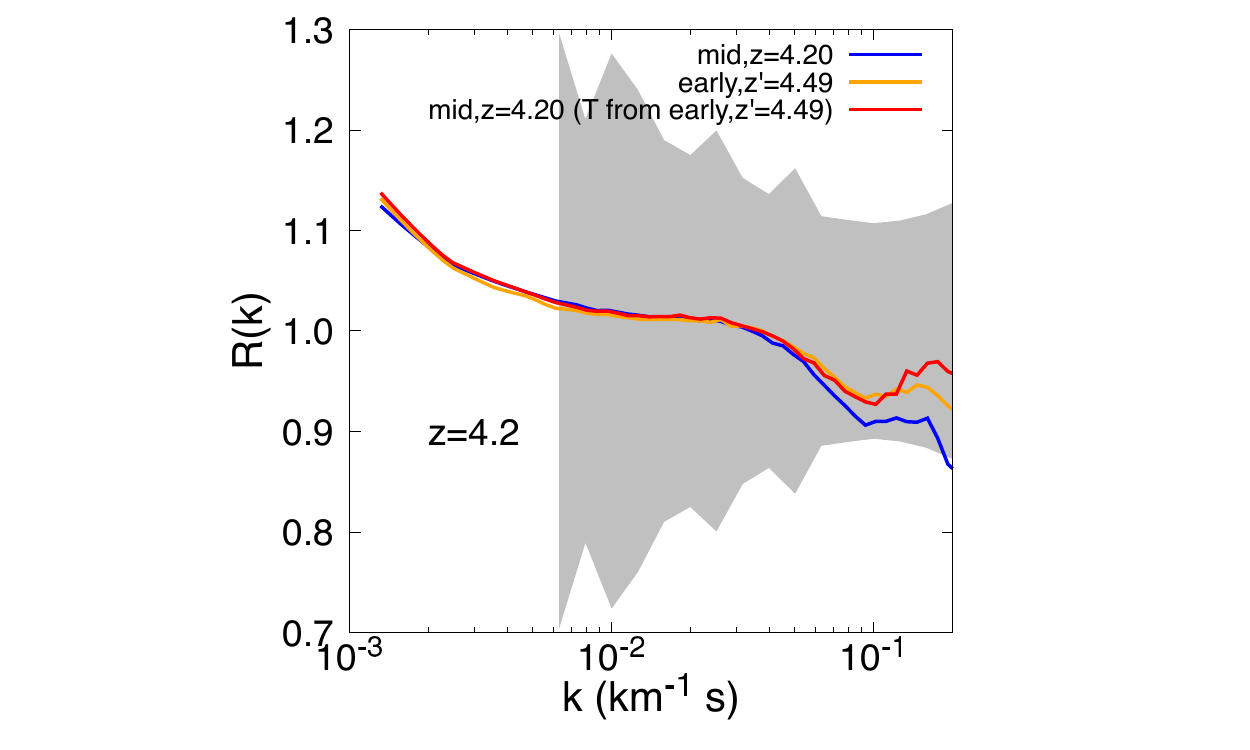} 
    \includegraphics[height=6cm,trim = 70 0 70 0]{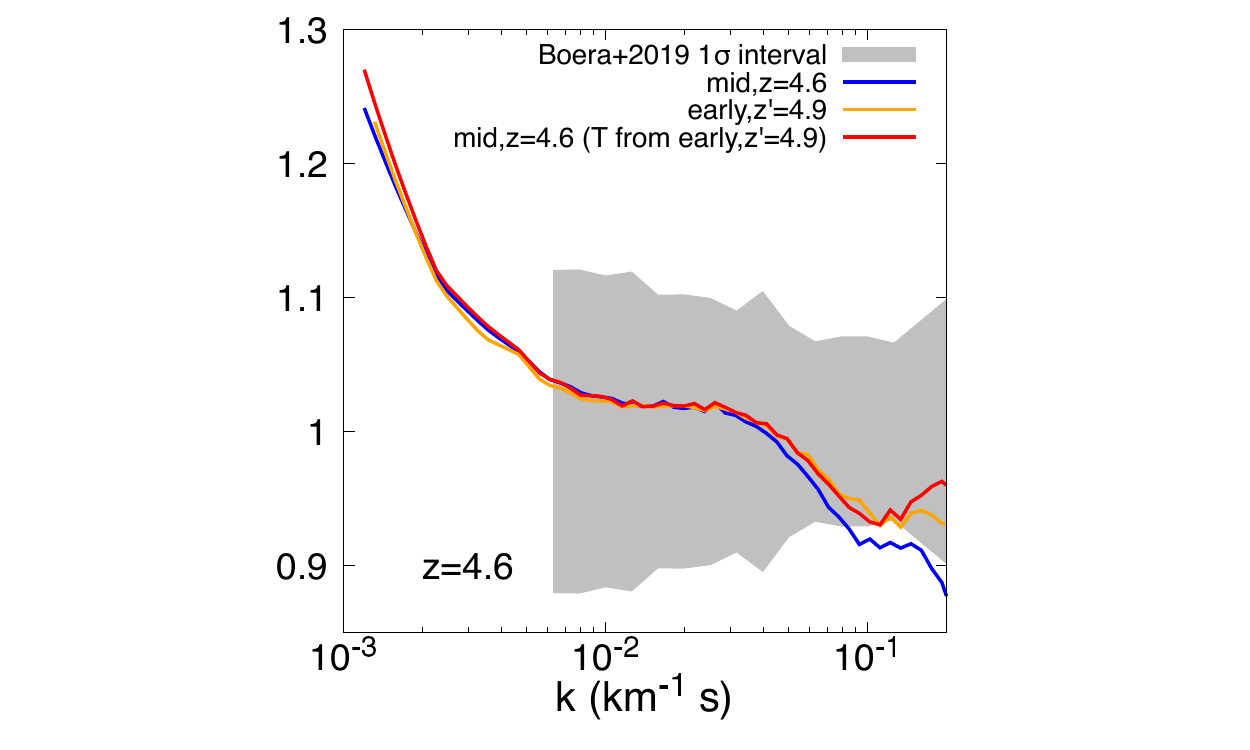}
    \includegraphics[height=6cm,trim = 80 0 100 0]{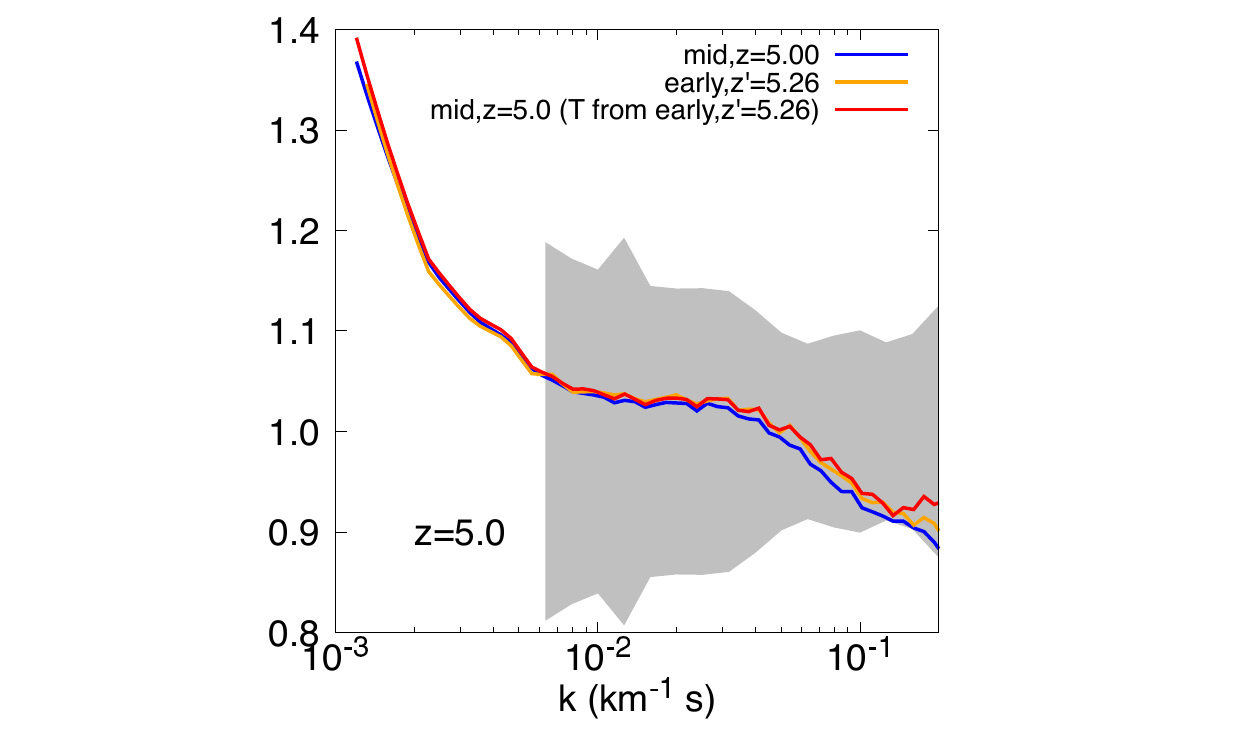} \\ 
    \vspace{-0.35cm}
    \caption{The ratio of the \Lya forest power spectrum from the RT-mid and Homog-mid simulations, $R\sm{mid}(k,z)$ (blue curves), in three redshift bins ($z=4.2,4.6,5.0)$, compared to $R\sm{early}(k,z^{\prime})$ (orange lines) at \textit{different} redshifts $z^{\prime}$. The redshift $z^{\prime}$ has been found by minimising the difference between $R\sm{mid}$ and $R\sm{early}$  at $k \leq 0.03\, \text{km}^{-1}\rm\,s$. This results in an excellent agreement between the large scale power predicted by the two reionisation models, with less than $1$ per cent difference between the two ratios at $k \leq 0.03\rm\,km^{-1}\,s$. This good agreement can be extended up to $k\sim 0.1 \rm\, km^{-1}\,s$ by substituting the gas temperatures in the RT-mid and Homog-mid simulations with those in the RT-early and Homog-early simulations, respectively (red lines), while leaving the original ionisation fraction unchanged.  This corrects for the remaining differences in the power spectra due to thermal broadening at small scales.  The shaded grey region shows the 1 $\sigma$ uncertainties on the power spectrum measurements presented by \citet{Boera2019}.  Note the different scale on the vertical axis of each panel.}
      \label{figure:interp_and_crosspol_zr67}
\end{figure*}

We now assess how different patchy reionisation models affect the \Lya forest power spectrum.  Alongside the RT-mid and Homog-mid models, we now use the simulations with the two additional reionisation histories displayed in Fig.~\ref{figure:global_props}:  a ``late'' model with reionisation\footnote{Recall that we formally define the end redshift of reionisation, $z_{\rm R}$, in the models as the redshift when the volume averaged \HI fraction first falls below $x_{\rm HI}=10^{-3}$.} ending at $z\sm{R} \simeq 5.3$, and an ``early'' model with $z\sm{R} \simeq 6.6$. In Fig. \ref{figure:ratio_Lya_zr53_zr60_zr67} the power spectrum ratio, $R(k)$, is displayed for all three models in redshift bins at $z=4.2,\,4.6$ and $5.0$.  The grey regions once again show the $1\sigma$ uncertainties on the power spectrum measurements from \citet{Boera2019}.  At large scales, $k\leq 0.03\rm\,km^{-1}\,s$, the extra power due to large-scale temperature flucuations is enhanced for a later end redshift to reionisation in all three redshift bins.  This is because the temperature fluctuations will begin to fade once the reionisation process has completed and the low density IGM adiabatically cools towards the thermal asymptote \citep[e.g][]{Theuns2002,HuiHaiman2003,McQuinnUpton2016}. The suppression of power at the smallest scales, on the other hand, shows no strong trend across the three redshift bins, although the early reionisation model has the smallest suppression in the small-scale power.  This is consistent with the early reionisation model being slightly colder than the other two models (see Fig.~\ref{figure:global_props}).  As discussed in Section~\ref{section:changes_Lya_forest}, differences between the hybrid-RT and homogeneous simulations at small scales are primarily due to thermal broadening (i.e. differences in the instantaneous gas temperature) and peculiar velocities associated with recently heated, expanding gas.  At large scales, the enhanced power is instead due to variations in the ionised hydrogen fraction associated with large scale temperature fluctuations.

Next, we consider whether the large scale enhancement of power at $k\leq 0.03 \rm\,km^{-1}\,s$ in the different reionisation histories can be emulated, simply by varying the redshift at which the power spectrum is measured from a \emph{single} reionisation model.  Since the large-scale temperature fluctuations fade predictably as the IGM cools following reionisation, we make the \emph{ansatz} that different reionisation histories should be equivalent in terms of their large scale power enhancement at similar time intervals following the completion of reionisation.  
We test this by comparing the power spectrum ratio, $R_{\rm early}(k)$, from the paired early reionisation models to the ratio from the mid reionisation models, $R_{\rm mid}(k)$. First, we compute $R_{\rm early}(k)$ on a redshift grid, using simulation outputs at intervals of $\Delta z=0.1$.  We linearly interpolate in redshift to obtain $R_{\rm early}(k)$ between these intervals.  Next, we find the redshift, $z^{\prime}$, at which $R_{\rm early}(k)$ best matches $R_{\rm mid}(k)$ on scales $k\leq 0.03\rm\,km^{-1}\,s$ at redshift $z$, where we use $z=4.2,\,4.6$ and $5.0$ as before.  In Fig. \ref{figure:interp_and_crosspol_zr67} we show the resulting ``match'' of $R_{\rm early}(k,z^{\prime})$ (orange curves) to $R_{\rm mid}(k,z)$ (blue curves) in the three redshift bins, where we find $z^{\prime}=4.49,\,4.90$ and $5.26$.  The ratios now show excellent agreement at \krange{}, with less than $1$ per cent difference between the two ratios.

At smaller scales, however, larger differences at the $\sim 5$ per cent level between the two ratios remain. We expect these differences to be mainly due to spatial variations in the thermal broadening kernel\footnote{In principle, any differences between $R(k,z^{\prime})$ and $R(k,z)$ on small-scales due to the \emph{hydrodynamical} response of the gas to heating (e.g. divergent peculiar velocities) should also be minimised at $z^{\prime}$.  The IGM in different reionisation models experience similar amounts of pressure smoothing when considered at approximately the same time interval after the initial heating \citep[see e.g.][for further discussion of this point]{Daloisio2019}}, as the models will not have exactly the same instantaneous gas temperatures at $z^{\prime}$ and $z$.   To verify this, we once again substitute the gas temperatures with those from other models, but this time substituting the Homog-early and RT-early temperatures at redshift $z'$ into the Homog-mid and RT-mid simulations at redshift $z$.  The result is shown by the red curves Fig. \ref{figure:interp_and_crosspol_zr67}, where we find that the good agreement now extends up to $k \sim 0.1 \text{ km}^{-1} \text{s}$.  We have also verified that a similar result holds on repeating this procedure for the late reionisation model ratio, $R_{\rm late}(k)$.   This implies that, in general, we may find $R(k,z^{\prime}) \simeq R\sm{mid}(k,z)$ for any $z\sm{R}$ within the range our simulations presently cover.   In this way we may emulate a ``patchy'' correction to the power spectrum simply by finding a best-fit relation for $z^{\prime}$ and $z$ as a function of $z_{\rm R}$.   This property is particularly useful for implementing a correction for inhomogeneous reionisation within existing grids of hydrodynamical simulations that use a spatially uniform ionising background.  

\begin{figure}
    \centering
      \includegraphics[trim=120 0 140 0, height=8cm]{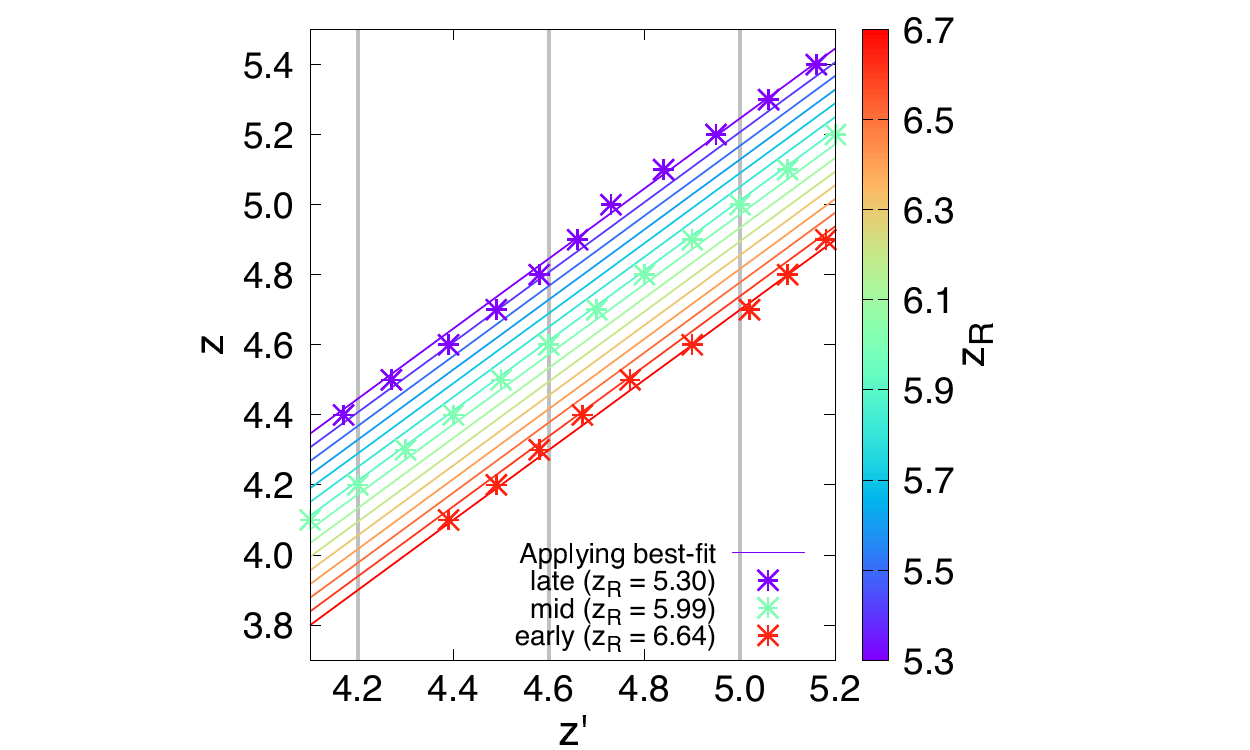}
      \vspace{-0.2cm}
    \caption{The redshifts, $z^{\prime}$, at which $R_{\rm early}(k,z^{\prime})$ (orange asterisks) and $R_{\rm late}(k,z^{\prime})$ (purple asterisks)  most closely match $R\sm{mid}(k,z)$ at redshift $z$ at $k \leq 0.03 \rm\,km^{-1}\,s$ (see text for details, as well as Fig. \ref{figure:interp_and_crosspol_zr67}).  The one-to-one identity relation for $R\sm{mid}(k,z)$ is also displayed (cyan asterisks).   The solid lines correspond to Eq.~(\ref{eqn:best_fit}), and show the best fit linear relation for $z$ as a function of both $z^{\prime}$ and the end redshift of reionisation, $z_{\rm R}$, where the colour of each line corresponds to the $z_{\rm R}$ shown in the colour bar (in $\Delta z_{\rm R} = 0.1$ steps, starting at $z_{\rm R}=5.3$).  The  vertical grey lines show the redshift bins that correspond to the power spectrum measurements presented by \citet{Boera2019}.}
    \label{figure:find_z60_zR_relation}
\end{figure}

\begin{figure}
    \centering
    \includegraphics[trim=50 0 0 0]{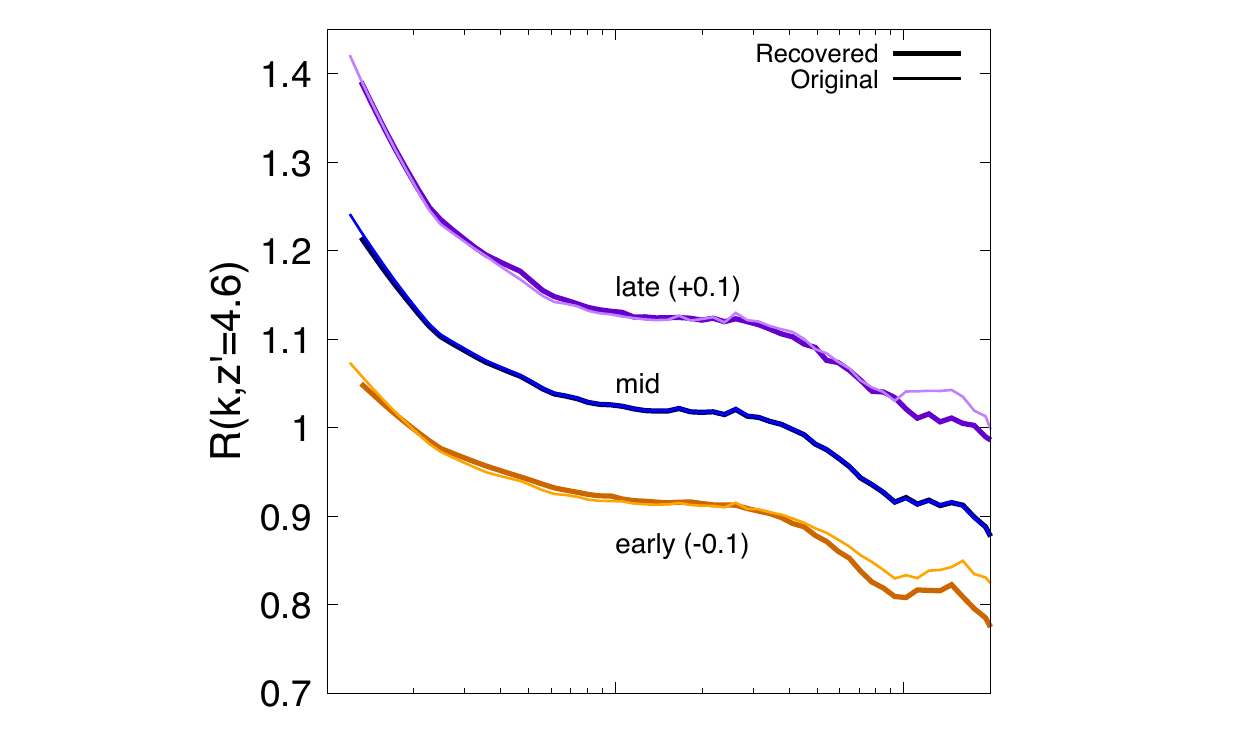}
    \includegraphics[width=6cm, trim = 63 50 62 45]{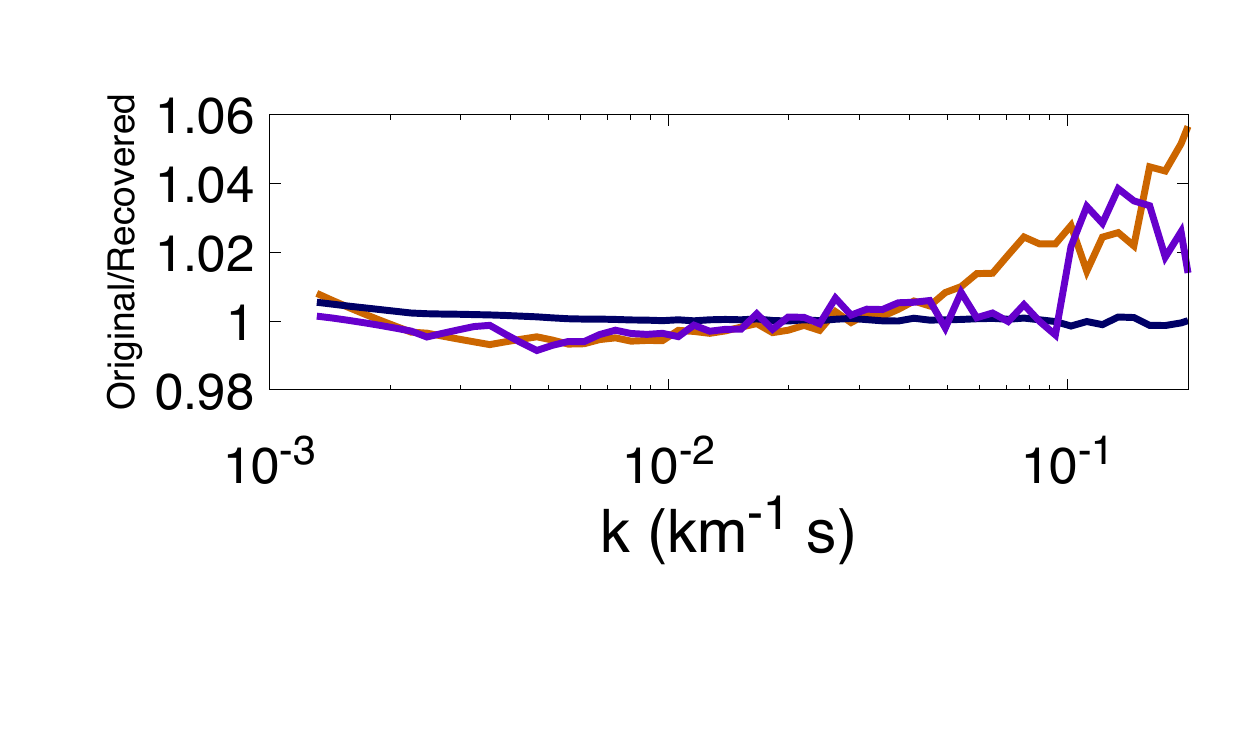}
    \vspace{0.1cm}
    \caption{{\it Upper panel:} Consistency test of the procedure we use for emulating a patchy reionisation correction to the \Lya forest power spectrum.    The test shows the recovery of $R\sm{zr}(k,z^{\prime})$ at $z^{\prime}=4.6$ for the late (orange curves), mid (blue curves), and early (purple curves) reionisation histories using only the power spectrum ratio from the mid model (see Table \ref{table:tabulated_pk}) and the best fit relation in Eq.~(\ref{eqn:best_fit}) to determine $z(z\sm{R},z')$.  Thick curves show the emulated ratio, while the thin curves show the original ratio from the simulations. For clarity, the early (late) models are shifted vertically by $-0.1$ ($+0.1$). {\it Lower panel:} The ratio of the original to recovered ratios shown in the upper panel. }
    \label{fig:apply_bf_tobins}
\end{figure}

We implement this correction within a Bayesian parameter inference framework in Section~\ref{section:MCMC} as follows.  In Fig. \ref{figure:find_z60_zR_relation}, we show all the $z^{\prime}-z$ pairs we have obtained from matching $R_{\rm early}(k)$ (green asterisks, $z_{\rm R}=5.3$) and $R_{\rm late}(k)$ (purple asterisks, $z_{\rm R}=6.6$) to $R_{\rm mid}(k)$, as well the one-to-one identity relation for $R_{\rm mid}(k)$ (cyan asterisks, $z_{\rm R}=6.0$).  Note that we limit ourselves to $z^{\prime} \leq 5.2$, in order to only consider redshifts where reionisation has completed in \text{all} models.  Furthermore, all our simulations currently finish at $z=4.1$, explaining why there are no data points at $z^{\prime}<4.4$ for the late reionisation model.  We then find the best-fit linear relation in the $z^{\prime}$--$z$ plane for \emph{each reionisation model}.  Furthermore, we assume there is an approximately linear increase in the best fit $z^{\prime}$--$z$ relation as a function of $z_{\rm R}$, over the range $5.3\leq z_{\rm R}\leq 6.7$.  The resulting linear relations are shown as the solid lines in Fig. \ref{figure:find_z60_zR_relation}, where the best fit for the redshift, $z$, at which $R_{\rm mid}(k)$ matches the power spectrum ratio for a model at $z^{\prime}$ with an end redshift of reionisation $z_{\rm R}$ is:
\begin{equation}
\label{eqn:best_fit}
    z(z^{\prime},z\sm{R}) = z^{\prime} + \xi z\sm{R} + \zeta, 
\end{equation}
where   $\xi =  - 0.390 \pm 0.007$ and $\zeta = 2.31 \pm 0.04$ for 1 $\sigma$ bootstrapped uncertainties on the best fit parameters. The inhomogeneous reionisation correction to the power spectrum is then $P_{\rm RT}(k,z^{\prime})=R_{\rm zr}(k,z^{\prime})P_{\rm homog}(k,z^{\prime})$, where $R_{\rm zr}(k,z^{\prime})=R_{\rm mid}(k,z)$, $R_{\rm mid}(k,z)$ is obtained by linearly interpolating the data in Table~\ref{table:tabulated_pk}, and $P_{\rm RT}(k,z^{\prime})$ and $P_{\rm homog}(k,z^{\prime})$ are, respectively, the corrected power spectrum and the original power spectrum from a simulation with a spatially uniform ionising background.

In Fig. \ref{fig:apply_bf_tobins} we perform a consistency check on this procedure.  We use  Eq.~(\ref{eqn:best_fit}) to obtain the redshift $z(z',z\sm{R})$, and hence $R\sm{mid}(k,z)$ for all three reionisation histories at $z^{\prime}=4.6$.  For wavenumbers \krange, the agreement is excellent, with a less than 1 per cent difference between the emulated (thick curves) and true (thin curves) ratios.  The differences at the smallest scales are up to 5 per cent, but as already discussed, this is explained primarily by spatial fluctuations in the thermal broadening in the models. These differences are furthermore well within the uncertainties on existing measurements of the \Lya forest power spectrum (see Fig.~\ref{figure:interp_and_crosspol_zr67}).

In summary, Table~\ref{table:tabulated_pk} combined with Eq.~(\ref{eqn:best_fit}) provide an approximate correction for inhomogeneous reionisation that can be applied to the \Lya forest power spectrum predicted by hydrodynamical simulations with a spatially uniform ionising background. A simple python script to calculate this correction is accessible at \url{https://github.com/marghemolaro/RT_1dps_correction.git}.  It is important to emphasise, however, that this correction is model dependent.  While we explore reionisation models that are broadly consistent with current observational constraints, our approach will not be applicable to reionisation histories that differ significantly from those displayed in Fig.~\ref{figure:global_props}, for reionisation models that finish well outside the range we have modelled, $5.3\leq z_{\rm R}\leq 6.7$. Furthermore, for simulation volumes that are much larger than we consider here, the accuracy of the power spectrum modelling may also benefit from an additional box size correction. Furthermore, note that our correction factor was obtained by comparing hybrid-RT and homogeneous UVB simulations that use a non-equilibrium thermo-chemistry solver. The application of our correction to grids of homogeneous UVB simulations that use an equilibrium solver (as is often the case in the literature) will therefore implicitly assume an initial spread in the temperature-density relation that differs from that predicted by equilibrium models, although such differences are expected to have a small effect on the 1D power spectrum \citep[see e.g.][]{Puchwein2015,Gaikwad2019}.
 
\begin{table*}
\caption{Tabulated values of $R\sm{mid}(z,k)$ (see e.g. the black curves in Fig.~\ref{figure:Ratio_Lyalpha_PS_wboot}). In combination with Eq.~(\ref{eqn:best_fit}), this table can be used to find an approximate correction to the \Lya forest power spectrum for the effects of inhomogeneous reionisation at redshift $z_{\rm R}$.  The corrected power spectrum is then $P_{\rm RT}(k,z^{\prime})=R_{\rm zr}(k,z^{\prime})P_{\rm homog}(k,z^{\prime})$, where $R_{\rm zr}(k,z^{\prime})=R_{\rm mid}(k,z)$. We have verified that linear interpolation between the tabulated redshift bins is sufficient for recovering the ratio for different reionisation histories to $1$ per cent accuracy for wavenumbers $k\leq 0.03\rm\,km^{-1}\,s$ (or equivalently, $\log(k/\rm km^{-1}\,s)\leq -1.5$) over the $z_{\rm R}$ range we consider.  However, at smaller scales,  $k \geq 0.03\rm\,km^{-1}\,s$, we expect the correction will only be accurate to within $\sim 5$ per cent due to differences associated with thermal broadening, although this is still well within the current measurement uncertainties.     Note that the $k$ bins are spaced following \citet{Boera2019}.}

\centering
\begin{tabular}{ c | c c c c c c c c c c c c c c}
 \hline
$\log(k/\rm km^{-1}\,s)$  & & & & & & & $z$ & & & & & & & \\
 & 4.1 & 4.2 & 4.3 & 4.4 & 4.5 & 4.6 & 4.7 & 4.8 & 4.9 & 5.0 & 5.1 & 5.2 & 5.3 & 5.4 \\
\hline
 -2.9  &  1.109  &  1.126  &  1.147  &  1.167  &  1.191  &  1.233  &  1.264  &  1.294  &  1.324  &  1.354  &  1.394  &  1.431  &  1.466  &  1.507 \\
 -2.8  &  1.092  &  1.105  &  1.122  &  1.138  &  1.157  &  1.182  &  1.205  &  1.228  &  1.250  &  1.272  &  1.300  &  1.328  &  1.353  &  1.383 \\
 -2.7  &  1.073  &  1.084  &  1.096  &  1.108  &  1.121  &  1.139  &  1.155  &  1.172  &  1.187  &  1.203  &  1.222  &  1.242  &  1.260  &  1.281 \\
-2.6  &  1.057  &  1.065  &  1.074  &  1.082  &  1.092  &  1.104  &  1.116  &  1.128  &  1.140  &  1.152  &  1.167  &  1.183  &  1.198  &  1.216 \\
-2.5  &  1.047  &  1.055  &  1.061  &  1.068  &  1.075  &  1.085  &  1.095  &  1.103  &  1.112  &  1.122  &  1.135  &  1.149  &  1.163  &  1.180 \\
-2.4  &  1.039  &  1.045  &  1.050  &  1.055  &  1.061  &  1.069  &  1.077  &  1.084  &  1.092  &  1.101  &  1.113  &  1.126  &  1.139  &  1.153 \\
-2.3  &  1.032  &  1.037  &  1.041  &  1.044  &  1.049  &  1.054  &  1.060  &  1.066  &  1.073  &  1.079  &  1.088  &  1.097  &  1.105  &  1.117 \\
-2.2  &  1.027  &  1.029  &  1.031  &  1.033  &  1.035  &  1.038  &  1.041  &  1.045  &  1.050  &  1.054  &  1.062  &  1.071  &  1.081  &  1.093 \\
-2.1  &  1.021  &  1.023  &  1.025  &  1.027  &  1.028  &  1.029  &  1.032  &  1.035  &  1.038  &  1.040  &  1.045  &  1.051  &  1.060  &  1.071 \\ 
-2.0  &  1.019  &  1.020  &  1.021  &  1.023  &  1.025  &  1.025  &  1.028  &  1.030  &  1.033  &  1.036  &  1.040  &  1.047  &  1.055  &  1.064 \\
-1.9  &  1.015  &  1.016  &  1.017  &  1.018  &  1.019  &  1.020  &  1.023  &  1.024  &  1.027  &  1.031  &  1.036  &  1.043  &  1.050  &  1.057 \\
-1.8  &  1.012  &  1.014  &  1.015  &  1.017  &  1.018  &  1.021  &  1.023  &  1.024  &  1.024  &  1.024  &  1.027  &  1.033  &  1.041  &  1.052 \\
-1.7  &  1.012  &  1.012  &  1.014  &  1.016  &  1.017  &  1.017  &  1.018  &  1.020  &  1.024  &  1.028  &  1.032  &  1.038  &  1.048  &  1.062 \\
-1.6  &  1.009  &  1.011  &  1.013  &  1.014  &  1.017  &  1.019  &  1.020  &  1.021  &  1.022  &  1.024  &  1.026  &  1.031  &  1.038  &  1.047 \\
-1.5  &  1.003  &  1.004  &  1.005  &  1.007  &  1.010  &  1.012  &  1.014  &  1.015  &  1.018  &  1.023  &  1.028  &  1.035  &  1.042  &  1.050 \\
-1.4 &  0.989 &  0.990 &  0.993 &  0.996 &  0.998  &  1.001  &  1.003  &  1.003  &  1.007  &  1.013  &  1.018  &  1.024  &  1.031  &  1.036 \\
-1.3 &  0.975 &  0.975 &  0.976 &  0.978 &  0.979 &  0.980 &  0.983 &  0.987 &  0.991 &  0.993 &  0.995 &  0.999  &  1.003  &  1.011 \\
-1.2 &  0.946 &  0.949 &  0.954 &  0.955 &  0.956 &  0.960 &  0.964 &  0.968 &  0.968 &  0.972 &  0.976 &  0.978 &  0.983 &  0.992 \\
-1.1 &  0.924 &  0.923 &  0.924 &  0.923 &  0.930 &  0.935 &  0.938 &  0.940 &  0.941 &  0.946 &  0.950 &  0.954 &  0.960 &  0.966 \\
-1.0 &  0.909 &  0.909 &  0.907 &  0.911 &  0.924 &  0.919 &  0.921 &  0.921 &  0.929 &  0.928 &  0.927 &  0.924 &  0.931 &  0.935 \\
-0.9 &  0.908 &  0.913 &  0.915 &  0.916 &  0.922 &  0.917 &  0.919 &  0.915 &  0.915 &  0.914 &  0.906 &  0.909 &  0.914 &  0.914 \\
-0.8 &  0.900 &  0.913 &  0.912 &  0.909 &  0.913 &  0.913 &  0.904 &  0.903 &  0.909 &  0.905 &  0.900 &  0.886 &  0.883 &  0.885 \\
-0.7 & 0.869 & 0.862 & 0.876 & 0.870 & 0.877 & 0.879 & 0.882 & 0.886 & 0.883 & 0.883 & 0.882 &  0.877  &  0.869 & 0.864 \\
       \hline
\end{tabular}

\label{table:tabulated_pk}
\end{table*}

\section{Recovery of thermal parameters from mock observations}
\label{section:MCMC}

A commonly used method for physical parameter recovery from \Lya forest power spectrum observations is Bayesian inference, where the data are compared with a grid of hydrodynamical simulations that span the relevant parameter space \citep{Viel2013,Irsic2017,Yeche2017,Boera2019,Palanque2020,Rogers2021}.  So far, aside from \citet{Onorbe2019} and \citet{Wu2019}, the parameter grids used in these studies have typically relied on simulations that assume homogeneous UV backgrounds.  In Section \ref{section:changes_Lya_forest}, however, we have shown that replacing a homogeneous UV background with a more realistic model for “patchy” reionisation can lead to significant differences in the resulting power spectrum.  Given that any observations may include such patchy effects, it is therefore important that we ask whether a grid of simulations using a homogeneous UV background is still able to accurately recover the underlying physical parameters from observations, or whether patchy reionisation effects on the power spectrum could significantly bias any constraints \citep[see e.g.][for a discussion of this point]{Hui2017}.

In order to answer this question, we proceed as follows.  First, we generate a set of mock \Lya forest data ($D\sm{mock}$) based on one of our patchy simulations. We then construct two grids of simulations for interpolating between the physical parameters of interest: one, \Ghomog{}, relying on homogeneous UV background simulations, and the other, \Grecov{}, additionally including the patchy correction described in Eq.~(\ref{eqn:best_fit}). We then perform a Bayesian inference analysis to assess if either model grid allows us to accurately recover the thermal parameters used to generate the mock data.

To perform our Bayesian inference analysis, we sample the parameter space using a Monte Carlo Markov Chain (MCMC) sampler, based on the Metropolis-Hastings algorithm \citep[e.g.][]{Irsic2017}. In Bayesian inference, the resulting distribution in the parameter space is directly proportional to the likelihood of the data given the input parameters, \emph{i.e.} the so-called Bayes' theorem,
\begin{equation}
    p(\theta|D) \propto {\cal L}(D|\theta) \pi(\theta),
\end{equation}
where ${\cal L}(D|\theta)$ is the likelihood, $\pi(\theta)$ contains the prior information on the parameters, and $p(\theta|D)$ is the posterior probability of the parameters obtained as the end result in our analysis. In order to sample the posterior distribution one needs to evaluate the likelihood and priors at each point in the parameter space. For the mock measurements of the \Lya forest power spectrum we use a Gaussian likelihood with the covariance matrix evaluated from the Homog-late simulation using a bootstrap method. We have also tested whether using either the homogeneous or Hybrid-RT simulations to generate the covariance matrix affects our results, and have verified that the resulting differences are negligible compared to the width of the posteriors. Different redshift bins in the analysis are treated as independent, and as such the Bayesian analysis at each of the three redshift bins we consider ($z=4.2$, $4.6$ and $5.0$) is independent.

\subsection{Thermal parameters and their priors}
\label{section:thermparams}

We now discuss the thermal parameters and priors we consider in our parameter recovery analysis.  Each of the (homogeneous UV background) simulations used for constructing our parameter grid features different heating rates at the onset of hydrogen reionisation (see the upper section of Table~\ref{table:summary_sims}).  These different heating rates are partly characterised by the cumulative energy per unit mass deposited into gas at the mean background density (see e.g. Fig.~\ref{figure:global_props}). This quantity, $u_0$, is highly correlated with the suppression of the \Lya forest power spectrum on small scales \citep{Nasir2016} and it acts as a proxy for the amount of pressure (or Jeans) smoothing induced by the integrated thermal history. Furthermore, \citet{Boera2019} (see their appendix J) have demonstrated that $u_{0}$  correlates very well with the physical scale used to characterise the physical extent of the pressure smoothing  \citep[e.g.][]{GnedinHui1998,Kulkarni2015}, with typical values of $u_0 \sim 6\rm\,eV\,m_{p}^{-1}$ corresponding to a smoothing scale $\lambda_{\rm p}\sim 60\;\mathrm{ckpc}$ at $z=4.2$.  It has also been shown that the small scale power in the \Lya forest at the three redshifts we consider here, $z= 4.2, 4.6$ and $5.0$, tightly correlates with $u_0$ integrated over a redshift range of $z=[4.2 - 12.0], [4.6 - 12.0]$ and $[6.0 - 13.0]$, respectively \citep{Boera2019}. Here we similarly adopt this definition, where the $u_0$ range covered by our parameter grid is $u_0=[4.02,21.1]$, $[3.65,21.1]$ and $[2.46,18.7] \;\mathrm{eV}\rm\,m_{\rm p}^{-1}$ at $z=4.2$, $4.6$ and $5.0$, respectively (see Table~\ref{table:summary_bayes_params}).

\begin{table}
\centering

\caption{Summary of the astrophysical parameters and priors used in our analysis of the \Lya forest power spectrum.}
\begin{tabular}{ l | l } 
\hline
Parameter & Prior (flat) \\
\hline
$\tau_{\rm eff}(z=4.2)$ & [0.3 - 1.8] $\times 1.01$ \\
$\tau_{\rm eff}(z=4.6)$ & [0.3 - 1.8] $\times 1.37$ \\
$\tau_{\rm eff}(z=5.0)$ & [0.3 - 1.8] $\times 1.92$ \\
$T_0(z=4.2)$ & [0.5 - 1.5] $\times 10^4 \;\mathrm{K}$ \\
$T_0(z=4.6)$ & [0.5 - 1.5] $\times 10^4 \;\mathrm{K}$ \\
$T_0(z=5.0)$ & [0.5 - 1.5] $\times 10^4 \;\mathrm{K}$ \\
$\gamma(z=4.2)$ & [1.0 - 1.7] \\
$\gamma(z=4.6)$ & [1.0 - 1.7] \\
$\gamma(z=5.0)$ & [1.0 - 1.7] \\
$u_0^{z=4.2}(4.2 - 12.0)$ & [4.02 - 21.1] $\;\mathrm{eV}\,\rm m_{\rm p}^{-1}$ \\
$u_0^{z=4.6}(4.6 - 12.0)$ & [3.65 - 21.1] $\;\mathrm{eV}\,\rm m_{\rm p}^{-1}$ \\
$u_0^{z=5.0}(6.0 - 13.0)$ & [2.46 - 18.7] $\;\mathrm{eV}\,\rm m_{\rm p}^{-1}$ \\
\hline
\end{tabular}
\label{table:summary_bayes_params}
\end{table}

However, $u_0$ alone is insufficient for describing the full range of possible IGM thermal histories. For instance, the same $u_0$ could be obtained if reionisation started early but the \emph{instantaneous} gas temperature of the IGM remained low (e.g. if reionisation were driven by ionising sources with soft spectra), or where reionisation occurred late but was driven by hard ionising sources that heated the IGM to much higher temperatures. Our choice of thermal parameters must therefore capture both the duration of the reionisation process as well as the instantaneous temperature. To decouple this information we therefore use two further parameters that  describe the temperature-density relation of the post reionisation IGM in the homogeneous UV background models\footnote{Note that a well defined power law temperature-density relation will not always apply in our patchy reionisation simulations, due to spatial fluctuations in the IGM temperature \citep[see e.g.][]{Keating2018}.  As already discussed earlier, however, the patchy correction given by Eq.~(\ref{eqn:best_fit}) effectively corrects for the effect that any departure from this power-law relation has on the \Lya forest power spectrum.}  -- the temperature at mean density, $T_0$, and the power-law index of the temperature-density relation, $\gamma$.  The relation between the (instantaneous) temperature and the density of the gas can thus be written as $T = T_0\Delta^{\gamma-1}$, where $\Delta = \rho/\langle \rho \rangle$ \citep{Hui1997,McQuinn2016}.  As for $u_{0}$, our assumed priors for $T_{0}$ and $\gamma$ are listed in Table~\ref{table:summary_bayes_params}; the prior ranges are chosen to encompass a physically plausible range of values.  Furthermore, since the goal of this analysis is to investigate potential shifts and biases in the recovery of these thermal parameters due to inhomogeneous reionisation, in the subsequent MCMC analysis we adopt agnostic flat priors on all three of these thermal parameters, $T_{0}$, $\gamma$, and $u_0$.

Lastly, the transmission in the \Lya forest also depends on the \HI photo-ionisation rate, $\Gamma_{\rm HI}$.  We therefore adopt the standard approach \citep[e.g.][]{Viel2004} of varying the effective optical depth in the simulated spectra, $\tau_{\rm eff}$, around the observed value for each of the redshift bins (see also the discussion in Section \ref{section:sim_power_spec}).  

\subsection[Emulator for the Lya forest power spectrum]{Emulator for the \Lya forest power spectrum}
\label{section:grids}

To evaluate the likelihood function used in our Bayesian inference analysis, we build an emulator that predicts the power spectrum of the \Lya forest transmitted flux at any given point in the parameter space.  Our emulator is based on the linear interpolation of a grid of simulated flux power spectra. This is a method that -- despite its simplicity -- works remarkably well and has been extensively utilised in earlier \Lya{} forest studies \citep{Viel2013,Irsic2017,Yeche2017,Palanque2020} and adapted in more sophisticated Gaussian process emulators \citep{Bird2019,Pedersen2021}.

We start by first constructing a grid of simulated flux power spectra from the suite of $12$  simulations in $20h^{-1}\rm\,cMpc$ boxes listed in the upper section of Table \ref{table:summary_sims} (for further details see Section~\ref{section::simulation}). In order to construct a sufficiently well sampled grid of models spanning the entire parameter range, we then post-process the $12$ initial simulations to achieve different parameter combinations for $u_{0},\,T_{0},\,\gamma$ and $\tau_{\rm eff}$. We refer to this interpolated grid as \Ghomog{}.  In order to interpolate the $T_0$--$\gamma$ plane, we follow the method described in \citet{Boera2019} and \citet{Gaikwad2020}. Briefly, we rotate and translate the gas particles in the temperature-density plane to obtain models with different $T_{0}$ and $\gamma$. This preserves the temperature-density cross-correlation coefficient, allowing one to inexpensively construct models with different thermal parameters on a finely spaced grid. We also include a  small correction for the smaller box size of these runs relative to our hybrid-RT models using the $40-2048$ model in the upper part of Table~\ref{table:summary_sims}. The size of the correction is at most 4\% at large scales, and below 2\% on average, over the scales of interest for this work.

Using this method, we construct a $15\times 10 \times 7=1050$ grid of parameter values on top of \emph{each} of the twelve $20h^{-1}\rm\,cMpc$ simulations in the upper section of Table~\ref{table:summary_sims}. The grid (\Ghomog{}) consists of $10$ values of $T_0$ spanning the range from $5,000\rm\,K$ to $15,000\rm\,K$  in steps of $1,000\rm\,K$ , $7$ values of $\gamma$ from $1.0$ to $1.7$ in steps of $0.1$, and $15$ different values of $\tau_{\rm eff}$ ranging from $0.3$ to $1.8$ times the value given by Eq.~(\ref{eqn:tau_trams}), in multiplicative steps of $0.1$. For $12$ simulations with different $u_{0}$ values, this gives a total of $12\times 1050=12\,600$ combinations on our parameter grid. 

We then adopt the patchy correction from Eq.~(\ref{eqn:best_fit}) in conjunction with the values in Table \ref{table:tabulated_pk} to construct a second ``recovered'' grid of models (\Grecov{}) that, while relying on the same homogeneous UV background simulations adopted in \Ghomog{}, now also corrects for the effect of patchy reionisation on the transmitted flux power spectra. The choice of $z\sm{R}$ is set by the end point of reionisation in each of the homogeneous models.  More specifically, when “correcting” the simulations in the top section of Table \ref{table:summary_sims} to obtain \Grecov{},  we adopt the $z\sm{R}$ values listed in column 6, and then use Eq.~(\ref{eqn:best_fit}) to modify the power spectrum appropriately. This allows us to construct, using the same method described above, an emulator that includes the effect of a patchy UV background. Note, however, that for the models with $z_{\rm R} \simeq 7.4$, this requires the uncertain extrapolation of the data in Table~\ref{table:tabulated_pk} to $z^{\prime}<4.1$.  We have therefore imposed a floor to Eq.~(\ref{eqn:best_fit}), such that we always assume $z^{\prime}=4.1$ if using a  $z\sm{R}$ values that give $z^{\prime}<4.1$ in Eq.~(\ref{eqn:best_fit}).  This may slightly overestimate the expected patchy correction for the $z_{\rm R}=7.4$ simulations, although note the patchy reionisation correction is already modest at $z^{\prime}=4.1$ and will become progressively smaller toward lower redshift.  As a check, we have also compared the results obtained from this approach to those obtained for a linear extrapolation of the data below $z^{\prime}=4.1$ in Table~\ref{table:tabulated_pk}. We find the differences are very small, and we do not expect this choice to impact on our conclusions.

\subsection[Generation of mock Lya forest power spectrum data]{Generation of mock \Lya forest power spectrum data}
\label{section:mockdata}

Our mock \Lya forest power spectrum data were constructed from our RT-late simulation. This is the reionisation model in which the effects of inhomogeneous reionisation on the power spectrum of the transmitted flux will be most prominent. The range of scales where we perform the power spectrum analysis is chosen to match  \citet{Boera2019}, $-2.2\leq \log(k/\rm km^{-1}\,s)\leq -0.7$, where the patchy correction is at the $5-10$ per cent level at scales of $k\simeq 0.1\rm km^{-1}\,s$, depending on the redshift and the reionisation model used.  We do not add any instrumental effects (e.g. noise or line broadening due to spectral resolution) to the mock data, although any detailed comparison to observed data will ultimately require this.

We construct two sets of data from RT-late, one with $\sim 10$ per cent relative errors (${D}\sm{mock}^{10\%}$), and one with $\sim 5$ per cent relative errors (${D}\sm{mock}^{5\%}$). The \citet{Boera2019} measurements typically have a scale dependent uncertainty between $10$--$25$ per cent (see e.g. Fig.~\ref{figure:interp_and_crosspol_zr67}) with a redshift path length of $dz=2.26$, $dz=5.63$ and $dz=2.41$ at $z=4.2$, $4.6$ and $5.0$, respectively. The covariance matrices for ${D}\sm{mock}^{10\%}$ (${D}\sm{mock}^{5\%}$)  were obtained by bootstrapping the mock data using single redshift snapshots with path lengths 
 $\times 2, \times 1, \times 1.6$ ($\times 9.5, \times 4.3, \times 7.3$) compared to those considered by \citet{Boera2019} in the $z=4.2,4.6,5.0$ redshift bins respectively. These path lengths were selected to achieve a mean uncertainty of 5$\pm$-1 per cent and 10 $\pm$ 1 per cent, respectively, for the $k$-bins under consideration. Hence, in the case of ${D}\sm{mock}^{10\%}$ we consider mocks that are slightly more precise than current observed data sets from high resolution data. On the other hand,  ${D}\sm{mock}^{5\%}$ tests the significantly improved precision that might be achieved by future measurements. This is likely to be achievable in the near future with a combination of the current-generation of large-scale surveys at low spectral resolution ($R\leq 5000$) such as DESI \citep{VargasMagana2019} and WEAVE-QSO \citep{Pieri2016}, and the increasing availability of homogeneous samples of high resolution quasar spectra ($R\geq 20\,000$) such as KODIAQ \citep{OMeara2021}, SQUAD \citep{Murphy2019} and XQR30 \citep{Bosman2021}.

\subsection{Results of MCMC analysis}
\label{section:bayes}

\begin{figure*}
\centering
%\addtocounter{figure}{-1}
\begin{subfigure}{.5\textwidth}
    \centering
    \includegraphics[scale=0.29]{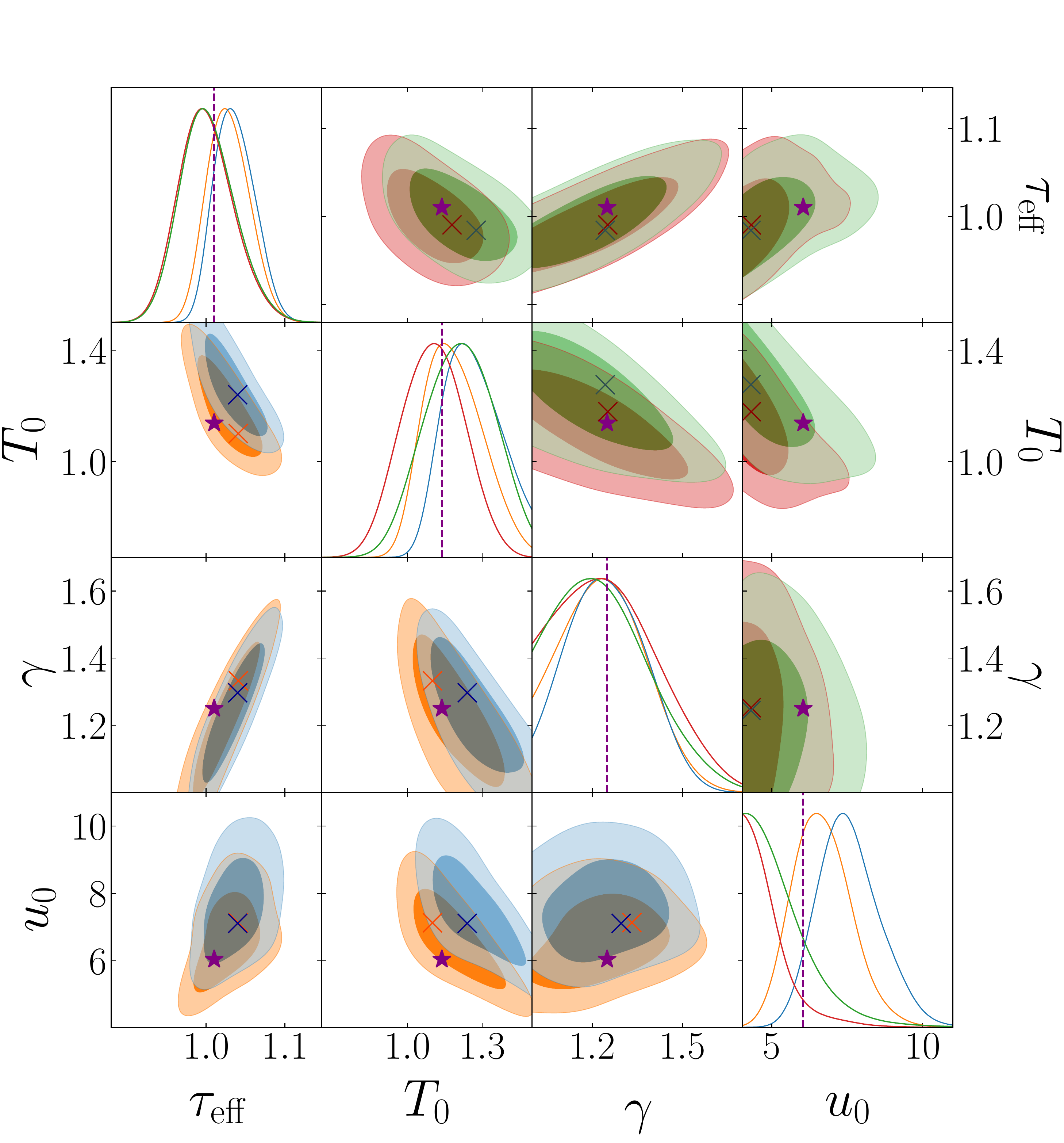}
    \caption{$z=4.2$}
    %\label{fig:sub1}
\end{subfigure}%
\begin{subfigure}{.5\textwidth}
    \centering
    \includegraphics[scale=0.29]{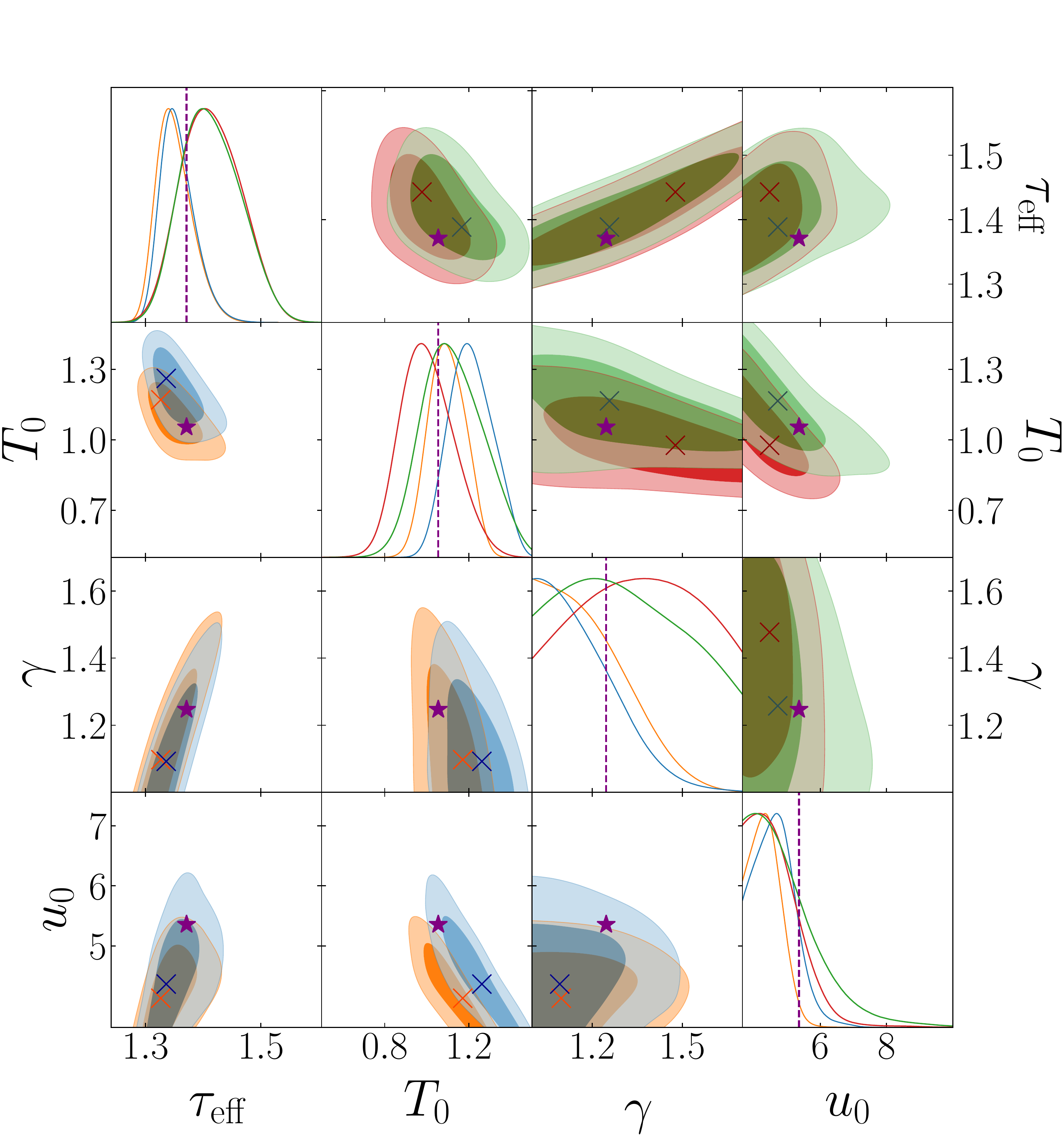}
    \caption{$z=4.6$}
    %\label{fig:sub2}
\end{subfigure}
\vskip\baselineskip
\begin{subfigure}{.5\textwidth}
    \centering
    \includegraphics[scale=0.3]{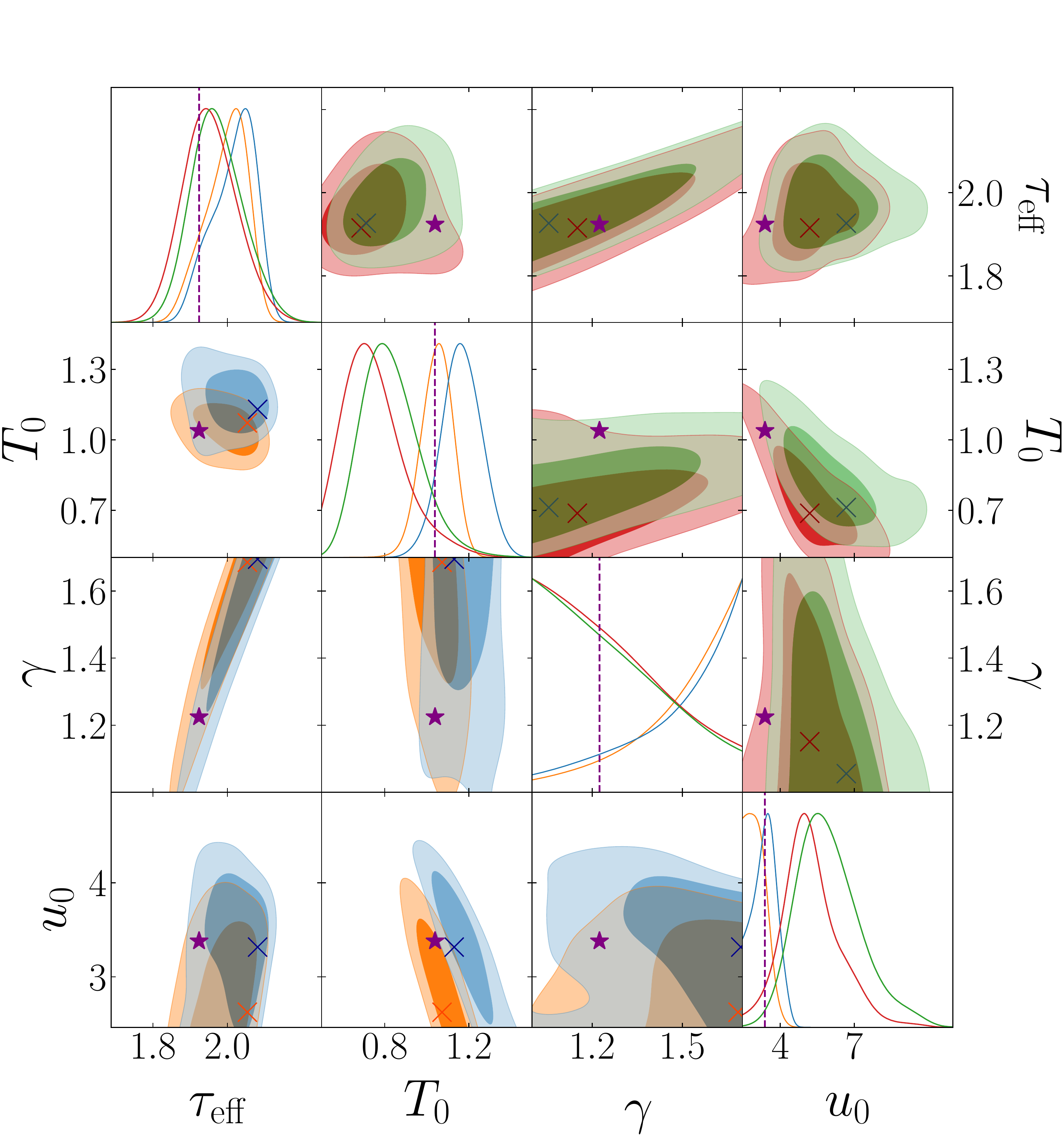}
    \caption{$z=5.0$}
    %\label{fig:sub3}
\end{subfigure}%
\begin{subfigure}{.5\textwidth}
    \centering
    \includegraphics[scale=1.0]{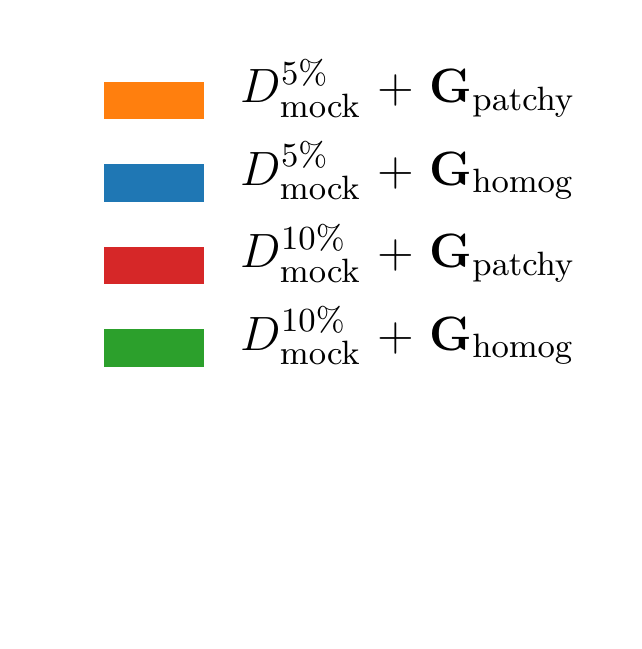}
    %\label{fig:sub3}
\end{subfigure}%
\caption{The one and two dimensional posterior distributions for $\tau_{\rm eff}$, $T_{0}$, $\gamma$ and $u_{0}$ from our analysis of mock \Lya forest power spectrum data drawn from the RT-late simulation at $z=4.2$ (upper left), $z=4.6$ (upper right) and $z=5.0$ (lower left).  The results are obtained using a grid of homogeneous UV background models (\Ghomog{}, blue and green contours) and when the patchy correction from Eq.~(\ref{eqn:best_fit}) was applied to each of the grid models (\Grecov{}, orange and red contours). These two analyses were performed on mock power spectrum data with a relative error of $\sim 10$ per cent ($D_{\rm mock}^{10\%}$, upper part of each panel), and $\sim 5$ per cent ($D_{\rm mock}^{5\%}$, lower part of each panel). The purple stars and vertical dashed lines correspond to the true parameter values. The crosses correspond to the best-fit parameters from the MCMC analysis with  (\Grecov{}, orange/red) and without (\Ghomog{}, blue/green) the patchy correction from Eq.~(\ref{eqn:best_fit}) applied to the parameter grid.}
\label{fig:mcmc_5p_panels}
\end{figure*}

The main results of our MCMC analysis are summarised in Fig.~\ref{fig:mcmc_5p_panels}. The three panels show the redshift bins $z = 4.2$ (top left), $4.6$ (top right) and $5.0$ (bottom). Each panel consists of $4\times 4$ sub-panels showing the marginalised two-dimensional or one-dimensional posterior distributions for each of the four parameters, $\tau_{\rm eff}$, $T_0$, $\gamma$ and $u_0$.  The sub-panels in the upper triangular section of each panel show the posteriors in the case where the relative errors on the mock \Lya forest power spectrum are $\sim 10$ per cent ($D_{\rm mock}^{10\%}$). The parameter estimation using either \Ghomog{} (green) or \Grecov{} (red) is almost identical, with the posteriors largely overlapping. The purple stars (and vertical dashed lines) in the sub-panels indicate the true parameter values used to construct the mocks. The true parameters are well recovered and are always within $1-2\sigma$ of the posterior distribution. The best-fit $\chi^2$ values are 44.2 and 46.1 for \Grecov{} and \Ghomog{} respectively, with 36 degrees of freedom. Similarly, the best-fit parameter values, indicated as a green or red cross for \Ghomog{} and \Grecov{}, respectively, are well captured within the posterior distributions.  Overall, this suggests that for $\sim 10$ per cent uncertainty on the flux power spectrum measurements -- which is similar or slightly better than the uncertainty on current high resolution ($R\sim 40,\,000$) data at $z>4$ \citep{Boera2019} --  the effects of patchy reionisation do not affect the recovery of the thermal state of the IGM \citep[see also][]{Wu2019}. 

Similar behaviour is observed in the lower section of each panel in Fig.~\ref{fig:mcmc_5p_panels}, where the results using \Ghomog (blue) and \Grecov{} (orange) and the mock data with 5 per cent relative errors ($D_{\rm mock}^{5\%}$) are displayed.  As expected, the posteriors are tighter compared to the $D_{\rm mock}^{10\%}$ case.   The true parameter values are again indicated as purple stars, showing that even when the uncertainty on the power spectrum is reduced by a factor of two, the parameters are still well recovered for \Grecov{} (orange contours).  The best fit \Grecov{} model has the $\chi^2$ value of 30.2 for 36 degrees of freedom.   Note, however, the best-fit values are sometimes shifted along a degeneracy axis with respect to the true parameter values.    While these correlations indicate that the parameters are not completely independent, they are nonetheless informative. There are three obvious correlations between $\tau_{\rm eff}$--$\gamma$, $T_0$--$\gamma$ and $T_0$--$u_0$. The first arises because the response of the power spectrum to $\tau_{\rm eff}$ or $\gamma$ is nearly scale independent, due to the power spectrum being sensitive to gas close to mean density at $z\simeq 4$--$5$.  The second parameter correlation, $T_0$--$\gamma$, arises because the \Lya{} forest is sensitive to gas over a narrow range of densities \citep[e.g.][]{Becker2011}.  This correlation therefore disappears when the typical overdensity probed by the \Lya{} forest becomes close to the mean density of the Universe.   Finally, the last parameter correlation is between $u_0$ and $T_0$, which is also the most pronounced in our analysis. This anti-correlation results from the small-scale power suppression in the \Lya{} forest by thermal broadening traced by $T_0$, and the pressure smoothing traced by $u_0$ \citep{Nasir2016,Garzilli2019,Wu2019}

\begin{figure*}
\centering
%\addtocounter{figure}{-1}
\begin{subfigure}{.5\textwidth}
    \centering
    \includegraphics[scale=0.24]{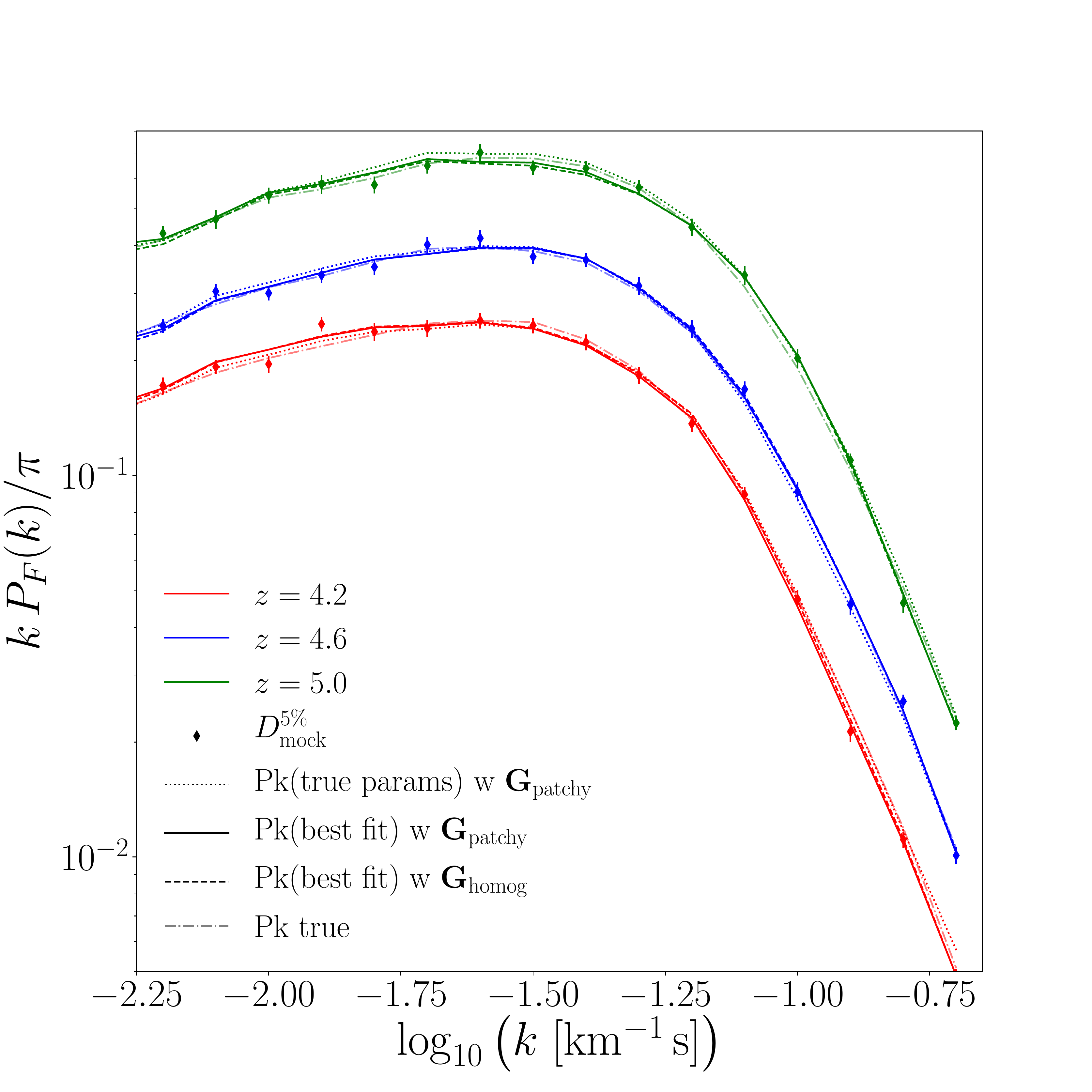}
    %\caption{$z=4.2$}
    %\label{fig:sub1}
\end{subfigure}%
\begin{subfigure}{.5\textwidth}
    \centering
    \includegraphics[scale=0.24]{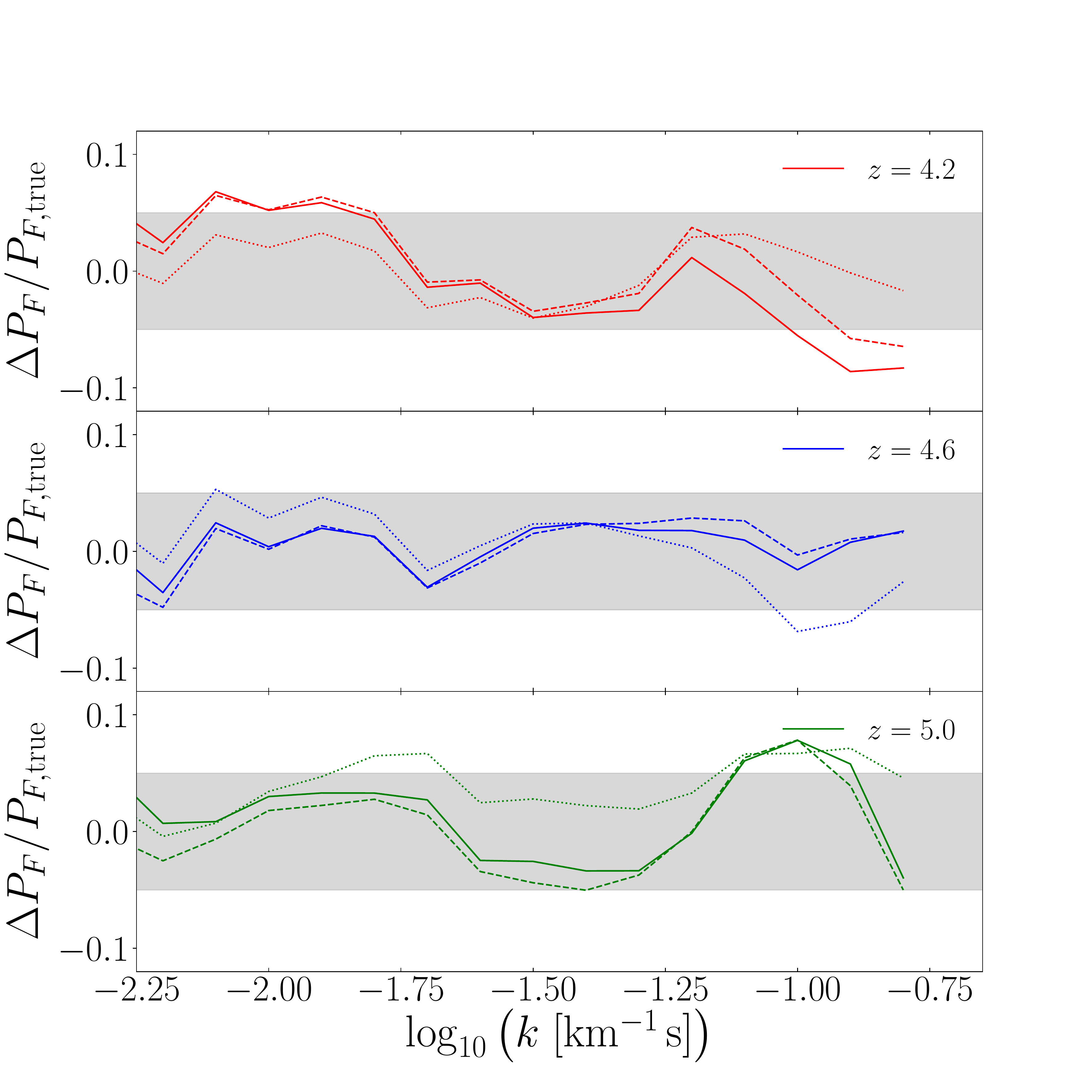}
    %\caption{$z=4.6$}
    %\label{fig:sub2}
\end{subfigure}
\caption{{\it Left panel:} the best-fit $P_{\rm F}(k)$ models for mocks with 5 per cent relative errors.  The points with error bars show the $D_{\rm mock}^{5\%}$ mock data, while the curves show the true underlying power from the RT-late simulation (dot-dashed) and the best-fit models with (\Grecov{}, solid) and without (\Ghomog{}, dashed) the patchy correction applied to the models.  The best-fit $\chi^2$/d.o.f. values are 30.2/36 and 32.0/36 for \Grecov{} and \Ghomog{} respectively.  For comparison, we also show the emulator prediction at the true parameter values (dotted).
{\it Right panel:} The residuals with respect to the true underlying model. The grey shading shows the range corresponding to a 5 per cent uncertainty.}
\label{fig:pk_bestfit}
\end{figure*}

For comparison, the best fit model for \Ghomog{} (blue contours) has the $\chi^2$ value of 32.0 for 36 degrees of freedom, but there are now up to $\sim 1\sigma$ shifts between the estimated and true parameters for \Ghomog{} (blue contours).  This implies that patchy reionisation effects should start to become important for \Lya forest power spectrum analyses at small scales, $k\sim 0.1\rm\,km^{-1}\,s$, if measurement uncertainties are at the $\sim 5$ per cent level.  At this level, however, several other effects may also become important in the future data sets, such as the accuracy of the power spectrum emulator \citep{Bird2019,Pedersen2021}, simulation initial conditions \citep{Bird2020}, and hydrodynamical methods \citep{Regan2007,Walther2021}.  Furthermore, the accuracy of the approximation for the patchy correction starts to approach (at most) the $5$ per cent level at the smallest scales (see Figure~\ref{fig:apply_bf_tobins}).  As an example, in the left panel of Fig.~\ref{fig:pk_bestfit} we show the best-fit power spectra obtained using \Grecov{} (solid) and \Ghomog{} (dashed) models for $D_{\rm mock}^{5\%}$ (i.e. the orange and blue contours in Fig.~\ref{fig:mcmc_5p_panels}).  These agree with the mock data very well  (recall their respective $\chi^2$ are $30.2$ and $32.0$ for \Grecov{} and \Ghomog{} respectively, for $36$ degrees of freedom). However the differences between the \emph{true power spectrum} and these two models is at around the same level as the difference between the true power spectrum and the emulator predicted power at the \emph{true parameter values} (the right panel of Fig.~\ref{fig:pk_bestfit} shows the ratio of the true power spectrum and emulated power spectrum for the true parameter values).  This implies that the emulator accuracy would need to be improved to analyse data with much less than a 5 per cent relative error on the power spectrum. 
  
\section{Conclusions}
\label{section::conclusions}

In this work we use simulations drawn from the Sherwood-Relics project (Puchwein et al. in prep) to model the effect of inhomogeneous reionisation on the 1D power spectrum of the \Lya forest transmitted flux in the  post-reionisation IGM at $4.2\leq z \leq 5$.  We use a novel hybrid approach for combining cosmological radiative transfer with hydrodynamical simulations that captures the patchy ionisation and thermal state of the IGM \emph{and} self consistently models the hydrodynamical response of the gas.  Our primary conclusions are summarised as follows:

\begin{itemize}
    \item Inhomogeneous reionisation suppresses the power spectrum at small scales, $k \sim 0.1$ \kms{},  by as much as $10$--$15$ per cent.  This is broadly consistent with \citet{Wu2019}, who used fully self-consistent radiation hydrodynamics simulations.  We note, however, that our hybrid-RT approach allows us to provide a factor of $\sim 13$ improvement in mass resolution and a factor of $\sim 4$ in volume relative to \citet{Wu2019}. We find this small-scale suppression is primarily driven by thermal broadening associated with the higher gas temperatures in the recently reionised regions, combined with the divergent peculiar velocity field that arises as the IGM dynamically responds to the increase in gas pressure.\\
   
    \item On larger scales, $k<0.03\rm\,km^{-1}\,s$, we also recover the (now well established) result that inhomogeneous reionisation boosts the power spectrum by $10$--$50$ per cent \citep[see also][]{Cen2009,Keating2018,Daloisio2019,Onorbe2019,Wu2019,Montero2020}.  This arises from the large scale variations in the IGM neutral fraction due to spatial fluctuations in the gas temperature, where gas that has been reionised recently is hotter and has less time to cool. The large scale variations in the neutral fraction, $x_{\rm HI}$, then arise because of the temperature dependence of the \HII recombination coefficient, where $x_{\rm HI}\propto T^{-0.72}$ for gas in ionisation equilibrium.  The boost to the large scale power increases with increasing redshift and/or for a later end redshift for reionisation. This is consistent with temperature fluctuations that gradually fade once reionisation is complete, since the low density IGM will adiabatically cool toward the thermal asymptote.\\ 
    
    \item  We use our hybrid-RT simulations to provide a ``patchy reionisation'' correction to the \Lya forest power spectrum on scales $-2.9\leq \log(k/\rm\ km^{-1}\,s)\leq -0.7$.  We provide a look-up table that enables this correction to be straightforwardly applied to existing grids of hydrodynamical simulations used in cosmological parameter inference frameworks at $4.1 \leq z \leq 5.4$ (see Table~\ref{table:tabulated_pk} and Eq.~(\ref{eqn:best_fit})). This correction has been derived from simulations that assume $5.3\leq z_{\rm R}\leq 6.7$.  We therefore caution against applying this correction to reionisation histories that differ significantly from those considered here. This correction reproduces our simulation results to within $1$ ($5$) per cent at scales $k\leq 0.03\rm\,km^{-1}\,s$ ($k\geq 0.03\rm\,km^{-1}\,s$). \\

    \item We use Bayesian parameter inference to assess the importance of any biases due to inhomogeneous reionisation for measurements of the IGM thermal state from the \Lya forest power spectrum.  We apply our patchy reionisation correction to a parameter grid of simulations with a homogeneous UV background using an MCMC analysis.   Over the scales probed by the \citet{Boera2019} power spectrum measurements,  $-2.2\leq \log(k/\rm km^{-1}\,s)\leq -0.7$, inhomogeneous reionisation does not introduce any significant bias in the inferred IGM parameters for mock data with a relative measurement uncertainty of $\sim 10$ per cent.  However, for relative uncertainties of $\sim 5$ per cent  -- as might be achieved by future observations --  there are  $\sim 1\sigma$ shifts between the estimated and true parameters due to the effect of inhomogeneous reionisation on the \Lya forest power spectrum.  Note, however, that several other effects will also become important at the $\sim 5$ per cent level, including the accuracy of the power spectrum emulator, simulation initial conditions, or the choice of hydrodynamics solver.\\
    
    \end{itemize}

In summary, this study demonstrates that the effect of inhomogeneous reionisation should be accounted for in future analyses of high precision measurements of the \Lya forest power spectrum that extend to large scales at $z>4$.  For this purpose, we have provided a look-up table that gives our (model dependent) reionisation correction for the \Lya forest power spectrum at $4.1\leq z \leq 5.4$ in a form that can be easily applied within other parameter inference frameworks that use similar models for reionisation.  Fortunately, we also find that hydrodynamical simulations that use a homogeneous UV background are still well suited to predicting the \Lya forest power spectrum at $4.2\leq z \leq 5$ on scales of $-2.2\leq \log(k/\rm km^{-1}\,s)\leq -0.7$, given the \emph{current} uncertainties ($\sim 10$ per cent) on high resolution ($R\sim 40,\,000$) observational data \citep{Boera2019}.  This is in good agreement with the earlier work of \citet{Wu2019}, who noted that inhomogeneous reionisation effects should not strongly bias existing constraints on alternative dark matter scenarios from the \Lya forest.  However, this picture will change if the uncertainties on the power spectrum measurements approach the $\sim 5$ per cent level at the scales that are most sensitive to the IGM thermal history or coldness of the dark matter, $k\sim 0.1\rm\,km^{-1}\,s$. This will also be the case for analyses of the \Lya forest power spectrum on larger scales, $k\leq 0.003\rm\,km^{-1}\,s $.  Large-scale spectroscopic surveys that will obtain huge numbers of low resolution spectra ($R\leq 5000$) such as DESI \citep{VargasMagana2019} or WEAVE-QSO \citep{Pieri2016} will push the power spectrum measurements to larger scales at $z\simeq 3$--$4$. These data will be sensitive to the remnant patchy heating associated with hydrogen reionisation \citep[e.g.][]{Cen2009,DAloisio2018fluc,Onorbe2019,Montero2020}.  Our results provide a first step toward developing a simple template for incorporating this into the parameter inference frameworks that will be used in forthcoming analyses of the \Lya forest power spectrum.

\section*{Acknowledgements}

The authors would like to thank the anonymous referee for very helpful comments and suggestions. The simulations used in this work were performed using the Joliot Curie supercomputer at the Tré Grand Centre de Calcul (TGCC) and the Cambridge Service for Data Driven Discovery (CSD3), part of which is operated by the University of Cambridge Research Computing on behalf of the STFC DiRAC HPC Facility (www.dirac.ac.uk).  We acknowledge the Partnership for Advanced Computing in Europe (PRACE) for awarding us time on Joliot Curie in the 16th call. The DiRAC component of CSD3 was funded by BEIS capital funding via STFC capital grants ST/P002307/1 and ST/R002452/1 and STFC operations grant ST/R00689X/1.  This work also used the DiRAC@Durham facility managed by the Institute for Computational Cosmology on behalf of the STFC DiRAC HPC Facility. The equipment was funded by BEIS capital funding via STFC capital grants ST/P002293/1 and ST/R002371/1, Durham University and STFC operations grant ST/R000832/1. DiRAC is part of the National e-Infrastructure.   MM and JSB are supported by STFC consolidated grant ST/T000171/1. VI is supported by the Kavli foundation.  MGH acknowledges support from the UKRI STFC (grant numbers ST/N000927/1 and ST/S000623/1). MGH and PG acknowledge support from the UKRI STFC (grant number ST/N000927/1). GK is partly supported by the Department of Atomic Energy, Government of India (research project RTI 4002), and by the Max-Planck-Gesellschaft through a Max Planck Partner Group. LCK was supported by the European Union’s Horizon 2020 research 
and innovation programme under the Marie Skłodowska-Curie grant agreement 
No. 885990. We thank Volker Springel for making \textsc{P-GADGET-3} available. We also thank Dominique Aubert for sharing the ATON code.

\section*{Data Availability}

All data and analysis code used in this work are available from the first author on reasonable request.  An open access preprint of the manuscript will be made available at arXiv.org.

\bibliographystyle{mnras}
\bibliography{references}

\appendix

\section{Illustration of the effect of physical quantities on \Lya absorption features} \label{app:example_spec}

\begin{figure*}
\includegraphics[width=0.85\textwidth,trim = 0 116 -0.2 80]{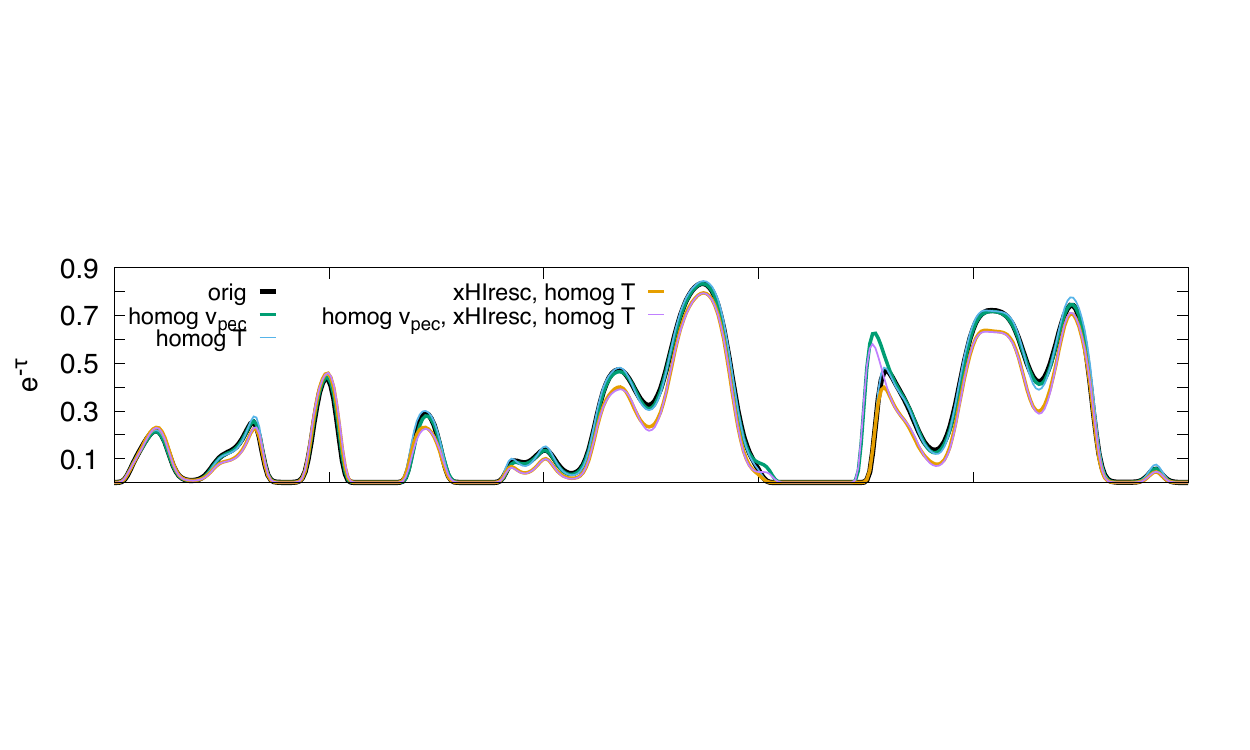}
\includegraphics[width=0.85\textwidth,trim = 4.9 114 0 40]{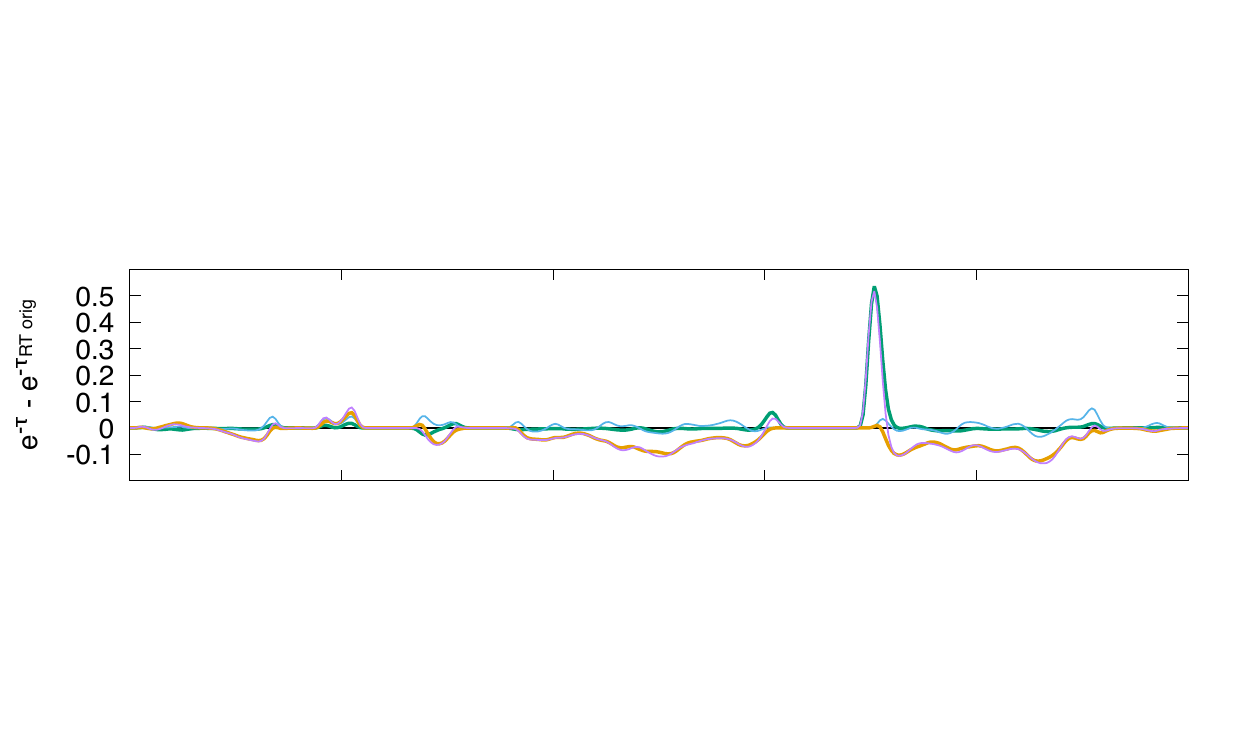}
\includegraphics[width=0.85\textwidth,trim = 0 115 -0.2 40.8]{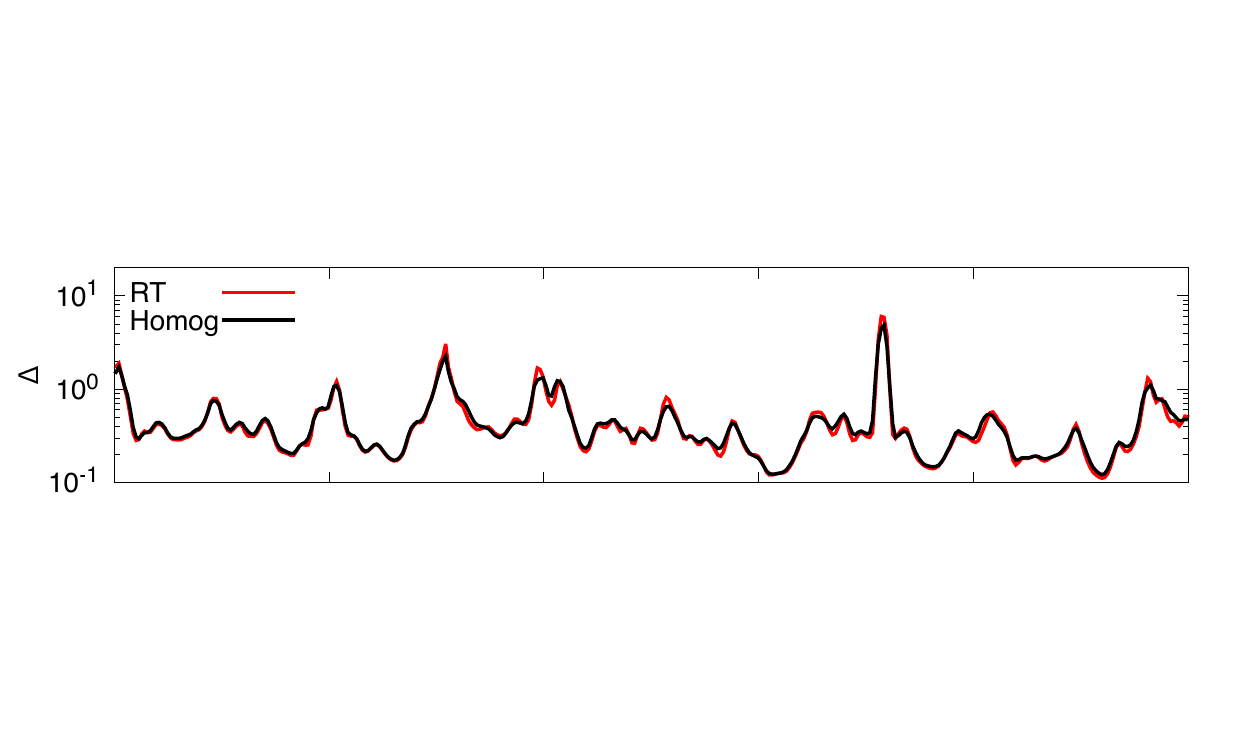}
\includegraphics[width=0.85\textwidth,trim = -4.9 115 -0.5 38.8]{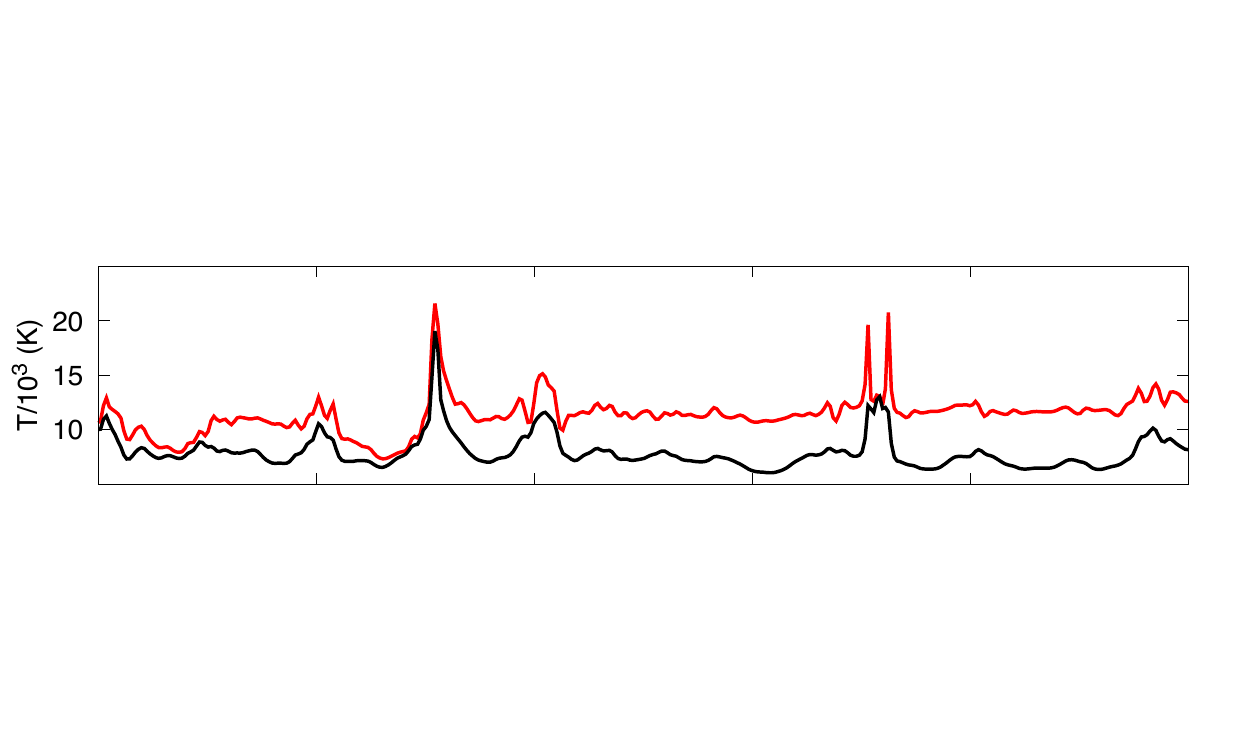}
\includegraphics[width=0.85\textwidth,trim = 0 115 -0.2 40]{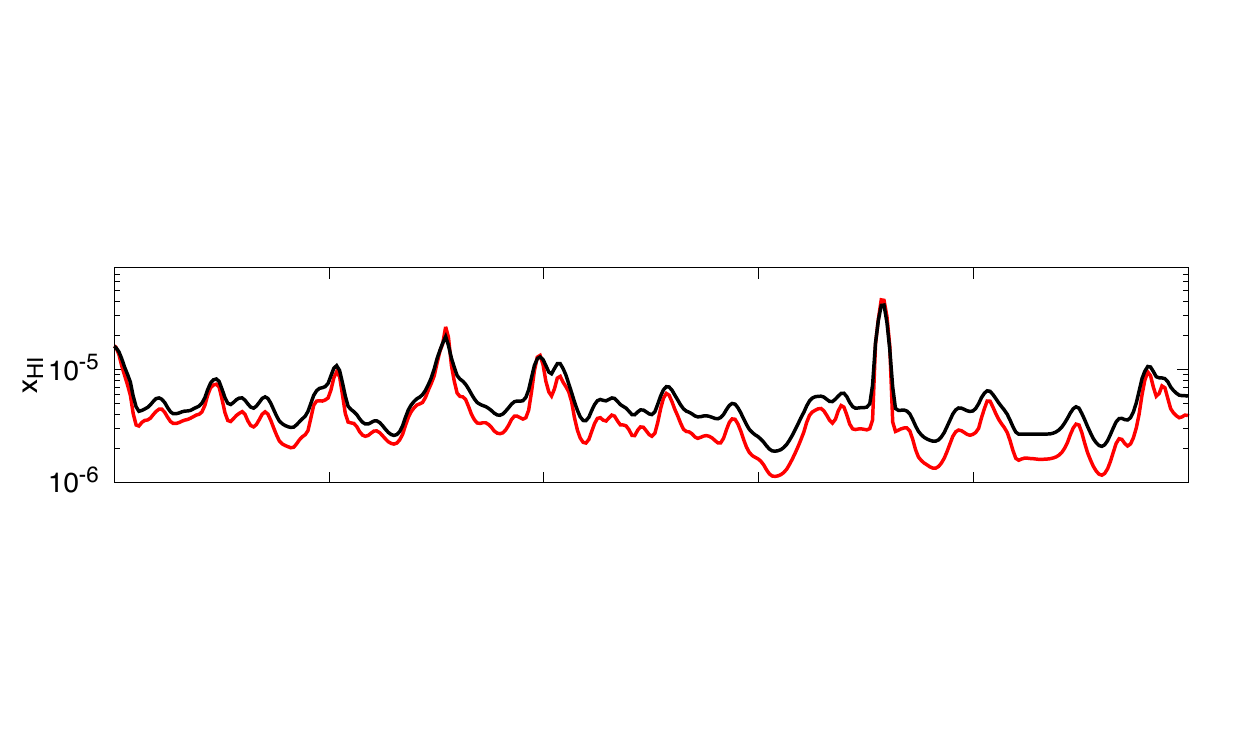}
\includegraphics[width=0.85\textwidth,trim = -5 103 -0.3 39]{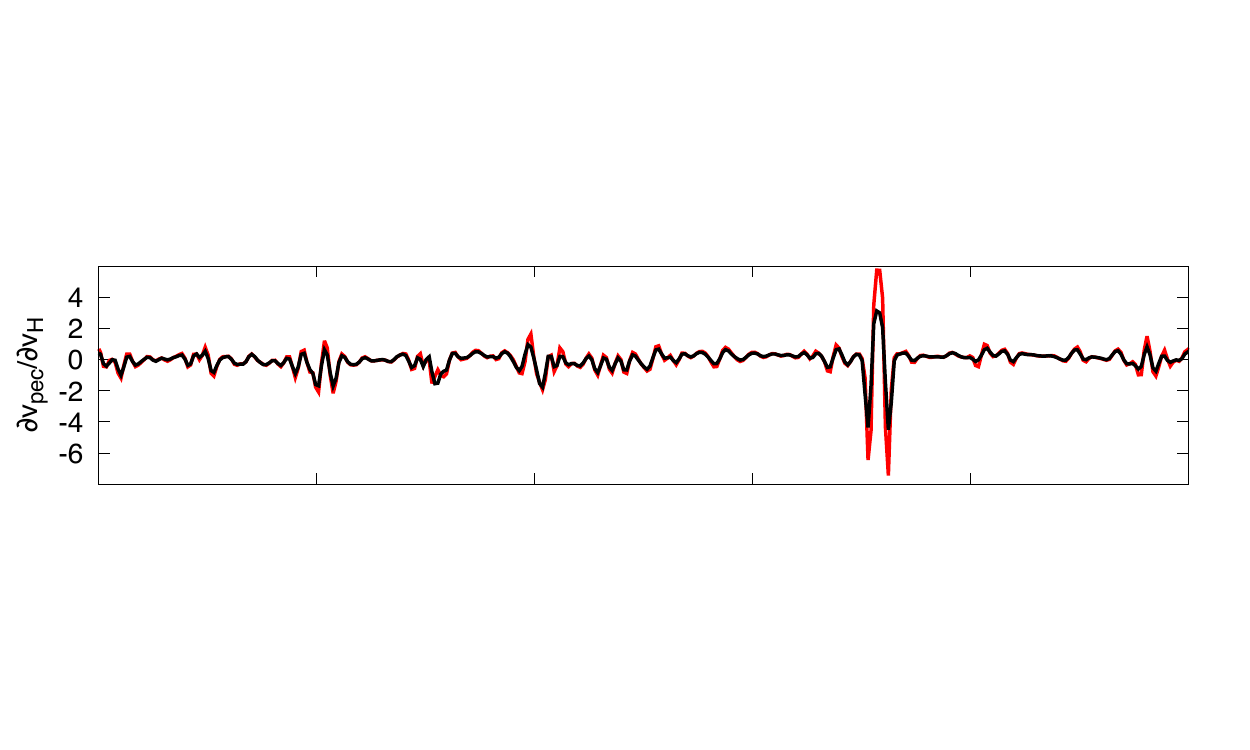}
\includegraphics[width=0.85\textwidth,trim = 4.9 70 0.3 40]{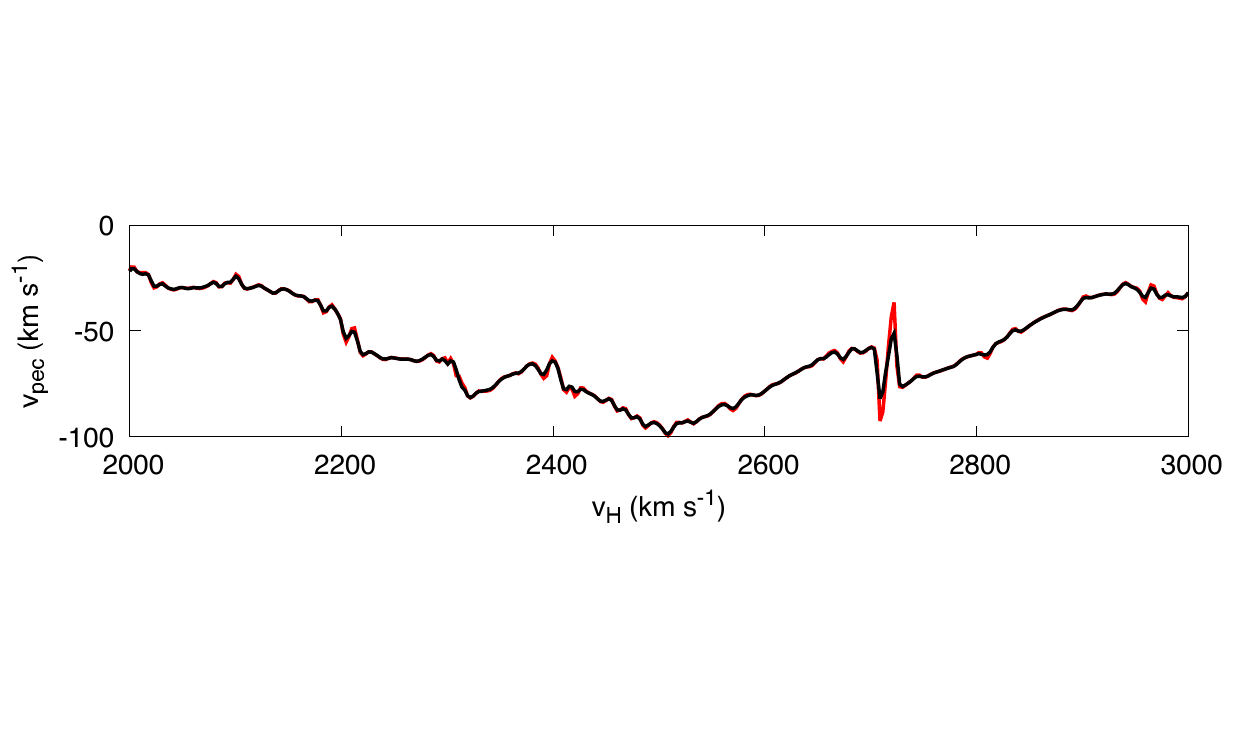}
\caption{\textit{Top panel}: The \Lya transmission along a section of a single line of sight for the cases shown in Fig.~\ref{figure:Ratio_Lyalpha_PS_wboot} at redshift $z=5$ (note the line styles used here also match those used in Fig.~\ref{figure:Ratio_Lyalpha_PS_wboot}).   The original spectrum from the RT-mid simulation is shown by the black curve.  \textit{Second panel}: the difference in the \Lya transmission between these cases and the original RT-mid spectrum for the same line of sight.  The remaining panels show the normalised gas density $\Delta$, gas temperature $T$, neutral fraction $x\sm{HI}$, peculiar velocity derivative $\VpecDx$, and peculiar velocity $v\sm{pec}$ profiles along the same segment for the RT-mid (red curves) and Homog-mid (black curves) simulations.}
\label{figure:los_one}
\end{figure*}

In Section \ref{section:changes_Lya_forest} we explored the effect that different physical quantities play in setting the shape of the ratio $R(k,z)=P_{\rm RT}(k,z)/P_{\rm homog}(k,z)$ by progressively replacing quantities in the RT-mid simulation with values from the paired Homog-mid simulation  (see Fig. \ref{figure:Ratio_Lyalpha_PS_wboot}).  Here we illustrate how this procedure affects the transmitted \Lya flux along a single line of sight. 

In the uppermost panel of Fig.~\ref{figure:los_one}, we show the \Lya transmission along a 1000 km s$^{-1}$ section of a single line of sight at redshift $z=5$ for the cases considered in Fig. \ref{figure:Ratio_Lyalpha_PS_wboot}.  The second panel shows the difference between each case and the original spectrum from the simulation (i.e. the black curve in the upper panel), where the line styles match those used in Fig.~\ref{figure:Ratio_Lyalpha_PS_wboot}. We also show the corresponding normalised density $\Delta$, gas temperature $T$, neutral hydrogen fraction $x\sm{HI}$, peculiar velocity derivative $\VpecDx$, and peculiar velocity $v\sm{pec}$, for both the RT-mid (red curves) and Homog-mid (black curves) models.  Note that, as was the case for Fig. \ref{figure:Ratio_Lyalpha_PS_wboot}, the mean transmission for all $5000$ lines of sight drawn from the simulations is rescaled to match the effective optical depth given by Eq.~(\ref{eqn:tau_trams}). The mean transmission along \textit{individual} lines of sight, however, can still vary around the mean of the ensemble.  Hence, the spectra displayed in Fig.~\ref{figure:los_one} can have a different mean transmission over the $1000\rm\,km\,s^{-1}$ range displayed.

In a few locations along the line of sight (e.g. from $2400\rm\,km\,s^{-1}$ to $2900\rm\,km\,s^{-1}$), there is a coherent decrease in the transmission for the cases that include a rescaling of the neutral hydrogen fraction, $x_{\rm HI}$, to account for differences in the temperature dependent recombination rate between the RT-mid and Homog-mid models (i.e. the purple and orange curves in Fig.~\ref{figure:los_one}).  As can be seen in the lower panels, these correspond to regions where the IGM is hotter - and therefore more highly ionised - in the RT-mid simulation.  When rescaling the $x\sm{HI}$ to correspond to the lower temperature in the Homog-mid simulation, the  substitution produces an increase in $x\sm{HI}$ since $x_{\rm HI}\propto T^{-0.7}$, leading to more absorption. The extended nature of these hot regions in the hybrid-RT runs, and the resulting boost to the transmission due to the more highly ionised hydrogen within them, leads to the increased power at large scales  (low-$k$ values) in the hybrid-RT simulations \citep[see also][]{Keating2018,Onorbe2019,Wu2019}.

On the other hand, when substituting the peculiar velocities (i.e. the green and purple curves) the largest visible differences in the transmission occur at $\sim 2700\rm\,km\,s^{-1}$, corresponding to a region where there is an enhancement in the divergence of the peculiar velocity field the RT-mid simulation.  This arises in regions that have been recently heated and are expanding (see Fig.~\ref{figure:delta_T_dvpec_zr60_distr}). The change in the peculiar velocities smooths the absorption and produces a localised decrease in the \Lya transmission.  This effect contributes toward the suppression of the power spectrum at small scales (high-$k$) seen Fig. \ref{figure:Ratio_Lyalpha_PS_wboot}.

Finally, the effect of differences in the thermal broadening kernel on the \Lya transmission is shown by the blue curve in Fig.~\ref{figure:los_one}.  The colder gas temperatures in the Homog-mid simulation results in less thermal broadening and hence slightly sharper absorption features compared to the original RT-mid spectrum (black curve).  Hence, the hot gas associated with large-scale temperature fluctuations can also suppress power on small scales, although to a lesser extent than the peculiar velocity field.   Note that the differences in the gas density, $\Delta$, due to spatial variations in the pressure smoothing scale are typically small.

\end{document}